\documentclass[floats,floatfix,amssymb,prd,twocolumn,superscriptaddress,nofootinbib]{revtex4-1}

\usepackage{subcaption}
\usepackage{ragged2e}
\DeclareCaptionJustification{justified}{\justifying}
\makeatletter

\newcommand{\subsetsim}{\mathrel{\mathpalette\subset@sim\relax}}
\newcommand{\subset@sim}[2]{%
  \vtop{\offinterlineskip\m@th
    \ialign{\hfil##\cr
      $#1\subset$\cr\noalign{\kern0.5pt}\scalebox{0.9}{$#1\sim$}\cr
    }%
  }%
}
\makeatother

\usepackage{amssymb,amsmath,verbatim,mathtools,needspace,enumitem,etoolbox,graphicx,physics,microtype,afterpage,bm}
\usepackage[dvipsnames, usenames]{xcolor}
\definecolor{linkcolor}{rgb}{0.0,0.3,0.5}
\usepackage{booktabs}

\definecolor{oucrimsonred}{rgb}{0.6, 0.0, 0.0}
\definecolor{persianblue}{rgb}{0.11, 0.22, 0.73}
\definecolor{forestgreen}{rgb}{0.13,0.35,0.13}
\usepackage[unicode, 
colorlinks=true, 
linkcolor=persianblue, 
citecolor=forestgreen, 
filecolor=persianblue,
urlcolor=oucrimsonred, 
pdfusetitle]{hyperref}

\usepackage[all]{hypcap}
\usepackage[T1]{fontenc}
\usepackage[utf8]{inputenc}
\usepackage{tabularx}
\usepackage{adjustbox}
\usepackage{float}
\usepackage{ulem}
\usepackage{xfrac}

\definecolor{azure}{rgb}{0.0, 0.5, 1.0}
\definecolor{VioletRed4}{rgb}{0.55, 0.13, .32}

\usepackage{multirow}
\usepackage{pifont}
\usepackage{fontawesome}
\usepackage{lmodern}

\usepackage{multirow}

\allowdisplaybreaks
\usepackage{tikz}
\usepackage{color}
\usepackage{framed}

\definecolor{rossos}{cmyk}{0,1,1,0.55}
\definecolor{bluscuro}{rgb}{0.15, 0.2, .85}
\definecolor{bluchiaro}{cmyk}{1,.3,0.,0.1}
\definecolor{ForestGreen}{rgb}{0.13, 0.55, 0.13}

\def\bea{\begin{eqnarray}}
\def\eea{\end{eqnarray}}

\def\d{{\mathrm{d}}}

\newcommand{\bs}{\begin{subequations}}
\newcommand{\es}{\end{subequations}}

\newcommand{\be}{\begin{equation}}
\newcommand{\ee}{\end{equation}}
\newcommand{\crm}{r_\mathrm{m}}
\newcommand{\cs}{c_\mathrm{s}}
\newcommand{\MH}{M_\mathrm{H}}
\newcommand{\RH}{R_\mathrm{H}}
\newcommand{\dc}{\delta_\mathrm{c}}
\newcommand{\ns}{n_\mathrm{s}}
\renewcommand{\d}{{\rm d}}
\newcommand{\rhob}{\rho_\mathrm{b}}

\newcommand{\lp}{\left (}
\newcommand{\rp}{\right )}

\def\lsim{\mathrel{\rlap{\lower4pt\hbox{\hskip0.5pt$\sim$}}
    \raise1pt\hbox{$<$}}}         
\def\gsim{\mathrel{\rlap{\lower4pt\hbox{\hskip0.5pt$\sim$}}
    \raise1pt\hbox{$>$}}}         

\makeatletter
\def\l@subsubsection#1#2{}
\makeatother

\newcommand{\infn}{INFN, Sezione di Roma, Piazzale Aldo Moro 2, 00185, Roma, Italy}

\begin{document}

\title{
Primordial black hole formation during the QCD phase transition:\\ 
threshold, mass distribution and abundance
}

\author{Ilia Musco}
\affiliation{\infn}

\author{Karsten Jedamzik}
\affiliation{Laboratoire Univers et Particules de Montpellier (LUPM),
Université de Montpellier (UMR-5299) CNRS, Place Eugène Bataillon
F-34095 Montpellier Cedex 05, France}

\author{Sam Young}
\affiliation{Instituut-Lorentz for Theoretical Physics, Leiden University,\\Niels Bohrweg 2, 2333 CA Leiden, The Netherlands}


\begin{abstract}
Primordial black hole (PBH) formation during cosmic phase transitions and annihilation periods, such as the QCD transition or the $e^+e^-$-annihilation, 
is thought to be particularly efficient due to a softening of the equation of state. We present a detailed numerical study of PBH formation during the
QCD epoch in order to derive an accurate PBH mass function. We also briefly consider PBH formation during the $e^+e^-$-annihilation epoch. Our
investigation confirms that, for nearly scale-invariant spectra, PBH abundances on the QCD scale are enhanced by a factor $\sim 10^3$ compared to a
purely radiation dominated Universe. For a power spectrum producing an (almost) scale-invariant PBH mass function outside of the transition, we find a 
peak mass of $M_{\rm pbh}\approx 1.9 M_{\odot}$ with a fraction $f\approx 1.5\times 10^{-2}$ of the PBHs having a mass of $M_{\rm pbh} > 10 M_{\odot}$, 
possibly contributing to the LIGO-Virgo black hole merger detections. We point out that the physics of PBH formation during the $e^+e^-$-annihilation epoch 
is more complex as it is very close to the epoch of neutrino decoupling. We argue that neutrinos free-streaming out of overdense regions may actually hinder
PBH formation.
\end{abstract}

\maketitle


\normalem

\section{Introduction}
The LIGO-Virgo collaboration joined later by KAGRA (LVK) \cite{LIGOScientific:2018mvr,LIGOScientific:2020ibl,LIGOScientific:2021djp}
has by now detected a large number ($\sim 90$) of black hole-black hole and neutron star-black hole mergers via the observation of 
gravitational wave emission during the final stages of coalescence. A few of these observed events fall into mass gaps, such as 
GW190814 where it was priorly predicted to not have any astrophysical candidates. There has been no detection so far of mergers 
with at least one member of the binary having a mass well below the Chandrasekhar mass, the lower limit for very compact astrophysical objects. Such a detection would unambiguously point to a non-astrophysical object, most likely a primordial black hole. 

It is well known that mildly non-linear, horizon size cosmological perturbations could collapse and form an apparent horizon, 
i.e. a primordial black hole (PBH, hereafter)~\cite{Zeldovich:1967,Hawking:1971ei,Carr:1975qj} (for a review see to~\cite{Khlopov:2008qy}).
When such collapse occurs during radiation domination in the early Universe, the dynamics is characterised by a competition between 
self-gravity and pressure forces, and observes the physics of critical phenomena~\cite{Niemeyer:1997mt,Niemeyer:1999ak,Musco:2004ak}.
When pre-existing energy density perturbations, such as believed to emerge from inflationary scenarios, are feature-less and almost scale-invariant, 
as observed in CMBR satellite missions, the equation of state (EoS) during the PBH formation epoch plays a crucial role. It has been 
argued~\cite{Chapline:1975tn,Jedamzik:1996mr,Jedamzik:1998hc} that PBH formation during the QCD epoch would be particularly
efficient due to a softening of the equation of state. At the time of that work, the QCD phase transition was believed to be of first order.
Fully general relativistic numerical simulations of PBH formation confirmed that PBHs form more easily during the QCD epoch~\cite{Jedamzik:1999am}, 
leading to a pronounced peak of PBHs on the $\sim 1M_{\odot}$ scale. Although the simulations were performed under the assumption of a 
first order transition, it was argued in~\cite{Jedamzik:1996mr} that any softening of the equation of state, even during other epochs,
would lead to a preferred scale in the PBH mass function. With advances in lattice gauge simulations it was possible to derive the 
zero chemical potential QCD and electroweak equation of state with high precision~\cite{Borsanyi:2016ksw,Bhattacharya:2014ara}. 
This equation of state, was recently used in approximate analytic calculations to derive the putative PBH mass 
function~\cite{Byrnes:2018clq,Carr:2019kxo,Sobrinho:2020cco}. This mass function indeed has a very well developed peak at 
$M\approx 1M_{\odot}$ and broader shoulder around $M\sim 30M_{\odot}$ due to pion annihilation.

It has been shown by now that PBHs, in the mass range probed by LVK, may only contribute a small fraction $f_{\rm pbh}\ll 1$ to the 
cosmological dark matter. For Gaussian initial conditions it was initially claimed that for $f_{\rm pbh} = 1$ the predicted merger rate largely 
surpasses that observed by Ligo~\cite{Sasaki:2016jop}, than shown that the existence of PBH binaries in dense clusters may change this 
conclusion~\cite{Raidal:2018bbj,Jedamzik:2020ypm,Young:2020scc}, to finally establish that even the small fraction of PBH binaries which never enter a PBH 
cluster still overproduces the merger rate (see e.g. \cite{Hutsi:2020sol}). It has been recently shown \cite{Juan:2022mir} 
that even in the case of $f_{\rm pbh}\ll 1$ a sizable contribution
to the LVK events is ruled out, though authors~\cite{Franciolini:2022tfm} which use the results of the present paper come to a different result. Similarly, initially it was claimed that 
microlensing constraints on compact dark matter in the Milky Way halo would be evaded by PBHs being in dense 
clusters~\cite{Clesse:2016vqa,Calcino:2018mwh,Carr:2019kxo}, which had been subsequently shown to be 
incorrect~\cite{Petac:2022rio,Gorton:2022fyb}. For non-Gaussian initial conditions, where PBHs are immediately born into clusters
of unknown density and size, merger rates are not known, but a combination mostly of microlensing - and Lyman-alpha - constraints~\cite{DeLuca:2022uvz} 
has been recently claimed to rule out $f_{\rm pbh} = 1$. Nevertheless, it seems still possible that PBH mergers contribute in part to the LVK 
observed signal and several authors have investigated this \cite{Bird:2016dcv,Sasaki:2016jop,
Eroshenko:2016hmn,Wang:2016ana,Ali-Haimoud:2017rtz,Chen:2018czv,Raidal:2018bbj,
Liu:2019rnx,Hutsi:2019hlw,Vaskonen:2019jpv,Gow:2019pok,Wu:2020drm,DeLuca:2020qqa,Jedamzik:2020omx,Hall:2020daa,Wong:2020yig,Kritos:2020wcl,Franciolini:2021xbq,
Bavera:2021wmw}.

The LVK collaboration is expected to significantly increase the data base on mergers during the observational runs O4 and O5. Such an extended 
data base on the binary population may allow a more detailed comparison between a putative PBH binary population and the data. A meaningful 
comparison may only be obtained when detailed results of the PBH mass function are known. Such detailed mass functions, are dependent on the 
characteristics of the initial perturbations, but also on the exact evolution and the final PBH mass of individual fluctuations. Whereas analytical results 
had been taken before~\cite{Byrnes:2018clq,Carr:2019kxo,Sobrinho:2020cco}, an accurate mass function can only be obtained via the numerical 
simulation of radiation fluctuations leading to PBHs, which we treat in the present paper. 

The outline of the paper is as follows: In Section \ref{sec:EoS} we summarize
the computation of the equation of state, making an important
comment concerning PBH formation during the $e^+e^-$ annihilation.
In Section \ref{sec:maths} we describe the mathematical aspects, with a detailed description of the 
initial condition, of the numerical results obtained in Section \ref{sec:Numerical_results}. Then in Section \ref{sec:Mass_distribution} we 
compute the mass distribution and the abundance of PBHs during the QCD transition.
Finally in Section \ref{sec:conclusions} we summarise our results drawing conclusions.

While finalizing our work a paper ~\cite{Escriva:2022bwe} 
of very similar spirit than ours has appeared. In the Appendix~\ref{sec:comparison} we comment and explain the substantial differences between our results.

\section{Equation of state in the early Universe}
\label{sec:EoS}
In the early Universe, between the end of the inflationary era and matter-radiation equality, the temperature decreases with cosmic expansion and the matter goes through several transitions, characterised by a non negligible softening of the equation of state. These include the electroweak
transition at temperature $T\approx 100\,$GeV, periods of
quark annihilation, the
QCD confinement transition at $T\approx 100\,$MeV and $e^+e^-$
annihilation at $T\approx 500\,$keV.
The chemical potential $\mu$ in the calculations of~\cite{Borsanyi:2016ksw,Bhattacharya:2014ara} is taken to be zero. This is an excellent approximation in the context of the early Universe due to the smallness of the cosmic baryon-to-photon ration which applies that $\mu \sim 10^{-10}$.

\subsection{The QCD and the electroweak transition}
Through detailed lattice gauge calculations, taking account of
realistic finite quark masses, it has become possible to calculate 
the equation of state at zero chemical potential for
the QCD transition in the early Universe. The ratio $w$ between pressure $p$ and total energy density $\rho$ of the medium is given by
\begin{equation}
w(T) \equiv \frac{p}{\rho} = \frac{4g_{*,s}(T)}{3g_{*}(T)} - 1\, ,
\end{equation}
where the functions $g_{*}(T)$ and $g_{*,s}(T)$ are defined by 
\begin{equation}
   g_{*}(T) = \frac{30\rho}{\pi^2T^4} 
    \quad \textrm{and} \quad 
   g_{*,s}(T) = \frac{45s}{2\pi^2T^3}\, ,
\end{equation}
with $s$ being the entropy density. The square of the speed of sound
$\cs^2 = \partial p/\partial \rho |_s$
may be computed via
\begin{equation}
    \cs^2(T) = \frac{4(4g_{*,s} + Tg_{*,s}^{\prime})}
                    {3(4g_{*} + Tg_{*}^{\prime})} - 1\, ,
    \label{eq:EoS}
\end{equation} 
where a prime denotes a derivative with respect to temperature.

These lattice calculations show clearly that the QCD quark-to-hadron 
transition is not a phase transition but a cross 
over~\cite{Bhattacharya:2014ara,Borsanyi:2016ksw}. 
Ref.~\cite{Borsanyi:2016ksw} has added to these calculations results 
from the literature concerning the electroweak transition to provide 
the cosmic equation of state between 
$T \simeq 280\,$GeV and $T \simeq 1\,$MeV. 

\begin{figure}[t!]
\centering
\vspace{-1cm}
\includegraphics[width=0.5\textwidth]{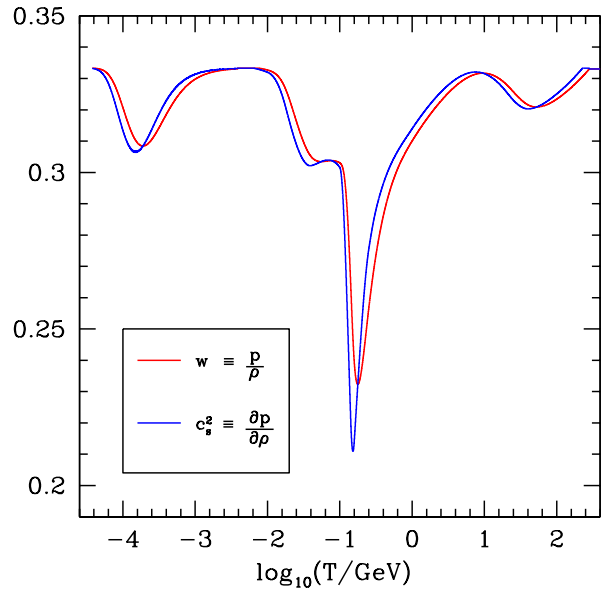}
\vspace{-2.3cm}
\caption{\label{fig:EOS} {\bf The EoS of the early Universe.}
The speed of sound squared, $\cs^2$ (blue)
and pressure over density, $w = p/\rho$ (red) are plotted as function of cosmic
temperature during the epochs of the electroweak transition, the QCD transition,
and the $e^+e^-$ annihilation.}
\end{figure}

\subsection{PBH formation during the $e^+e^-$ annihilation}
 We have extended the equation of state to include the $e^+e^-$ annihilation epoch: results 
 are shown in Fig.~\ref{fig:EOS}, showing the speed of sound squared $\cs^2$ and $p/\rho$. 
 It is clearly seen that $\cs^2$ and w drop each time the number of relativistic degrees of 
 freedom changes in the early Universe, with the change during the QCD transition, though 
 still small, the most pronounced.


It has been noted \cite{Jedamzik:1996mr,Carr:2019kxo}, 
that the decrease of $w$ and $\cs^2$
during the $e^+e^-$ annihilation epoch should lead to an enhancement of the abundance of PBHs on the \mbox{$M_{\rm pbh}\sim 10^5M_{\odot}$} scale. We argue here that this may not necessarily be the case, as the cosmic $e^+e^-$ annihilation epoch is quite different from the QCD epoch and the electroweak epoch. At temperatures $T\approx 1-2\,$MeV, shortly before the $e^+e^-$ annihilation, with a maximum reduction of $w$ at $T\approx 200\,$keV, neutrinos decouple from the
Universe. In particular, since there interactions with the rest of the plasma
freeze out, neutrinos can free-stream out of overdense regions, effectively
reducing the overdensity. One may schematically write for the overdensity
\begin{equation}
\frac{\delta\rho}{\rho} = \frac{\sum\limits_i \delta\rho_i}{\sum\limits_i \langle\rho_i\rangle} \, ,
\end{equation}
where brackets denote cosmic average and the index $i$ runs over particle species. For adiabatic perturbations one has 
$\delta\rho_i = K\langle\rho_i\rangle$ with $K$ a quantity independent of
species, such that $\delta\rho/\rho = K$. However, the free-streaming of the neutrinos destroy the adiabaticity of the perturbation since 
$\delta\rho_{\nu}\approx 0$.
Having $\langle\rho_{\nu}\rangle/\langle\rho_{\rm tot}\rangle = g_{\nu}/g_{\rm tot} = 5.25/10.75 \approx 0.5$ one may estimate that the original perturbation
has only approximately half of the overdensity after neutrino free-streaming.
On the other hand, the critical threshold for PBH formation will not reduce
by as much as a factor of two due to the $e^+e^-$ equation of state, such that
the argument would imply that formation of PBHs during any epoch after neutrino coupling is highly suppressed for scale invariant perturbation spectra. Seen from the
opposite point of view, even if neutrinos are initially homogeneous, the
density perturbation in photons and $e^+e^-$ would gravitationally attract
neutrinos. Since those neutrinos do not even exert pressure, a further 
reduction of w to $w\sim 0.33/2$ would occur, favoring PBH formation. We tend
to think the first argument is dominant, suppression of PBH formation.
However, only a dedicated simulation of PBH formation with a fluid and free-streaming component could definitely answer this question. Such a study
is beyond the scope of the current paper.

\section{Mathematical Formulation}
\label{sec:maths}
In this section we briefly review the mathematical formalism used to study PBH formation. For more details the reader is referred to~\cite{Musco:2018rwt}. Particular attention is given to the initial conditions used in Section \ref{sec:Numerical_results} for the numerical simulations, discussing the differences with respect to the standard case of a radiation dominated Universe.

\subsection{Curvature perturbation in comoving gauge}
PBHs are formed from the collapse of large-amplitude non-linear cosmological perturbations. 
In the standard scenario of adiabatic perturbations these are sourced by a geometrical term, i.e. the curvature perturbation $\zeta$, appearing as a perturbation in the Friedmann-Lemaitre-Robertson-Walker (FLRW) metric, written in the comoving uniform-density gauge as
\begin{equation}
    \mathrm{d}s^2 = -\mathrm{d}t^2 + a^2(t)\exp\left(2\zeta\right)\mathrm{d}\mathbf{x}^2,
    \label{eq:metric_FLRW}
\end{equation}
where $a(t)$ is the scale factor as a function of the cosmic time $t$. 

We are working in spherical symmetry, which is well justified in this context, because of the large amplitude of the energy density peaks collapsing into PBHs~\cite{bbks}. This allows us to consider a simple diagonal form of the 3+1 decomposition of the metric, following the Misner-Sharp-Hernandez (MSH) formulation~\cite{Misner:1964je}, based on the cosmic time metric
\begin{equation}
    \mathrm{d}s^2 = -A^2(r,t)\mathrm{d}t^2 + B^2(r,t)\mathrm{d}r^2 + 
    R^2(r,t) \mathrm{d}\Omega^2
    \label{eq:metric_MS}
\end{equation}
where the radial coordinate $r$ is taken to be comoving with the fluid\footnote{In the comoving gauge we are considering the four-velocity of the fluid as being equal to the unit normal vector orthogonal to the hypersurface of constant cosmic time $t$, namely $u^\mu = n^\mu$.}. 
The metric coefficient $R$ is the so called areal radius while \mbox{$\mathrm{d}\Omega^2 = \mathrm{d}\theta^2 + \sin^2\theta \,\mathrm{d}\phi^2$} is the element of a 2-sphere. 

In the MSH formalism (see Appendix \ref{sec:MSH} for more details) it is useful to introduce two differential operators:
 \begin{equation}
D_t \equiv \frac{1}{A} \frac{\partial}{\partial t}  
\ \ \ \ \textrm{and} \ \ \ \
D_r \equiv \frac{1}{B} \frac{\partial}{\partial r}  
\label{eq:operators} \ ,
\end{equation}
corresponding to derivatives with respect to proper time and radial proper distance. Applying these to the areal radius $R$ two additional quantities are defined:
 \begin{equation}
U \equiv D_t R = \displaystyle{\frac{1}{A} \frac{\partial R}{\partial t}}   
\ \ \ \ \textrm{and} \ \ \ \
\Gamma \equiv D_r R = \displaystyle{\frac{1}{B} \frac{\partial R}{\partial r}} 
\label{eq:U&Gamma} \ ,
\end{equation}
 with $U$ being the radial component of four-velocity in the ``Eulerian'' (non-comoving) frame,  measuring the velocity of the fluid with respect to the centre of the sphere, where $R$ is used as the radial coordinate. In the homogeneous and isotropic FLRW Universe $U$ is simply given by the Hubble law $U=HR$ where $R(r,t) = a(t)r$. The quantity $\Gamma$ instead gives a measure of the spatial curvature, and in FLRW one gets $\Gamma^2 = 1 - Kr^2$ where $K=0,\pm1$. 

In general $U$ and $\Gamma$ are related to the Misner-Sharp-Hernandez mass $M$ by an algebraic expression
 \begin{equation}
 \Gamma^2 = 1 + U^2 - \frac{2M}{R} \ ,
 \label{eq:Hamiltonian}
\end{equation}
corresponding to the Hamiltonian constraint. The quantity $M$ is  measuring the total mass contained within a sphere of radius $R$ 
\begin{equation}
    M = \int_0^R 4\pi \rho x^2\mathrm{d}x 
    = \frac{R}{2} \left( 1 - \nabla^\mu R \nabla_\mu R \right)
    \label{eq:MS_mass}
\end{equation}
where the second equality is an alternative covariant expression to define the Misner Sharp mass $M$.

\subsection{Gradient expansion}
\label{sec:gradient_expansion}
On superhorizon scales, when the length scale of the perturbation is much larger than the cosmological horizon, the curvature perturbation $\zeta$ for adiabatic perturbations is time independent, being only a function of the comoving coordinate $r$. In this regime the FLRW metric of equation \eqref{eq:metric_FLRW} is written as
\begin{equation}
    \mathrm{d}s^2 = -\mathrm{d}t^2 + a^2(t)\,e^{2\zeta(r)}
    \left[\mathrm{d}r^2 + r^2\mathrm{d}\Omega^2\right] 
    \label{eq:asymptotic}
\end{equation}
corresponding to the asymptotic limit, $t\to0$, of the cosmic time metric  \eqref{eq:metric_MS}. For our purpose, the comoving curvature perturbation $\zeta$ is chosen as initial condition, assumed to result from the dynamics of a prior inflationary epoch, or an equivalent phase generating a power spectrum of cosmological adiabatic perturbations. 

Using the definition of $\Gamma$ given by \eqref{eq:U&Gamma} one can compute the zeroth order of the Hamiltonian constraint 
\begin{equation}
    \Gamma = 1+r\zeta^\prime(r) \,,
\end{equation}
showing that in the super horizon regime the deviation from a spatially flat Universe is proportional to $r\zeta^\prime(r)$, consistent with the freedom of re-scaling the scale factor when adding a constant to the value of $\zeta$. In Section~\ref{sec:compaction_function} we define consistently how to measure the perturbation amplitude.

Using the gradient expansion approach~\cite{Shibata:1999zs,Tomita:1975kj,Salopek:1990jq,Polnarev:2006aa,Harada:2015yda}, it is possible to expand the MSH variables (see Appendix 
\ref{sec:MSH}) as power series of a small parameter $\epsilon \ll 1$, up
to the first non-zero order. In the expanding FLRW Universe $\epsilon$ is 
conveniently identified with the ratio between the Hubble radius $\RH$ and the 
length scale of the perturbation $R_\textrm{m}$
\begin{equation}
   \epsilon \equiv \frac{\RH}{R_\mathrm{m}} = \frac{1}{aH\tilde{r}_\mathrm{m}}
    \label{eq:epsilon}
\end{equation}
where $\tilde{r}_\mathrm{m}=\crm e^{\zeta(\crm)}$ is identified by the peak of the compaction function, defined later in Section~\ref{sec:compaction_function}. The comoving coordinate  $\tilde{r}\equiv re^{\zeta(r)}$ takes into account that the curvature profile enters in the asymptotic metric \eqref{eq:asymptotic} at zeroth order, modifying the comoving coordinate with respect to the FLRW solution.

The time evolution of the gradient expansion approach is equivalent to 
linear perturbation theory, but allows having a non linear amplitude of the 
curvature perturbations if the spacetime is sufficiently smooth on the 
scale of the perturbation (see~\cite{Lyth:2004gb}). This is equivalent to 
saying that pressure gradients are small when $\epsilon \ll 1$ and are not 
playing an important role in the evolution of the perturbation.

In this regime the energy density contrast $\delta\rho/\rhob$ for adiabatic perturbations can be written as~\cite{Yoo:2020dkz}
\begin{equation} 
\frac{\delta\rho}{\rho_b}(r,t) = - 
\frac{4}{3} \Phi
\left(\frac{1}{aH}\right)^2 e^{-5\zeta(r)/2} \nabla^2 e^{\zeta(r)/2},
\label{eqn:non-linear}
\end{equation}
where the function $\Phi(t)$ depends on the equation of state of the 
Universe and is obtained by solving 
the following equation
\begin{equation}
    \RH\frac{\d\Phi}{\d\RH} + \frac{5+3 w}{3(1+w)} \Phi - 1 = 0
    \label{eq:Phi}
\end{equation}
integrated from past infinity (i.e. $t,\RH \to 0$) to the time when the 
amplitude of the perturbation is computed. When the equation of state is 
characterised by a constant value $\bar w$, one has
${\d\Phi}/{\d t} = 0$, and the equation \eqref{eq:Phi} is analytically solved by
\begin{equation}\label{solPhi}
\bar \Phi = \frac{3(1+ \bar w)}{(5+3 \bar w)},
\end{equation}
yielding $\bar \Phi=2/3$ for a radiation fluid with $\bar w = 1/3$.
The value of $\Phi$, is measuring to which extent the curvature profile $\zeta$, appearing in the left hand side of Einstein equations, is affecting the stress energy tensor on the right hand side. Note that $\Phi=0$ if $w=-1$, consistent with the cosmological constant that cannot be perturbed. The function $\Phi$ is analogous to the coefficient appearing in the Bardeen potential\footnote{In cosmological linear perturbation theory, the Bardeen Potential $\Psi$ at super horizon scale is given by
$\Psi(r) = - \frac{3(1+w)}{5+3w}\zeta(r)$\,.}.

The full calculation of the gradient expansion, applied to the MSH equations of 
Appendix~\ref{sec:MSH}, gives the quasi-homogeneous solution summarised  in 
Appendix~\ref{sec:QHS}. This is used as initial conditions for the numerical simulations discussed later in Section~\ref{sec:Numerical_results}.  

\subsection{The perturbation amplitude}
\label{sec:compaction_function}
Determining whether a cosmological perturbation characterized by an overdensity is able to form a PBH depends on the perturbation amplitude $\delta$: if it is larger than a threshold $\dc$, an apparent horizon will form, satisfying the condition for the formation of a marginally trapped surface $R(r,t)=2M(r,t)$. From equation~\eqref{eq:Hamiltonian} this corresponds to the condition $\Gamma^2=U^2$, which has two possible solutions: 
\begin{itemize}
    \item $\Gamma = U$ : this is the condition for the cosmological horizon $\RH=1/H$, which is also an apparent horizon, within an expanding region of the FLRW Universe ($U>0$).
    \item $\Gamma = - U$ : this is the condition for the formation of the apparent horizon for a black hole, within a collapsing region ($U<0$).
\end{itemize}
The mathematical properties of a marginally trapped surface have been discussed in detail in~\cite{Helou:2016xyu}.

Focusing on the evolution of the collapse of a cosmological perturbation when the amplitude $\delta>\dc$, the gravitation potential overcomes the pressure gradients and an apparent horizon appears leading to the formation of a PBH. When $\delta<\dc$ instead, the perturbation is dispersed by pressure forces into the expanding universe. The perturbation amplitude $\delta$ is measured at the peak of the compaction function~\cite{Shibata:1999zs,Musco:2018rwt} defined as
\be
\label{a}
\mathcal{C} \equiv 2\frac{M(r,t)-M_\mathrm{b}(r,t)}{R(r,t)} \,,
\ee
where the numerator is given by the difference between the Misner-Sharp mass within a sphere of radius $R(r,t)$, and the background mass \mbox{$M_\mathrm{b}(r,t)=4\pi \rhob(r,t)R^3(r,t)/3$} within the same areal radius, 
but calculated with respect to a spatially flat FLRW metric. 

According to the gradient expansion approach, on superhorizon scales (i.e. $\epsilon \ll 1$) the compaction function is time independent, as is $\zeta(r)$, and is given by
\be \label{comp}
\mathcal{C}(r) = - \Phi r\zeta'(r) \left[ 2+r\zeta'(r) \right] \,.
\ee
As shown in \cite{Musco:2018rwt}, the comoving length scale of the perturbation is consistently defined by $r=\crm$, where the 
compaction function reaches its maximum (i.e. \mbox{$\mathcal{C}'(\crm) = 0$}), which gives
\bea
\label{rm_condition} 
\zeta'(\crm)+\crm\zeta''(\crm)=0 \,.
\eea

As shown explicitly in \cite{Musco:2018rwt}, the compaction function is related to the energy density profile by the integration of the energy density profile: 
\begin{equation}
    \mathcal{C}(r) = \frac{\tilde{r}^2}{\tilde{r}^2_\mathrm{m}} \delta(r,t_H) 
\end{equation}
where $t_H$ is defined as the cosmological horizon crossing time, when $aH\tilde{r}_\mathrm{m}=1$ ($\epsilon=1$). Although in this regime the gradient expansion approximation is not very accurate, and the horizon crossing defined in 
this way is only a linear extrapolation, this provides a well defined criterion for consistently measuring the amplitude of different 
perturbations, understanding how the threshold is varying because of the different initial curvature profiles (see~\cite{Musco:2018rwt} 
for more details).  

The overdensity $\delta(r,t)$ is defined as the mass excess of the energy density averaged over the spherical volume of radius $R(r,t)$, that is
\begin{equation}
    \delta(r,t) \equiv \frac{4\pi}{V} \int_0^R 
    \frac{\delta\rho}{\rhob} R^2 \d R \simeq 
    \frac{3}{\tilde{r}^3} \int_0^{\tilde{r}} \frac{\delta\rho}{\rhob} \tilde{r}^2 \d \tilde{r} \,.
    \label{eqn:density}
\end{equation}
 The second equality is obtained by neglecting the higher order terms in 
 $\epsilon$, approximating \mbox{$R \simeq a(t)\tilde{r}$}, which allows making a simple integration over the comoving volume of radius $r$. The amplitude $\delta$ of the perturbation is therefore defined as the excess of mass averaged over a spherical volume of radius $R_\mathrm{m}$ using a top-hat window function, computed at the time of the cosmological horizon crossing $t_H$. This quantity is equivalent to the peak amplitude of the compaction function measured on superhorizon scales. 

Looking at \eqref{comp} the perturbation amplitude $\delta$ can be written in terms of a variable with Gaussian statistics linearly related to the curvature perturbation \mbox{$\delta_l \equiv -2\Phi\, r\zeta^\prime(r_m)$} 
\be \label{eq:delta_l}
\delta =
\lp  \delta_l - \frac{1}{4 \Phi} \delta_l^2\rp .  
\ee
 This expression will be very useful later on when we are going to compute the mass distribution of PBHs in Section~\ref{sec:Mass_distribution}. The derivation of this formula relies on the assumption of spherical symmetry, which is generally taken to be the case for PBH formation - where the peaks that form PBHs are rare, and therefore expected to be close to spherically symmetric \cite{Bardeen:1985tr}. However, it has recently been considered that, whilst peaks in $\delta$ where PBHs form are high, and rare, this may not correspond to high peaks in $\zeta$ \cite{Young:2022phe}.

\section{Numerical results}
\label{sec:Numerical_results}
In this section we discuss the numerical results obtained with a numerical code developed
by one of us in the past \cite{Musco:2004ak} and abundantly used to study the formation of 
PBH formation when the Universe is radiation dominated~\cite{Musco:2008hv, Musco:2012au,Musco:2018rwt}. The code has been fully described previously and therefore just 
a very brief outline of it will be given here. 
The EoS to describe the QCD phase transition described as in Section~\ref{sec:EoS} has been implemented into the code using the relation between the energy density and the temperature, with $w$ and $\cs^2$ being now functions of the energy density $\rho$.

\subsection{Numerical scheme}
\label{sec:numerical_scheme}
The numerical scheme which we are using is a Lagrangian hydrodynamics code with 
the grid designed for calculations in an expanding cosmological background. The basic
grid uses logarithmic spacing in a mass-type comoving coordinate, allowing it to reach
out to very large radii while giving finer resolution at small radii.

The initial data follow from the quasi-homogeneous solution, fully described in 
Appendix~\ref{sec:QHS}, specified on a spacelike slice at constant initial cosmic time 
$t_\mathrm{i}$ with $a(t_\mathrm{i})\crm=10\RH$ ($\epsilon=10^{-1}$). The outer edge of the grid has 
been placed at $90\RH$, to ensure that there is no causal contact between the boundary 
and the perturbed region during the time of the calculations. The initial data is then 
evolved using the Misner-Sharp-Hernandez equations to generate a second set of initial data on a null slice which is then evolved using the Hernandez-Misner equations~\cite{Hernandez:1966zia} to follow the subsequent evolution leading to possible black hole formation. 

In this formulation, each outgoing null slice is labeled with a time coordinate $u$, which takes a constant value everywhere on the slice, and the formation of the apparent horizon is moved to $u\to\infty$ because of the increasing 
redshift of the null rays emitted by the collapsing shells. Moving along an outgoing null ray one has
\begin{equation}
    A\mathrm{d}t = B\mathrm{d}r \,,
\end{equation}
and the observer null time $u$ is defined as
\begin{equation}
    F\mathrm{d}u = A\mathrm{d}t - B\mathrm{d}r 
    \label{eq:F}
\end{equation}
where $F$ needs to be determined from the integration of equation~\eqref{eq:F}. In terms of this, the so called {\it observer time} metric, which is no longer diagonal, becomes 
\begin{equation}
    \mathrm{d}s^2 = - F^2 \mathrm{d}u^2 - 2FB \mathrm{d}r \mathrm{d}u + R^2 \mathrm{d}\Omega^2
\end{equation}
The operators equivalent to equation~\eqref{eq:operators} are replaced by
     \begin{equation}
D_t \equiv \frac{1}{F} \frac{\partial}{\partial u}  
\ \ \ \ \textrm{and} \ \ \ \
D_k \equiv \frac{1}{B} \frac{\partial}{\partial r} = D_t + D_r
\label{operators} \ ,
\end{equation}
where $D_k$ is the radial derivative in the null slice. 

\begin{figure}[t!]
\centering
\vspace{-1.cm}
\includegraphics[width=0.5\textwidth]{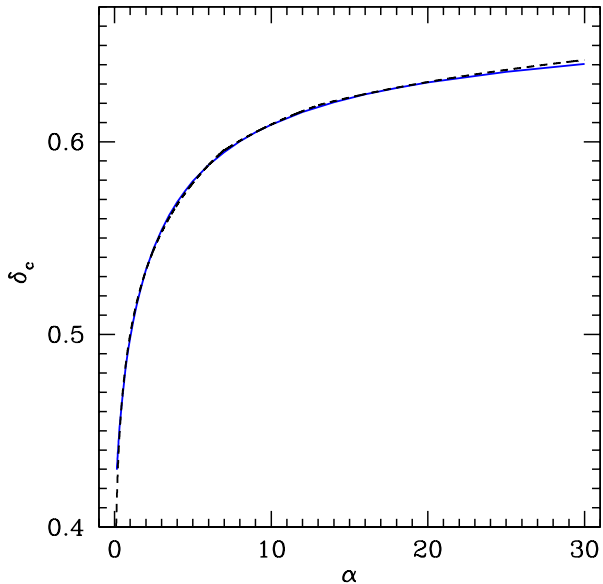}
\vspace{-2.3cm}
\caption{{\bf The threshold for PBHs during the radiation dominated Universe.} These plots show the behavior of the threshold $\dc$ when $w=1/3$, as functions of the shape parameter $\alpha$. The blue line represents the numerical results while the analytic fit given by \eqref{eq:delta_c_analytical_Musco} is plotted with a dashed line.}
\label{fig:delta_c_rad}
\end{figure}

During the evolution in the observer time coordinate (see~\cite{Musco:2004ak} for more details about the equations) the grid is modified with an adaptive mesh refinement scheme (AMR), built on top of the initial logarithmic grid, to provide sufficient resolution to follow black hole formation down to extremely small values of $\delta-\dc$.

In this reference frame the formation of the apparent horizon is obtained asymptotically for $u\to\infty$ synchronized with  with the condition $R(r,t)\to 2M(r,t)$, avoiding the formation of the singularity inside the black hole. In practice the simulations are stopped when $(1-2M/R) < 10^{-3}$, which allows to compute the mass of the black hole with very good accuracy, because the further evolution is negligible. For more details about the properties of apparent horizons one can see~\cite{Helou:2016xyu}.

\begin{figure*}[t!]
	\centering
	\vspace{-1cm}
	\includegraphics[width=0.495\textwidth]{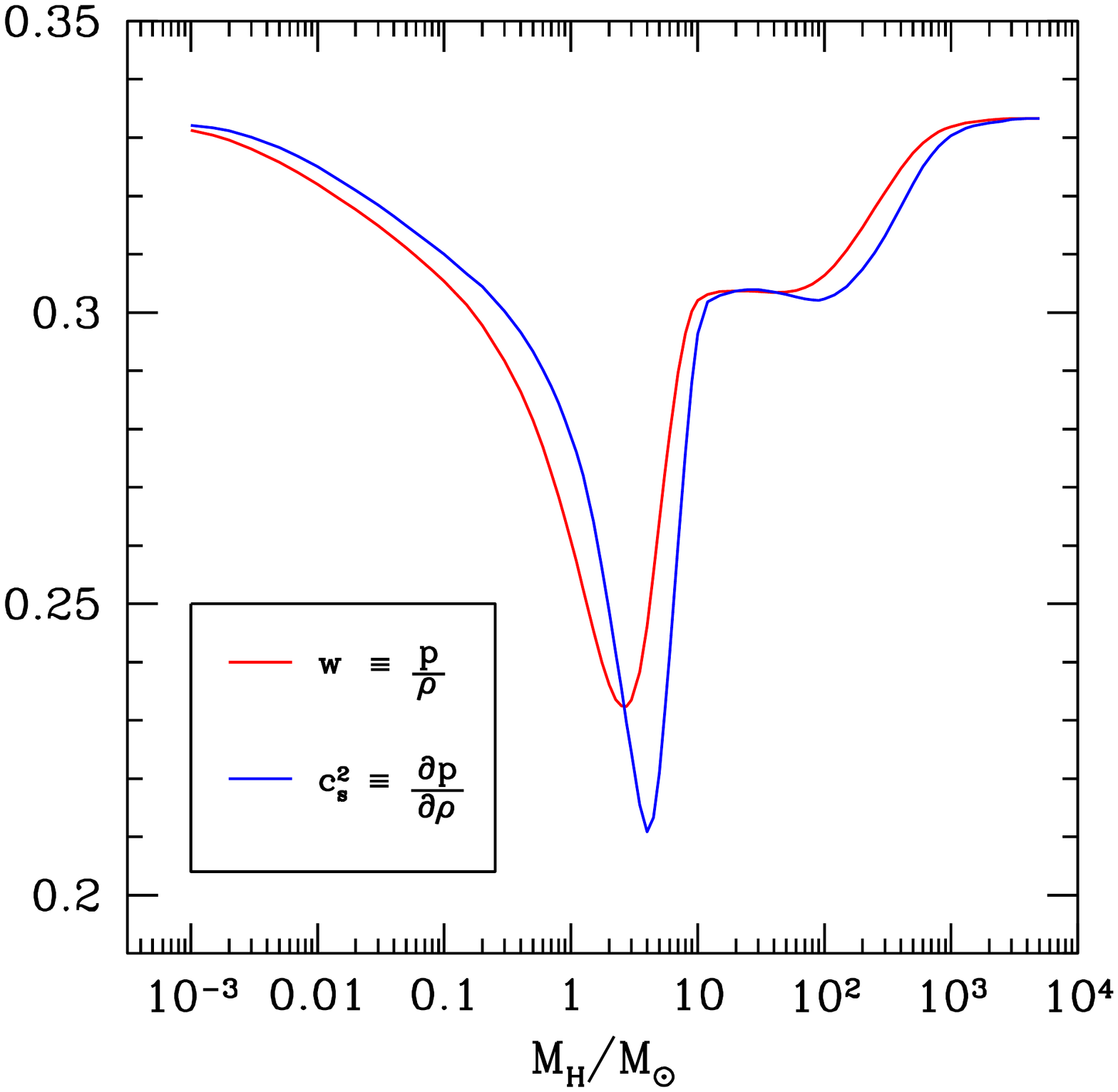}
	\includegraphics[width=0.495\textwidth]{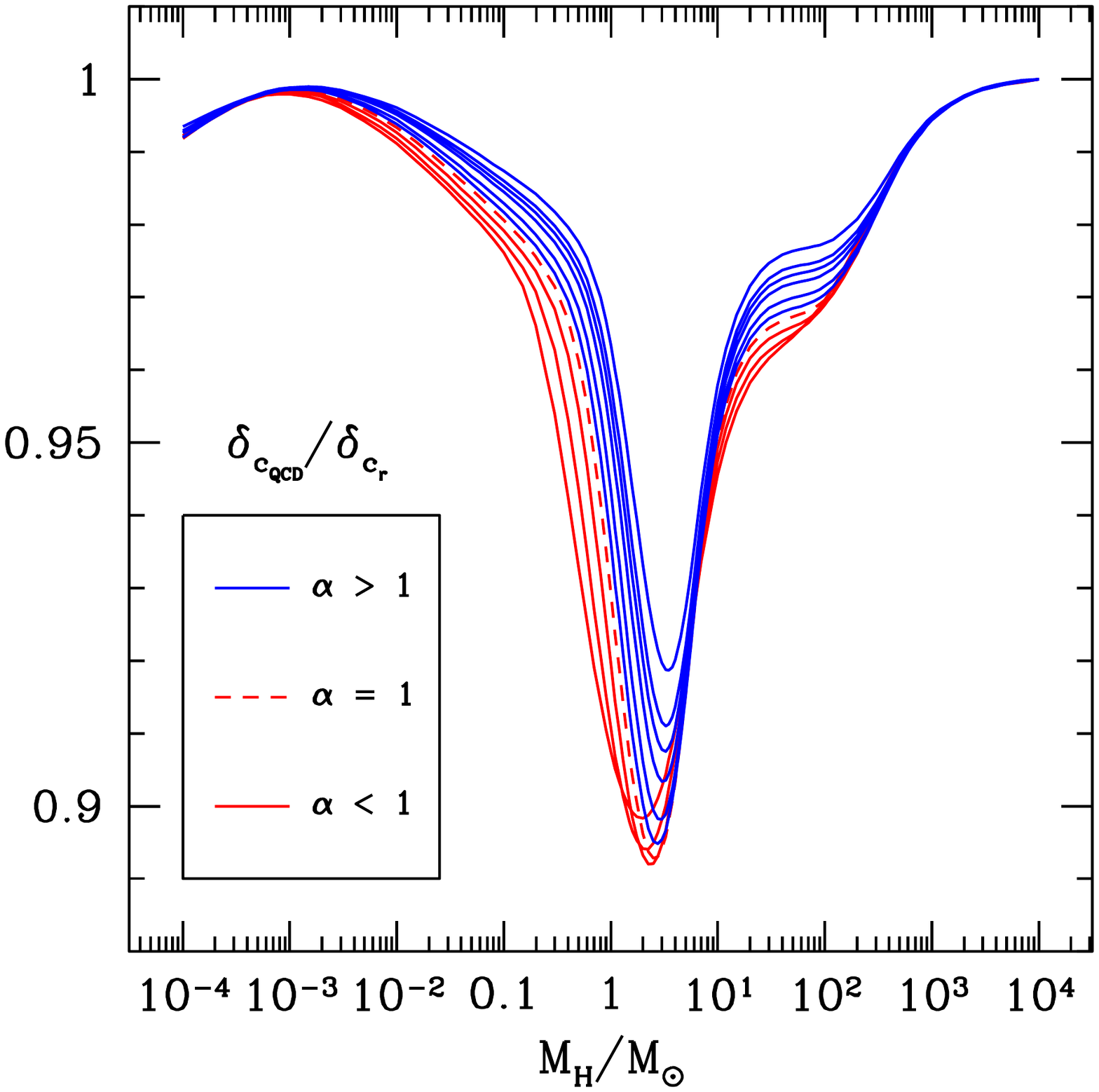}
\vspace{-2.3cm}
\caption{{\bf The threshold for PBHs during the QCD transition.} The left plot shows the behaviour of the equation of state during the QCD transition, plotting the ratio between the pressure and the energy density $w=p/\rho$ and the sound speed squared $\cs^2$ in terms of the cosmological horizon mass normalised by one solar mass. The right plot shows the corresponding evolution of the threshold $\dc$ during the QCD transition for different values of the shape parameter $\alpha$, normalised with respect to the value of $\dc$ when $w=1/3$.}
\label{fig:delta_c_QCD}
\end{figure*}

\subsection{The threshold for PBH formation}
\label{sec:Threshold}
As seen in \cite{Musco:2018rwt,Escriva:2019phb,Musco:2020jjb} the value of the threshold for PBHs depends on the shape of the cosmological perturbation, falling within the range $2/5 \leq \dc \leq 2/3$ for a radiation dominate Universe, with the corresponding threshold for the Gaussian component $\delta_l$,  within the range $0.49 \lesssim \delta_{l\mathrm{c}} \leq 4/3$. The shape dependence can be parameterised by a dimensionless parameter $\alpha$ 
\be\label{eq:alpha}
\alpha \equiv -\frac{\tilde{r}^2_\mathrm{m}\mathcal{C}^{\prime\prime}(\tilde{r}_\mathrm{m})}{4\mathcal{C}(\tilde{r}_\mathrm{m})} = -\frac{\crm^2\mathcal{C}^{\prime\prime}(\crm)}{4\mathcal{C}(\crm)\left[1-\frac{3}{2}\mathcal{C}(\crm)\right]}  .
\ee
where $\tilde{r}=re^{\zeta(r)}$ has been defined earlier in \eqref{eq:epsilon}. The second derivatives are done with respect to $\tilde r$ and $r$ respectively: this is showing that the peak amplitude of the compaction function does not cancel out with the peak amplitude of the second derivative, when the second derivative is computed with respect to $r$ instead of $\tilde r$ (see~\cite{Musco:2020jjb} for a more detailed discussion).

The shape parameter is measuring the width of the compaction function at the peak, where the apparent horizon is going to form if $\delta>\dc$. 
For larger values of  $\alpha$ the peak of the compaction function becomes narrower, with a sharp transition from the density of the central region within $\crm$ and the outer region, whereas for smaller values of $\alpha$ the transition is smoother. This affects the efficiency of the pressure gradients trying to prevent the black hole forming, and explains why $\dc$ increases for larger values of $\alpha$, as shown in Figure~\ref{fig:delta_c_rad}. 

In general, the collapse is mainly affected by the matter distribution inside the region forming the black hole, characterized just by the shape parameter $\alpha$, plus small corrections induced by the particular configuration of the tail outside this region~\cite{Musco:2018rwt}. 
The shape parameter of the average profile shape can be computed from the shape of the power spectrum of the cosmological perturbation if $\zeta$ is a Gaussian variable \cite{Musco:2020jjb}, apart from some possible variation depending on the effects of the sub-horizon modes that could affect the collapse. 

According to the numerical simulations, in a radiation dominated Universe there is a simple analytic relation to compute the threshold for PBH formation as a function of the shape parameter $\alpha$, corresponding to 
the numerical fit given by~\cite{Musco:2020jjb}:
\be \label{eq:delta_c_analytical_Musco}
\delta_\mathrm{c}= 
\begin{cases}
\alpha^{0.047}-0.50 \quad\quad\quad 0.1 \lesssim\alpha\lesssim 7 \\
\alpha^{0.035}-0.475 \quad\quad\quad 7 \lesssim\alpha\lesssim 13 \\
\alpha^{0.026}-0.45 \quad\quad\quad 13 \lesssim\alpha\lesssim 30
\end{cases}
\ee
In Figure~\ref{fig:delta_c_rad} we show how the numerical behaviour of $\dc$, plotted with a blue line, is very well fitted by \eqref{eq:delta_c_analytical_Musco}, plotted with a dashed line\footnote{The numerical results are well described also by an analytic expression~\cite{Escriva:2019phb} written in terms of Gamma functions, equivalent to \eqref{eq:delta_c_analytical_Musco}.}. 


\begin{figure*}[t!]
	\centering
	\vspace{-1.cm}
	\includegraphics[width=0.495\textwidth]{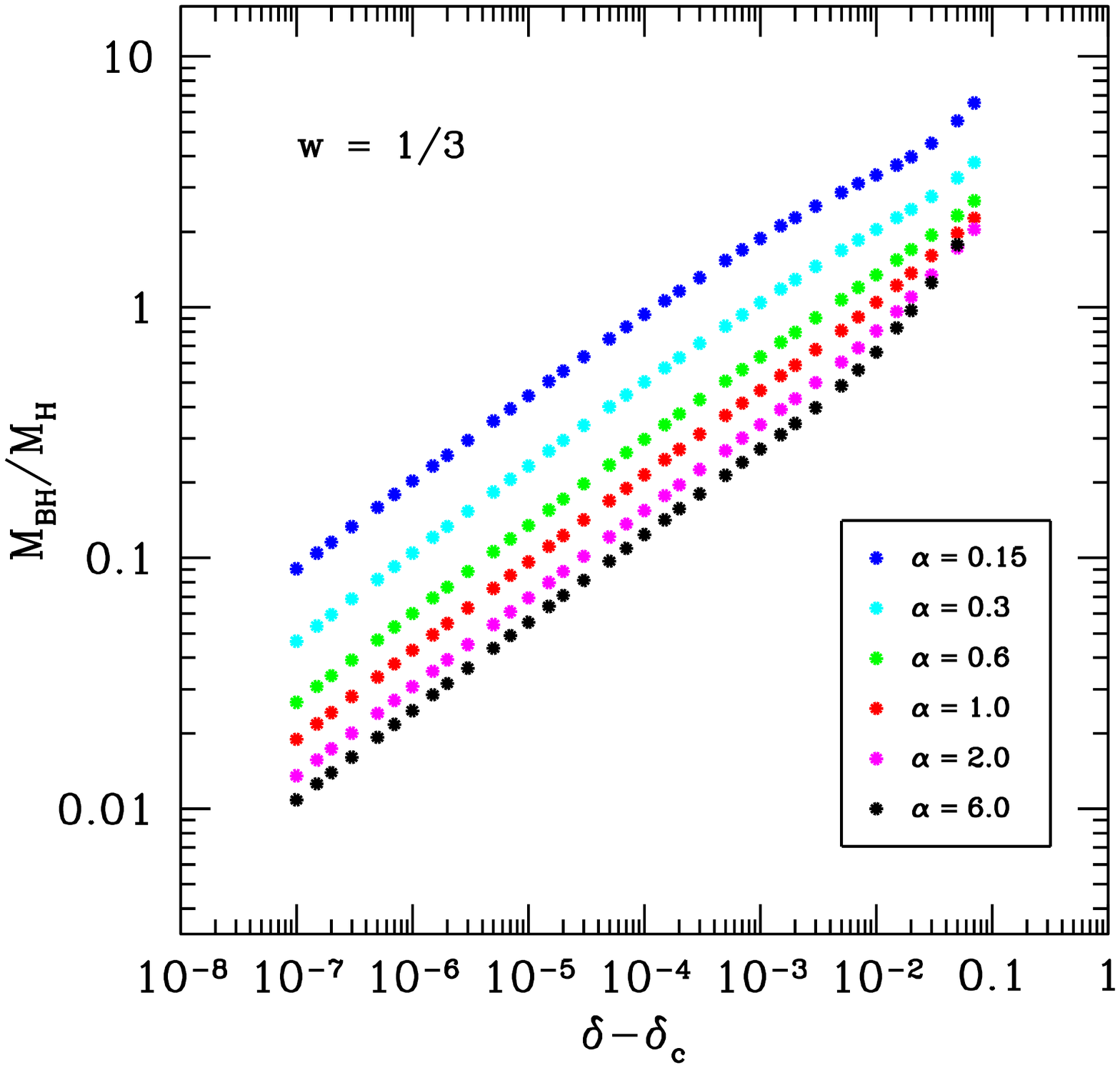}
	\includegraphics[width=0.495\textwidth]{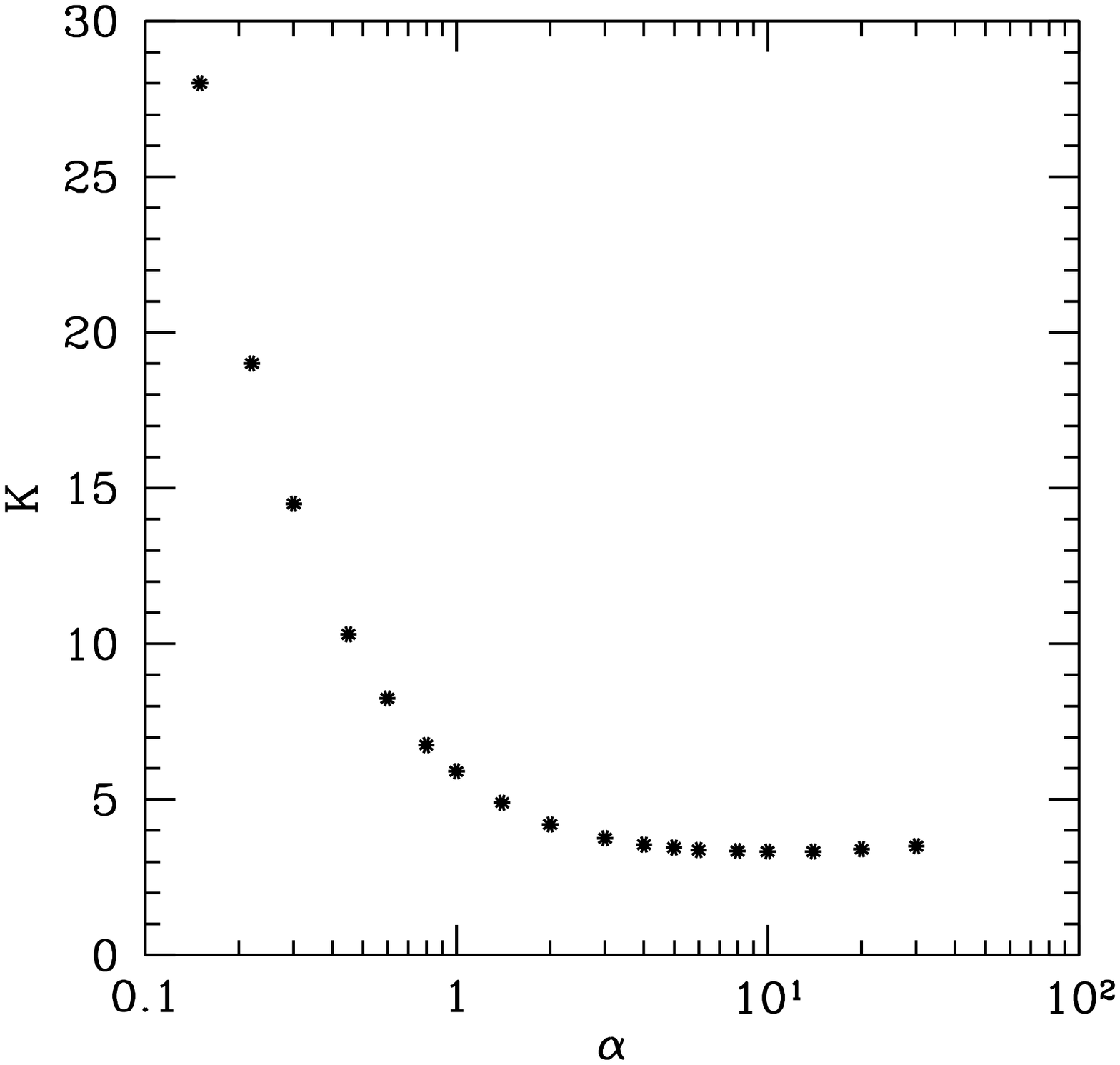}
\vspace{-2.3cm}
\caption{{\bf The mass of PBHs for a radiation dominated Universe.} This figure shows the critical collapse behaviour of equation \eqref{eqn:criticalScaling} for $w=1/3$: in the left panel we plot the scaling laws for different values of $\alpha$ obtained from the results of the numerical simulations, while in the right panel we show the corresponding dependence of the scaling law parameter $K$ as function of $\alpha$.}
\label{fig:scaling_rad}
\end{figure*}

The varying equation of state (EoS) during the QCD epoch introduces an intrinsic scale which translates into a dependence of the threshold on the cosmic epoch when collapse occurs. This can be conveniently parameterised with $\MH$, the mass of the cosmological horizon at horizon crossing. In the left panel of Figure~\ref{fig:delta_c_QCD} we focus on the behaviour of the equation of state, plotting $w$ and $\cs^2$ as functions of $\MH/M_\odot$. 
The largest deviation from a pure radiation EoS occurs during confinement of quarks and gluons to hadrons at $T\approx 200\,$ MeV,
associated with a horizon mass $\MH/M_\odot\simeq 2.5 - 4$. However, the
existence of strongly interacting matter influences the EoS over
a large range of horizon masses between    $\MH/M_\odot\sim10^{-3}$ and $\MH/M_\odot\sim10^4$. This is partially due to increasing interaction strengths as one approaches the QCD crossover from higher temperatures and partially due to various annihilation epochs, heavier quarks for smaller horizon masses and pions for larger horizon masses.

\begin{figure*}[t!]
	\centering
	\vspace{-1.cm}
	\includegraphics[width=0.495\textwidth]{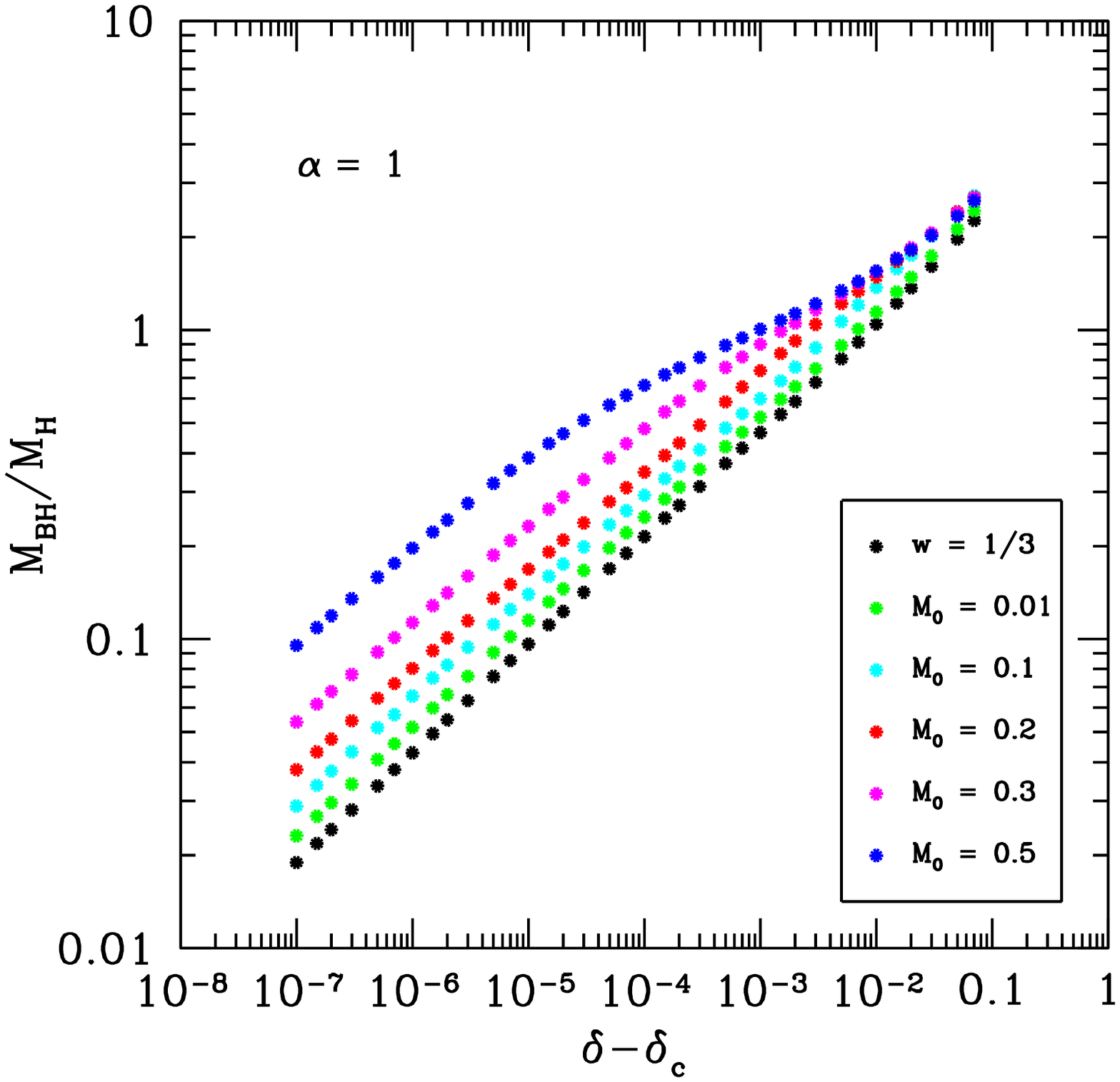}
	\includegraphics[width=0.495\textwidth]{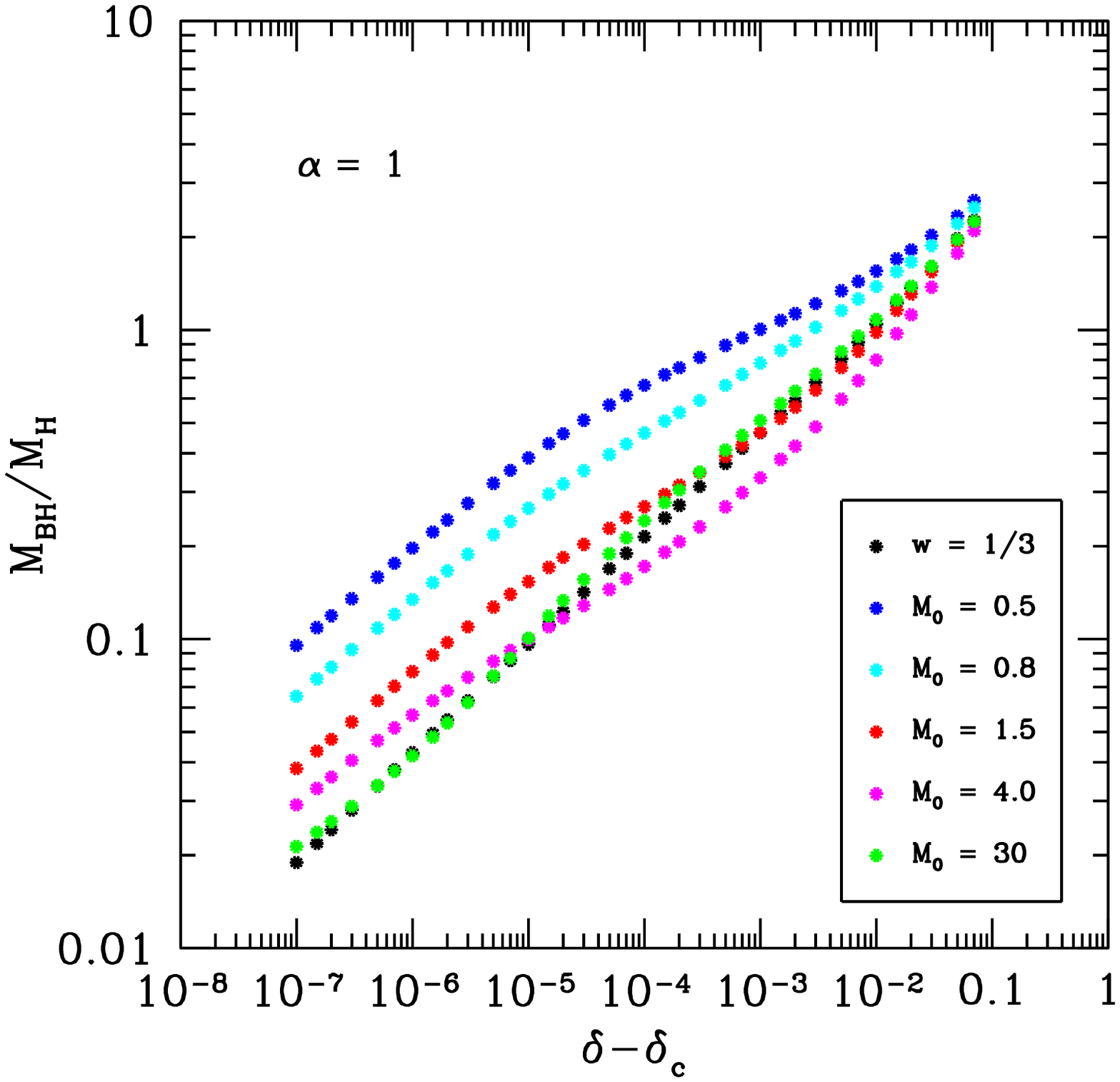}{\vspace{-3cm}}
    \includegraphics[width=0.495\textwidth]{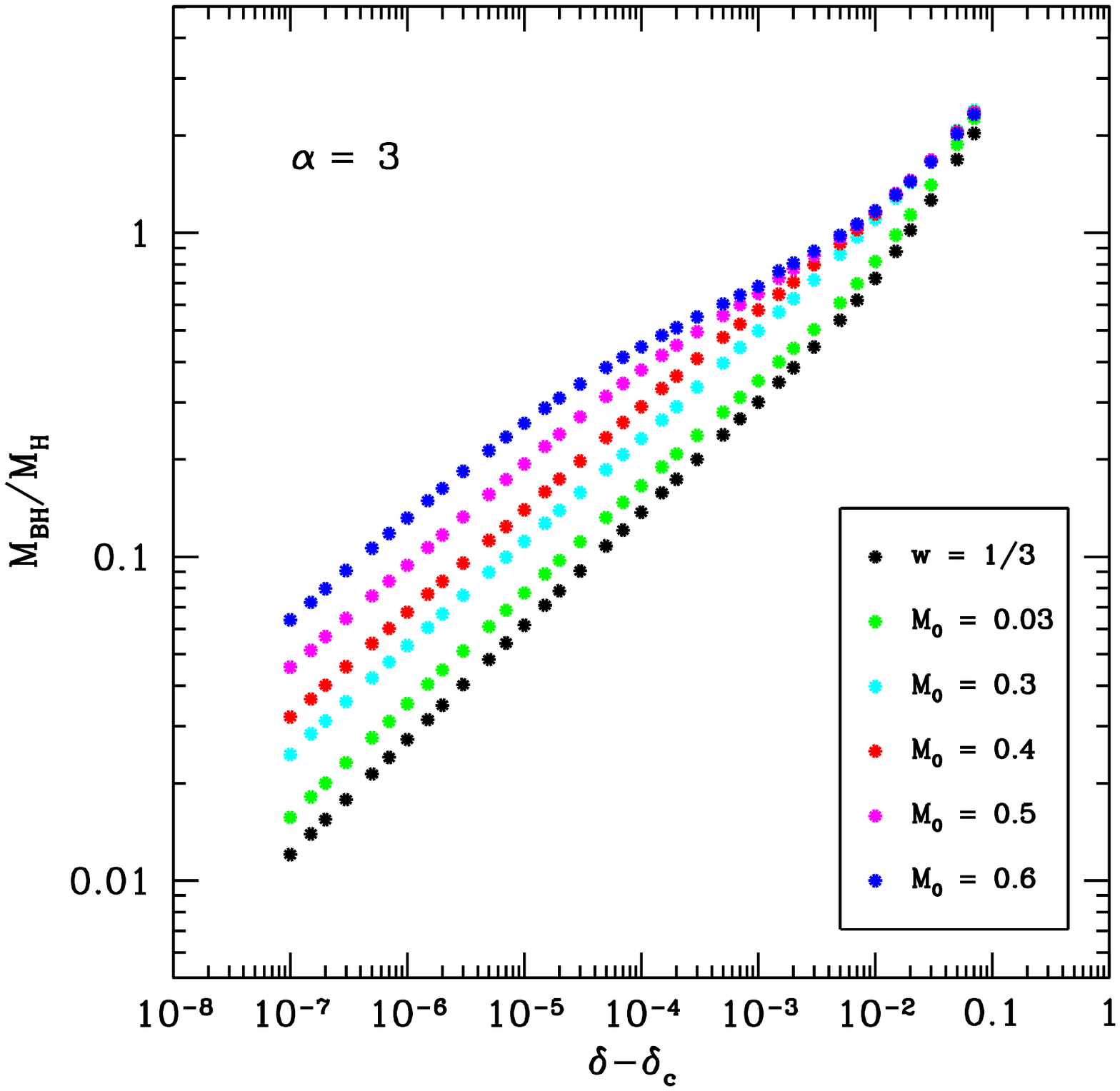}
	\includegraphics[width=0.495\textwidth]{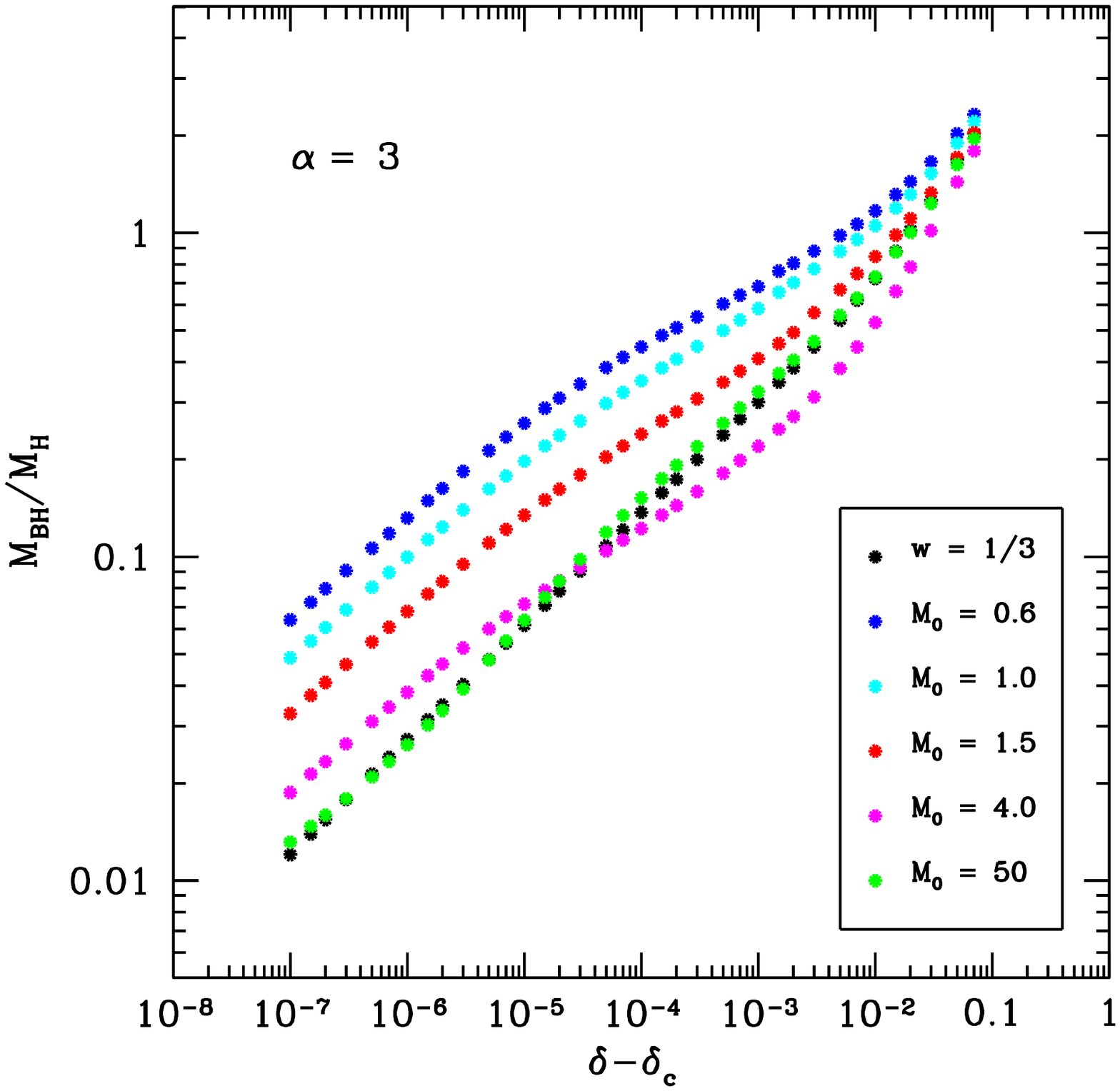}
\vspace{-2.3cm}
\caption{{\bf The mass of PBHs during the QCD transition.} This figure shows how the critical collapse is modified during the QCD transition, plotting the scaling law for $\alpha=1$ (top panels) and $\alpha=3$ (bottom panels). Left (right) panels show the scaling relation for perturbations crossing into the horizon before (after) the confinement transition at $T\approx 200\,$GeV. They are labeled by $M_0\equiv \MH/M_\odot$. For reference, the black dots show the scaling law when $w=1/3$.}
\label{fig:scaling_QCD}
\end{figure*}

The right panel of Figure~\ref{fig:delta_c_QCD} shows the corresponding behaviour of the threshold $\dc$ for different values of the shape parameter. The threshold is normalised  with respect to the corresponding value $\delta_\mathrm{c,r}$ when the universe is radiation dominated, given by \eqref{eq:delta_c_analytical_Musco}. Very large values of $\alpha$ are not consistent with the shape of the power spectrum, because a very peaked spectrum like a Dirac delta gives 
$\alpha\simeq6.33$ corresponding to $\dc\simeq0.59$, and therefore we are not calculating the threshold for very large value of $\alpha$,  with the last blue line of Figure~\ref{fig:delta_c_QCD} obtained for $\alpha=8$.
For $\alpha<1$, shown with red lines in the range $0.15\leq\alpha\leq1$, the minimum value of the threshold is slightly decreasing for increasing values of $\alpha$, with the smallest value reached for $\alpha\simeq1$, shown in Figure~\ref{fig:delta_c_QCD} with a dashed line.

 Looking at the qualitative behaviour, as expected one can observe it following that for the equation of state in the left plot. The minimum value of the threshold is reached when $1\lesssim \MH/M_\odot\lesssim3$, slowly increasing for larger values of $\alpha$. The shape parameter also affects the relative change of the threshold, with a variation larger than $10\%$ if $\alpha\lesssim1$ while for larger values of the shape parameter the relative change of the threshold is a bit lower, up to $8\%$ for $\alpha=8$. This is consistent with the increasing effect of the pressure gradients, becoming stronger for larger value of $\alpha$, when the threshold is also larger. 
 
 These results are basically consistent with what ihas been found recently by Escriva et al. in \cite{Escriva:2022bwe}. However we obtain a larger deviation of the threshold, up to $25\%$ more. The origin of this difference is not clear, but we note that the parameterisation of the curvature profile used in \cite{Escriva:2022bwe} in terms of the shape parameter is different from this paper. In Appendix~\ref{sec:comparison} we compare the results and the methodologies of these two papers, discussing the different physical conclusions obtained.

 \begin{figure*}[t!]
	\centering
	\vspace{-1.cm}
	\includegraphics[width=0.495\textwidth]{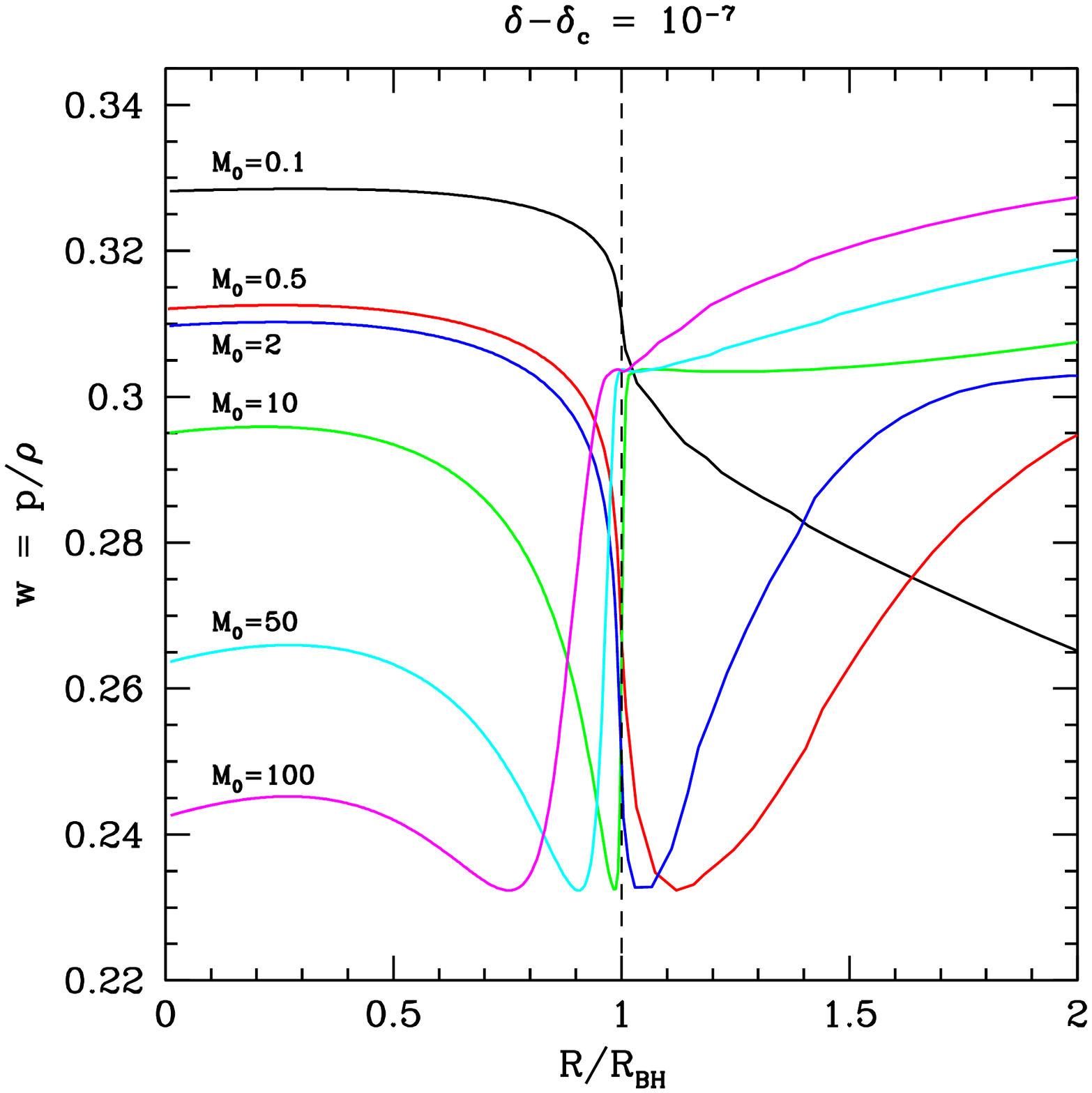}
	\includegraphics[width=0.495\textwidth]{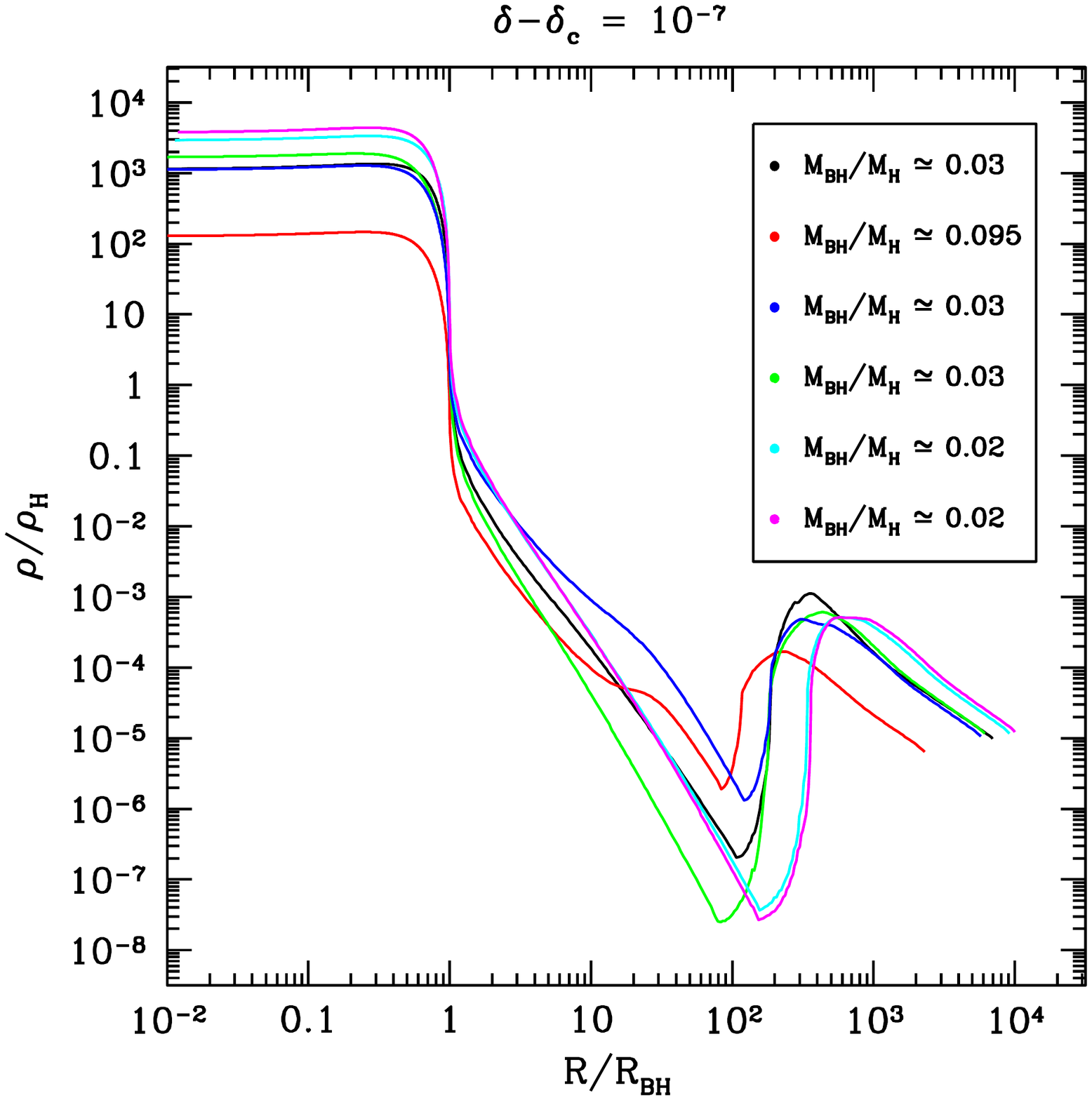}
\vspace{-2.3cm}
\caption{{\bf Profiles of $w=p/\rho$ (left panel) and $\rho$ (right panel) when PBHs are formed}. These are shown in the observer time coordinate, with all of the lines having an initial amplitude $(\delta-\dc)=10^{-7}$, while varying the horizon crossing time, labeled by $M_0=\MH/M_{\odot}$. In the right panel we also give the values of the PBH masses showing that smaller values of the mass correspond to higher values of the energy density in the inner region inside the horizon.}
\label{fig:e&w}
\end{figure*}

\subsection{The mass of PBHs: scaling law relation}
\label{sec:scaling}
One of the present authors has shown~\cite{Niemeyer:1997mt,Niemeyer:1999ak} that the mass spectrum of PBHs is characterised  by the scaling law relation of critical collapse, given by 
\begin{equation}
    M_\mathrm{PBH} = K M_\mathrm H \left(\delta-\dc \right)^\gamma\,.
    \label{eqn:criticalScaling}
\end{equation}
The cosmological horizon mass $\MH$ identifies the epoch of the Universe when the cosmological perturbations collapsing into PBHs are crossing the cosmological horizon, while $\gamma$ is a parameter depending only on the equation of state~\cite{Neilsen:1998qc}, with $\gamma\simeq0.36$ when $w=1/3$. The other two parameters $K$ and $\dc$ also depend on the effects of pressure gradients, i.e. the equation of state,  described by the value of $w$ and $\cs^2$, and the shape of the initial configuration, identified by the shape parameter $\alpha$.

The nature of critical collapse in the context of PBHs has been then intensively investigated by one of the authors of this work \cite{Musco:2008hv,Musco:2012au} and in the left panel of Figure~\ref{fig:scaling_rad} one can observe the scaling law behaviour for different values of $\alpha$: for $M_\mathrm{PBH}< M_\mathrm H$ the exponent $\gamma\simeq0.36$ is constant, while for larger values deviation
from scaling are visible. This is because the critical collapse is characterised by a self similar behaviour~\cite{Musco:2012au}, which is scale free. 

In the right panel of Figure~\ref{fig:scaling_rad} we show how $K$ is varying with $\alpha$\footnote{The behaviour obtained in Figure~\ref{fig:scaling_rad} is similar to that observed in \cite{Escriva:2021pmf}, but some significant differences may be noticed. These are likely due to same reasons mentioned earlier regarding the discrepacy in the threshold.}. It is interesting to notice that for $\alpha\gtrsim6$ the value of $K$ is almost constant, approximately $3.4$. This is consistent with the Dirac delta limit of the shape of the power spectrum, corresponding to $\alpha\simeq6.33$.

In \cite{Franciolini:2022tfm}, it has been shown that a nearly scale invariant power spectrum, with a spectral index $\ns\simeq1$, leads to 
perturbations with a shape parameter $\alpha\simeq3$, having a value of the threshold $\dc\simeq0.55$ when $w = 1/3$. For this reason we are computing the scaling law relation for the mass of PBHs formed during the QCD transition only for $1\lesssim\alpha\lesssim5$, enough to describe a wide range of spectral indices, consistent with the cosmological power spectrum obtained for different models of the very early Universe. 

In Figure~\ref{fig:scaling_QCD} we show the mass of PBHs as function of $(\delta-\dc)$ for $\alpha=1$ and $\alpha=3$. The scaling law is now also a function of the horizon scale when the perturbation crosses the cosmological horizon, parameterised by the dimensionless parameter $M_0\equiv \MH/M_\odot$. The critical behaviour, keeping $\gamma$ constant and $K$ depending only on the shape is still preserved when the collapse is very critical, with $(\delta-\dc)\lesssim10^{-4}$ and $M_{\mathrm{BH}}\lesssim0.1\MH$. 

For $\alpha=1$, the largest deviation of the scaling law, with respect to radiation, is reached when $M_0\simeq0.5$, while for larger values of $\alpha$ this is reached slightly later ($M_0\simeq0.6$ for $\alpha=3$, $M_0=0.8$ for $\alpha=5$). This slight delay observed for larger values of $\alpha$ is consistent with the small shift of the minimum of the threshold towards larger value of the masses observed in Figure~\ref{fig:delta_c_QCD}. The value of $M_0$ for which one has the largest deviation of the scaling law is smaller than the one for which we have found the minimum of the threshold. This is due to the time taken for PBHs to form after cosmological horizon crossing:
for $\alpha=1$ the minimum
occurs at $M_0\simeq2.5$ while for $\alpha=3$ the minimum is observed at $M_0\simeq3$, as shown in Figure~\ref{fig:delta_c_QCD}. Afterwards, the variation of $\gamma$ and $K$ describes the scaling law slowly coming back to the scaling of radiation, consistent with behaviour of the threshold when PBHs form after the transition, with some oscillations of $\gamma$ and $K$ before reaching the end of the transition. This is shown in the right panel of Figure~\ref{fig:scaling_QCD}.

 \begin{figure*}[t!]
	\centering
	\vspace{-0.4cm}
	\includegraphics[width=0.329\textwidth]{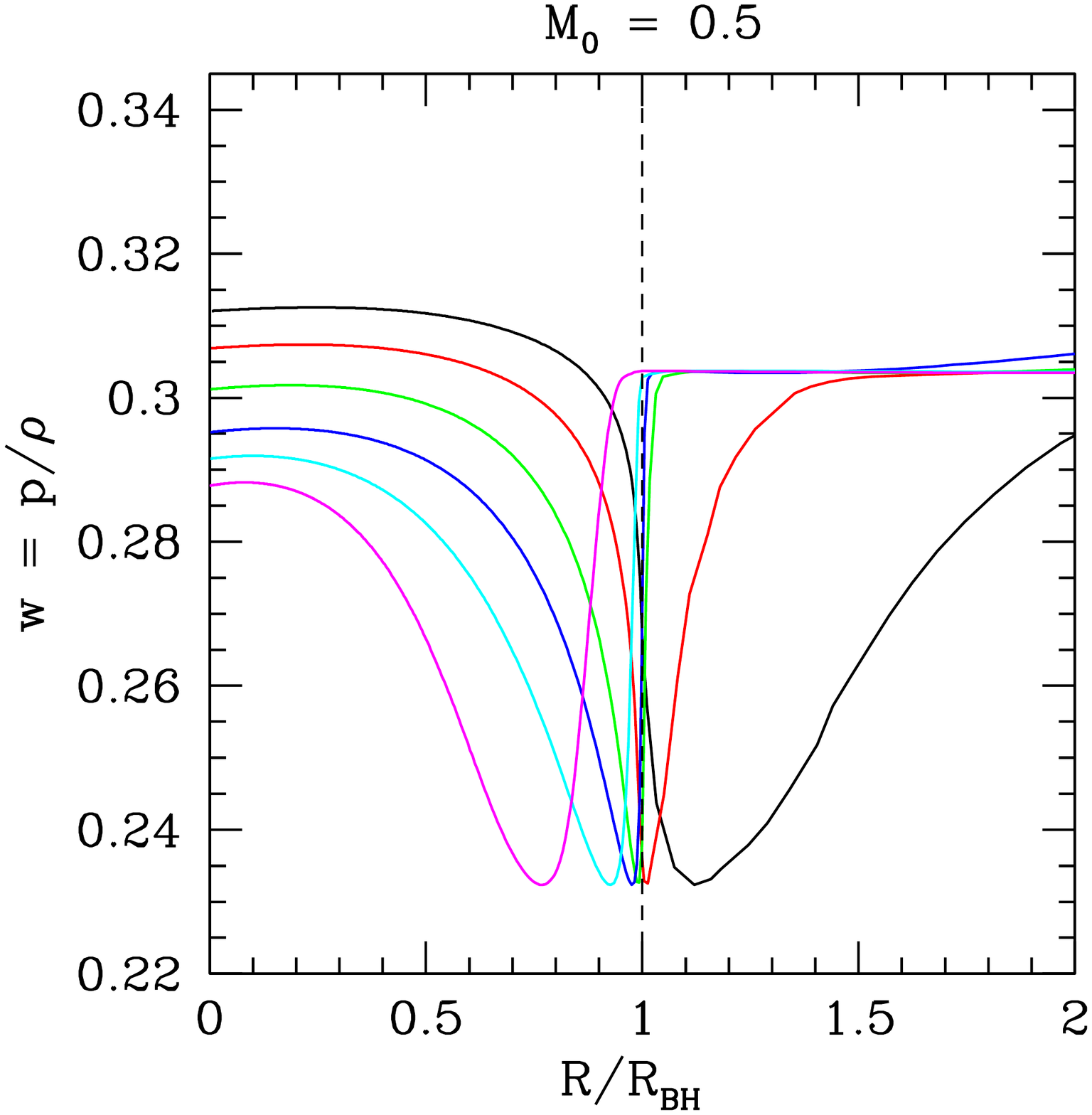}
	\includegraphics[width=0.329\textwidth]{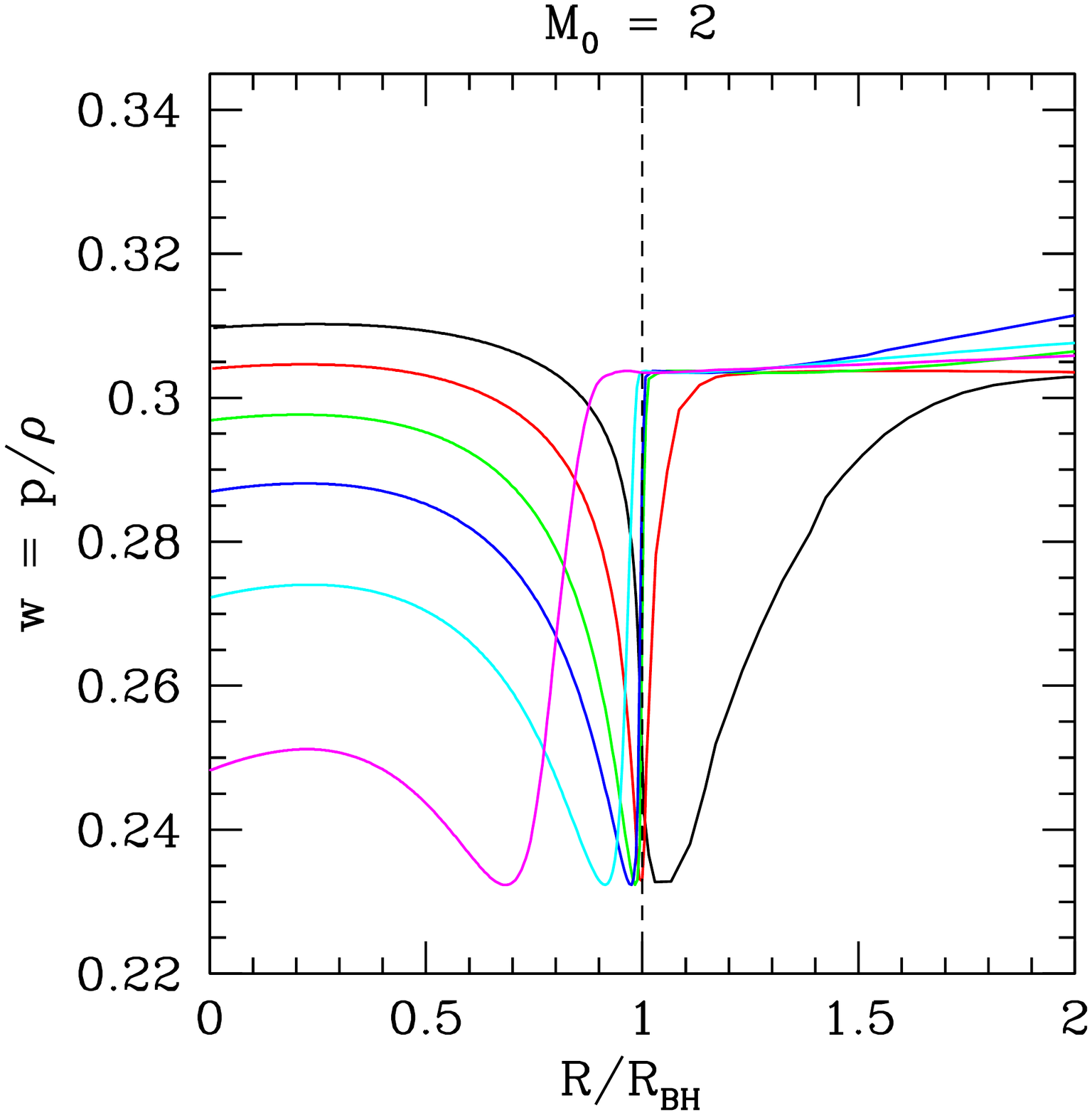}
    \includegraphics[width=0.329\textwidth]{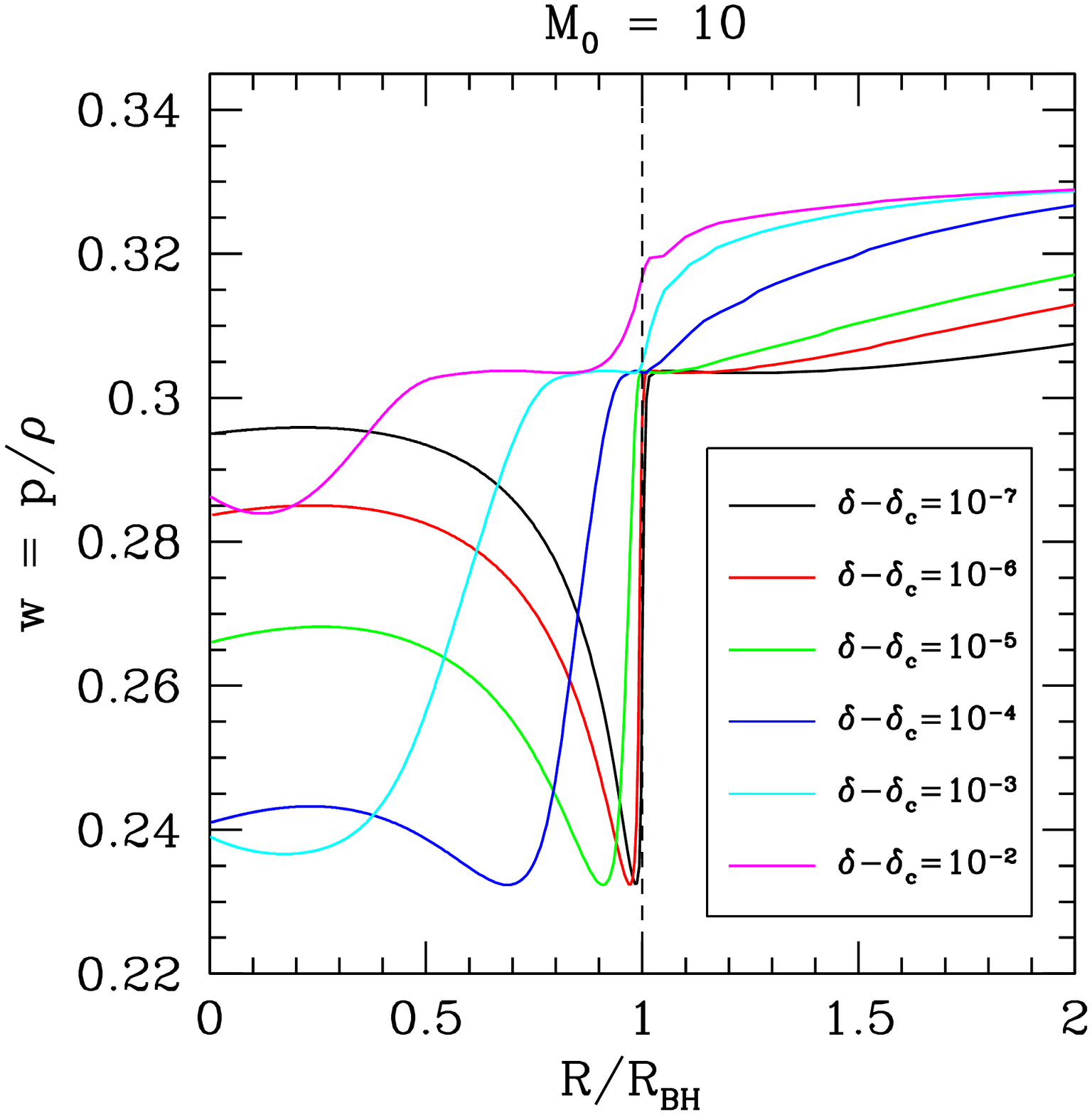}
\vspace{-1.5cm}
\caption{ {\bf Profiles of $w=p/\rho$ for $M_0=0.5$ (left panel), $M_0=2$ (central panel) and $M_0=10$ (right panel) when PBHs are formed}. These are computed in the observer time coordinate, varying the critical amplitude of the perturbation. The different colors refer to different values of $(\delta-\dc)$.}
\label{fig:w_M0}
\end{figure*}

In Figure~\ref{fig:e&w} we show the profiles of the ratio $w=p/\rho$ and the energy density $\rho$, respectively in the left and right panels, for PBHs forming at different values of $M_0$, while keeping the same value for $\delta-\dc=10^{-7}$. Note that the snapshot is shown in 
observer time (i.e. on out-going null geodesics).
The areal radius $R$ on the horizontal axis is normalised with respect to the black hole radius $R_{\mathrm{BH}}$, demonstrating that the apparent horizon forms approximately where the medium is in the middle of the QCD transition. It is quite remarkable to notice that this is happening for a broad range of masses. For initial conditions very close to the critical one, i.e. 
for very small $(\delta-\dc)$, there is a fine balance between pressure forces and gravity keeping fluctuations in near-quasi equilibrium for
several dynamical times~\cite{Musco:2008hv,Musco:2012au}. It is hence not too surprising that this equilibrium includes the apparent horizon being at the minimum value of $w$. This implies that for fluctuations entering
the horizon not too far from the transition, the pressure minimum of the
transition is an attractor solution. 

 In the right panel of Figure~\ref{fig:e&w} we show the corresponding profiles of the energy density normalised with respect to the value at horizon crossing, using the same colour coding for $M_0$ as in the left panel. It is interesting to point out that, for $(\delta-\dc)=10^{-7}$, the value of the energy density outside the apparent horizon is roughly of the same order as the energy density of the Universe at cosmological horizon crossing of the perturbation. In Figure~\ref{fig:w_M0} we show the behaviour of $w$ while keeping $M_0$ constant for different values of $\delta-\dc$, showing three sample cases: before, during, and after the QCD transition. As the collapse becomes less and less critical, i.e. for increasing $(\delta-\dc)$ , the apparent horizon forms more quickly, and the condition that the medium is at the depth of the phase transition outside the horizon is reached with less accuracy.  

\section{Mass distribution and cosmological abundance}
\label{sec:Mass_distribution}
In this section, we will apply the results of the simulations describing the formation of PBHs to the calculation of the PBH mass function and abundance. It is desirable to compute the PBH abundance and mass function directly from $\zeta$, which appears in the FLRW metric, equation \eqref{eq:asymptotic}. To perform the calculation, we will follow the method outlined in \cite{Young:2019yug}, applying peaks theory and accounting for the non-linearity between the curvature perturbation $\zeta$ and the density - and will only briefly summarise the method here. We will also make the standard assumption that $\zeta$ follows a Gaussian distribution, although it has been argued that inflationary models which predict a large PBH abundance typically also predict a non-Gaussian distribution \cite{Figueroa:2020jkf,Biagetti:2021eep}, which can have a large impact on the PBH abundance and mass function (see e.g. \cite{Young:2022phe,Ferrante:2022mui,Gow:2022jfb} for recent discussions of the effect of non-Gaussianities on the PBH abundance). 

\subsection{PBH abundance in peaks theory}

As discussed in section \ref{sec:compaction_function}, we will consider that PBHs form at sufficiently large peaks in the (smoothed) density \cite{Young:2014ana,Young:2019osy}. From Equation~\eqref{eq:delta_l} the smoothed density can be expressed as
\begin{equation}
    \delta(\mathbf{x},t) =\delta_l(\mathbf{x},t)-\frac{1}{4\Phi}\delta_l^2(\mathbf{x},t),
    \label{eqn:deltaFull}
\end{equation}
where $\delta_l\equiv2\Phi r\zeta'$ is linearly related to the curvature perturbation $\zeta$.

We define the $i^\mathrm{th}$-order moments of the power spectrum $\sigma_i^2$ as
\begin{equation}
    \sigma_i^2(r) = \frac{4}{9}\Phi^2 \int\limits_0^\infty \frac{\mathrm{d}k}{k}\left(k r \right)^4 k^{2i} \tilde{W}^2(k,r)T^2(k,r_H)\mathcal{P}_\zeta(k),
\end{equation}
where $\mathcal{P}_\zeta(k)$ is the power spectrum of $\zeta$ and $\tilde{W}(k,r)$ and $T(l,r)$ are, respectively, the window function and (linear) transfer function in Fourier space, given by
\begin{equation}
    \tilde{W}(k,r) = 3\frac{\sin(kr)-kr\cos(kr)}{(kr)^3},
\end{equation}
\begin{equation}
    T(k,r) = 3\frac{\sin(kr/\sqrt{3})-kr\cos(kr/\sqrt{3})/\sqrt{3}}{(kr/\sqrt{3})^3},
\end{equation}
where we note that, as we will evaluate the density at horizon crossing\footnote{we here use a linear extrapolation from super-horizon scales; and the non-linear effects close to horizon-crossing have been discussed recently in \cite{Musco:2020jjb}}, we have set the smoothing scale equal to the horizon scale $r=(aH)^{-1}$.

For a Gaussian distribution, the number density of peaks in the range $\delta_l\rightarrow \delta_l+\mathrm{d}\delta_l$ is given, using peak theory \cite{Bardeen:1985tr}, by
\begin{equation}
n(\delta_l,r) = \frac{1}{3^{3/2}(2\pi)^2}\left(\frac{\sigma_{1}}{\sigma_0}\right)^3 \left(\frac{\delta_l}{\sigma_0} \right)^3\exp\left( -\frac{\delta_l^2}{2\sigma_0^2} \right).
\label{eqn:numberDensity}
\end{equation}
The fraction of the Universe collapsing to form PBHs from perturbations of a single scale $r$ is then given by integrating the number density of peaks over the range of values of $\delta_l$ that form PBHs:
\begin{equation}
    \beta(r) = \frac{4}{3}\pi r^3 \int\limits_{\delta_{l\mathrm{c}}}^{4/3}\mathrm{d}\delta_l \frac{M_\mathrm{PBH}(\delta_l,\MH(r))}{\MH(r)}n(\delta_l,r),
    \label{eqn:beta}
\end{equation}
where $M_\mathrm{PBH}$ and $M_\mathrm{H}$ are the PBH mass and horizon mass respectively. The lower limit corresponds to the critical value for PBH formation of the linear component of the compaction, $\delta_{l\mathrm{c}} = 2\Phi(1-\sqrt{1-\dc/\Phi})$, whilst the upper limit corresponds to the highest value for type I perturbations\footnote{for type I perturbations, the areal radius $R$ increases monotonically with coordinate radius $r$, whilst this is not true for type II perturbations, corresponding to larger values of $\delta_l$. The collapse of type II perturbations has not been well studied, although it is expected that they do form PBHs \cite{Kopp:2010sh}. Since the abundance of such perturbations is exponentially suppressed, and has negligible impact on the PBH abundance, we neglect type II perturbations.}.

Typically, it is assumed either that PBHs form with a fixed fraction of the horizon mass, or that the PBH mass follows the critical scaling relationship described in section \ref{sec:scaling}, given by equation \eqref{eqn:criticalScaling}.
 In this paper, we go beyond previous studies and make use of numerical results from the simulations, as seen in Figure \ref{fig:scaling_QCD}, in order to accurately determine the PBH mass from initial conditions, accounting for the scale and amplitude of the initial perturbation, as well as the varying equation of state during the QCD phase transition. Where necessary, we will take the values $K=4$ and $\gamma=0.36$ to make comparisons with calculations using the scaling relationship.
 
\begin{figure*}[t!]
 \centering
 \vspace{-1.cm}
  \includegraphics[width=0.495\textwidth]{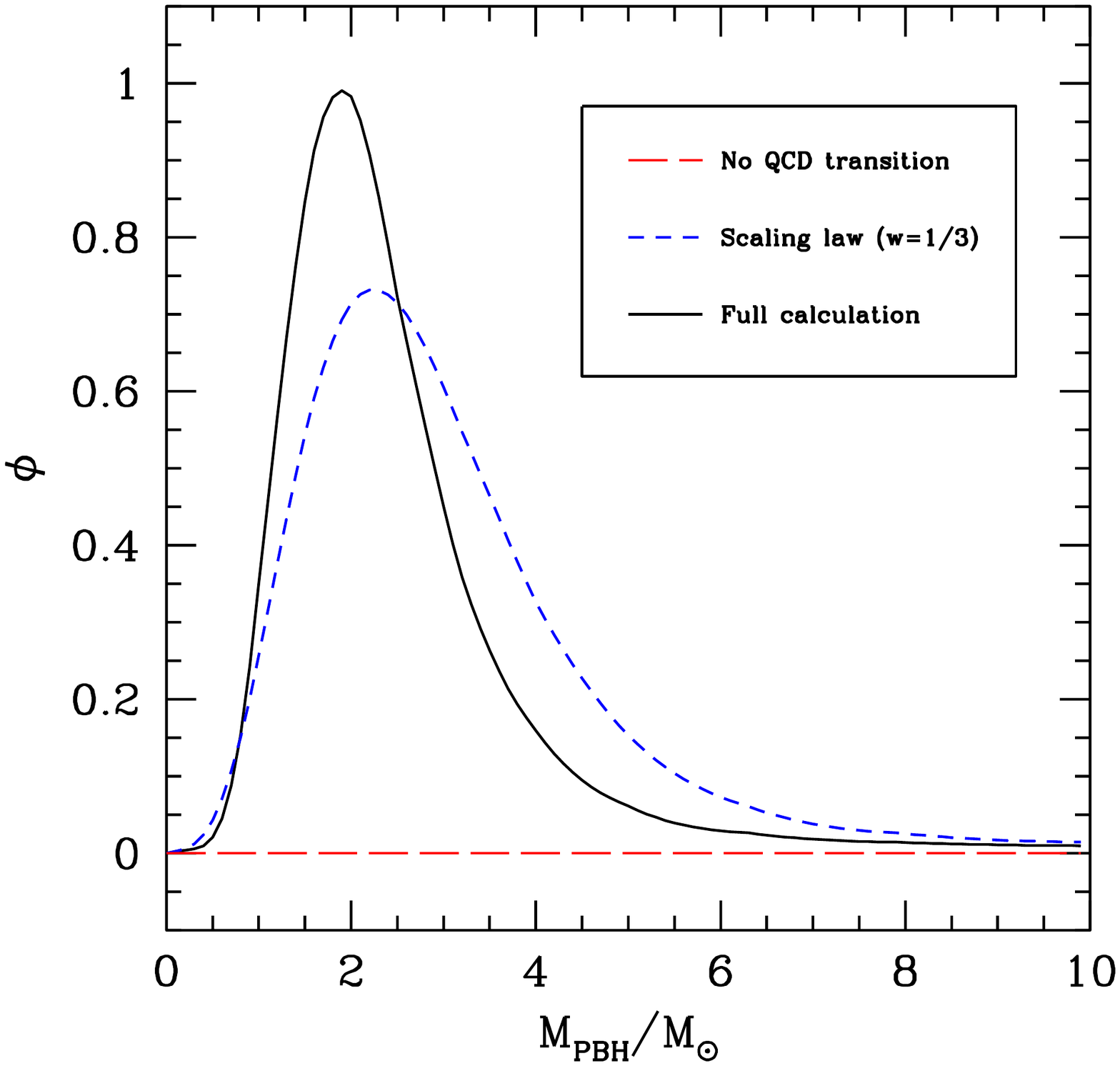}
  \includegraphics[width=0.495\textwidth]{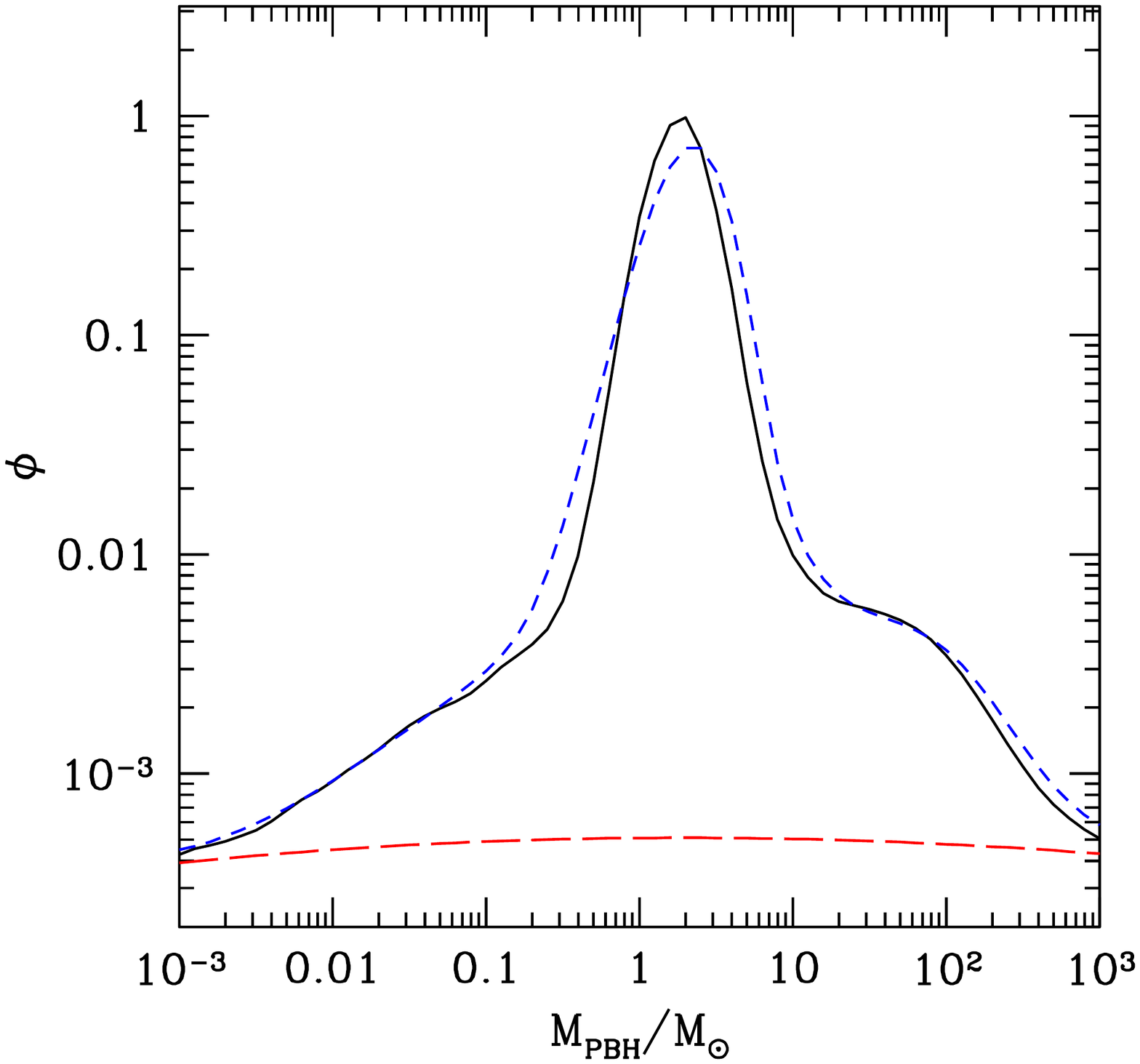}
    \vspace{-2.3cm}
  \caption{ {\bf The PBH mass function during the QCD transition.} The mass function $\phi(M_\mathrm{PBH})$ is shown for $\alpha=3$, for three different cases: ignoring the phase transition, using the scaling law relationship for radiation, and finally the full calculation using the masses given by the simulations. The same data is shown on both plots, with linear axes on the left and logarithmic axes on the right. 
For the full calculation, the total PBH abundance is $f_\mathrm{PBH}=1$ (by choice), whilst we have $f_\mathrm{PBH}\approx 0.93$ when the scaling law is used (for the same choices of parameters).}
  \label{fig:mass function}
  \vspace{1cm}
\end{figure*}

The total abundance of PBHs is then calculated by integrating over the range of scales at which PBHs form, and can be expressed as a fraction of the dark matter composed of PBHs,
\begin{equation}
    f_\mathrm{PBH} = \frac{\Omega_\mathrm{PBH}\left( t_\mathrm{eq}\right)}{\Omega_\mathrm{CDM}\left( t_\mathrm{eq}\right)} = \frac{1}{\Omega_\mathrm{CDM}} \int\limits_{r_\mathrm{min}}^{r_\mathrm{max}} \frac{\mathrm{d}r}{r} \left(\frac{r_\mathrm{eq}}{r}\right)\beta(r),
    \label{eqn:fpbh}
\end{equation}
where $r_\mathrm{min}$ and $r_\mathrm{min}$ correspond to the minimum and maximum scales, respectively, at which PBHs are considered to form. The $(r_\mathrm{eq}/r)$ term accounts for the evolution of the PBH density parameter between formation and the time of matter-radiation equality, where we have assumed radiation domination for the duration. As is more typically done, equation \eqref{eqn:fpbh} can also be expressed as an integral over the horizon mass,
\begin{equation}
    f_\mathrm{PBH} = \frac{1}{\Omega_\mathrm{CDM}\left( t_\mathrm{eq}\right)} \int\limits_{M_\mathrm{min}}^{M_\mathrm{max}} \frac{\mathrm{d}\MH}{\MH} \left(\frac{M_\mathrm{eq}}{\MH}\right)^{1/2}\beta(\MH),
    \label{eqn:fpbh2}
\end{equation}
where $M_\mathrm{eq}=2.8\times 10^{17}M_\odot$ is the horizon mass at the time of matter-radiation equality. The horizon mass $\MH$ can be related to the horizon scale $r$ as \cite{Nakama:2016gzw}
\begin{equation}
    \MH = M_\mathrm{eq}\left( \frac{g_*}{10.75} \right)^{-1/6} \left(\frac{r}{r_\mathrm{eq}}\right)^2,
    \label{eqn:MH}
\end{equation}
where $g_*$ is the number of relativistic degrees of freedom (although we neglect this effect due to the extremely weak dependence). Inverting this gives the horizon scale as a function of the horizon mass, $r=r\left(\MH\right)$.

Finally, we define the PBH mass function as the derivative of $f_\mathrm{PBH}$,
\begin{equation}
    \phi(M_\mathrm{PBH}) = \frac{\mathrm{d}f_\mathrm{PBH}}{\mathrm{d} \ln M_\mathrm{PBH}}.
\end{equation}
The final PBH mass $M_\mathrm{PBH}$ is a function of the initial perturbation amplitude $\delta_l$ (and vice versa), allowing $\beta$, equation \eqref{eqn:beta}, to be expressed as an integral over the PBH mass. This allows us to write the final expression for the mass function as
\begin{multline}
    \phi(M_\mathrm{PBH}) = \frac{M_\mathrm{PBH}}{\Omega_\mathrm{CDM}\left( t_\mathrm{eq} \right)} \int\limits_{M_\mathrm{min}}^{M_\mathrm{max}}  \frac{\mathrm{d}\MH}{\MH} \left(\frac{M_\mathrm{eq}}{\MH} \right)^{1/2} \\
    \times\frac{4\pi r^3}{3} \frac{\mathrm{d}\delta_l}{\mathrm{d}M_\mathrm{PBH}} \frac{M_\mathrm{PBH}}{\MH} n\left( \delta_l ,\MH \right),
\end{multline}
where the perturbation amplitude $\delta_l$ at scale $\MH$ required to form a PBH of mass $M_\mathrm{PBH}$ is calculated numerically from the simulation data, $\delta_l=\delta_l\left(M_\mathrm{PBH},\MH\right)$, and the scale $r$ is a function of the horizon mass, as given by equation \eqref{eqn:MH}. The derivative is also calculated numerically as a function of PBH mass and horizon scale, $\mathrm{d}\delta_l/\mathrm{d}M_\mathrm{PBH}=\mathrm{d}\delta_l/\mathrm{d}M_\mathrm{PBH}\left(M_\mathrm{PBH},\MH\right)$. 
In principle, the integration limits $M_\mathrm{min}$ and $M_\mathrm{max}$ should span all scales
\footnote{There is a maximum perturbation amplitude, arising from the non-linearity of $\delta$, equation \eqref{eqn:deltaFull} (during radiation domination, $\delta_\mathrm{max}=2/3$, corresponding to $\delta_l=4/3$). This means that there is a maximum mass PBH mass that can form at a specified horizon mass (which then gives a minimum horizon mass at which PBH of given mass will form). However, the formation of PBHs from type II perturbations, $\delta_l>4/3$, is not well understood and it may be possible for more massive PBHs to form, although this is typically neglected from the calculation.}.
However, in practice, since PBHs typically form with a mass close to the horizon mass, a smaller integration range can be used, and in our case we fix the integration limits to the range of values for which data is available from the simulations (see Figure \ref{fig:scaling_QCD}). The final results are not sensitive to the specific values used.

\begin{figure*}[t!]
\vspace{-1cm}
 \centering
  \includegraphics[width=0.495\textwidth]{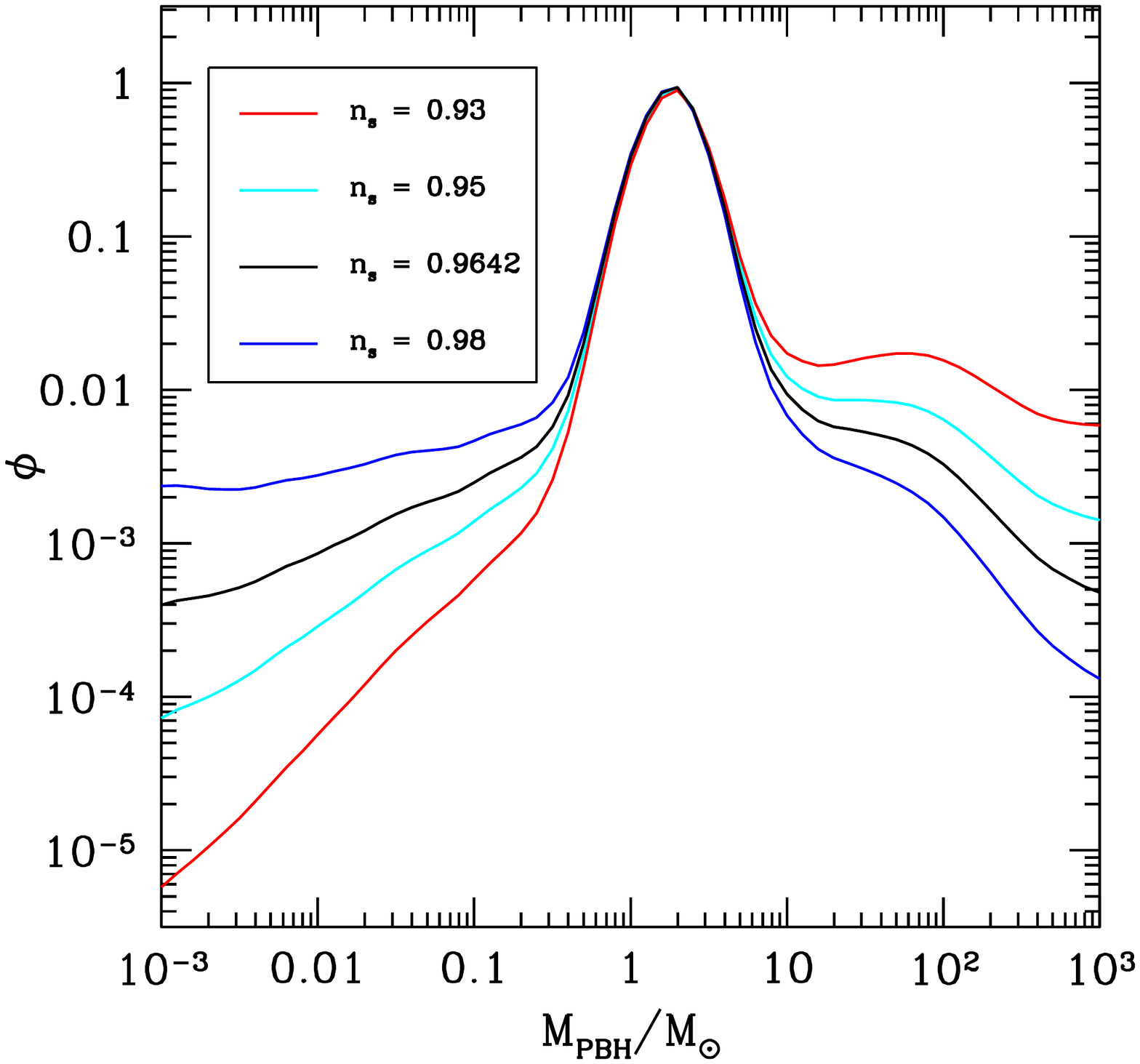}
  \includegraphics[width=0.495\textwidth]{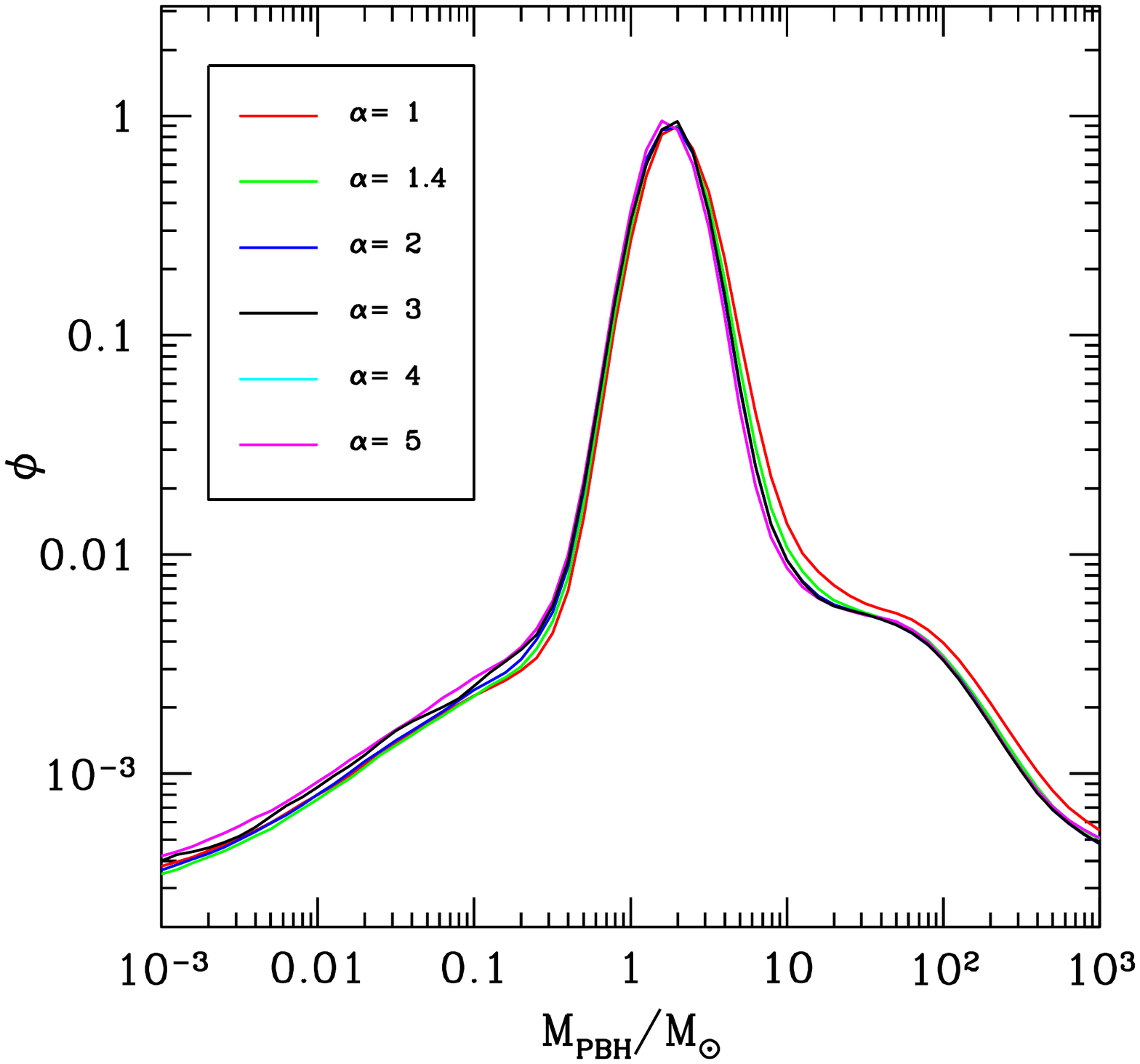}
  \vspace{-2.3cm}
  \caption{ {\bf The effect of the spectral index and the shape profile on the mass function.} The mass function $\phi(M_\mathrm{PBH})$ is shown for different values of $n_s$ with $\alpha=3$ (left panel), and different values of $\alpha$ with $n_s=0.9642$, the same value as used in Figure~\ref{fig:mass function} (right panel). The amplitude of the power spectrum is fixed in each case to give $f_\mathrm{PBH}=1$. In the left panel the logarithmic axes allow showing the significant changes to the tail of the distribution, while the peak is almost the same. In the right panel we can appreciate the minor effect of the shape on the whole profile of the mass function.}
  \label{fig:mass_function_ns_alpha}
\end{figure*}

\subsection{PBH abundance from a power law power spectrum}

In this paper, we will limit ourselves to the discussion of PBHs formed when the power spectrum follows a simple power law\footnote{Making use of the results from this paper, reference \cite{Franciolini:2022tfm} makes a fuller comparison of different forms for the power spectrum, and the the reader is directed there for further discussion.}
\begin{equation}
    \mathcal{P}_\zeta(k) = \mathcal{A}\left( \frac{k}{k_*} \right)^{n_s-1},
\end{equation}
where $\mathcal{A}$ is the amplitude of the power spectrum, $n_s$ is the spectral index and $k_*$ is the pivot scale. Note that, although it takes the same form, this power spectrum is separate from that measured on CMB scales and is here used only to describe the power spectrum on the much smaller scales at which we consider PBH formation.
We will consider that $\mathcal{A}$ and $n_s$ are free parameters in our model. The particular value chosen for the pivot scale $k_*$ is arbitrary (but does affect the overall normalisation of the power spectrum),
and for convenience, we will take it to correspond with the QCD phase transition, where the critical value $\dc$ takes its minimum value at horizon mass $M_*$ (see Figure \ref{fig:delta_c_QCD}). We therefore take $k_*$ to be given by $k_*=2\pi/r_*=4.442\times 10^6\mathrm{Mpc}^{-1}$.

In order to fully explore the effects of the transition, and for concreteness, we will consider parameters such that the following 2 conditions are met:
\begin{enumerate}
    \item If the effect of the phase transition is neglected, PBH formation is close to scale invariant at the scale of the phase transition, $M_*$. We define this such that the derivative of the total PBH formation rate is zero at the time when the horizon mass is $M_\mathrm{H}=M_*$:
    \begin{equation}
        \left.\frac{\mathrm{d}}{\mathrm{d}\MH}\left(\MH^{-1/2}\beta\right)\right|_{M_{*}} = 0.
    \end{equation}
The factor $\MH^{-1/2}$ arises from the redshift term (as seen in equation \eqref{eqn:fpbh2}). This means that, as much as possible, any features seen in the mass function are due to the QCD transition, rather than to features in the primordial power spectrum. 
    \item The overall abundance of PBHs formed in the range $0.01 M_\odot<\MH<100 M_\odot$ make sup the DM abundance, $f_\mathrm{PBH}=1$. The calculation is not sensitive the values of these cut-offs, as the mass function is strongly suppressed around these scales. However, the PBH abundance does eventually diverge if no cut-offs are included.
\end{enumerate}
Applying these conditions, we initially choose the parameters $\mathcal{A}=9.76\times 10^{-3}$, and $n_s=0.964$. For this paper, we will only consider this power spectrum in order to compute the effect of the phase transition. A separate paper uses the results presented here to compare the mass function from varying power spectra, and to compare these to the LIGO-Virgo black hole masses \cite{Franciolini:2022tfm}.

Figure \ref{fig:mass function} shows the mass function $\phi$ for PBHs formed during the phase transition, for profile shapes corresponding to $\alpha=3$ - which is close to the expected value for a broad power spectrum. The red dashed line shows the predicted mass function if there was no phase transition - and we see that it is (approximately) scale-invariant over the scales considered. The blue dashed line shows the mass function that would be calculated using the data for the changing threshold value $\dc$ during the phase transition, but using the critical scaling relationship given by \eqref{eqn:criticalScaling} - as used by e.g. reference \cite{Escriva:2022bwe}. Finally the solid line shows the mass spectrum given by the full calculation.

We can see that accounting for the phase transition increases PBH abundance by a factor $\mathcal{O}(10^3)$, and that the effect is dominated by the change in the critical value. Accounting correctly for the mass of PBHs produced during the transition results in a more peaked mass function, which peaks at a marginally lower value ($1.9M_\odot$ instead of $2.2M_\odot$), and has an effect on the total abundance of order 0.2. For most current purposes, therefore, we consider that it is sufficient to account only for the changing threshold value. However, should more precision be required (such as comparing the mass function to the masses of the LIGO-Virgo black holes) then it is advisable to utilise the full data set from the simulations describing the PBH mass.

We consider the effect of changing the spectral index of the power spectrum in the left panel of Figure \ref{fig:mass_function_ns_alpha}. In this paper, we limit ourselves to the consideration of power spectra which, in the absence of the phase transition, predicts a mass function close to flat. For the values of the spectral index considered, $n_s = 0.93,0.95,0.9642,0.98$, we see that the peak of the mass function is largely unaffected, although there are significant changes to the tails of the distribution. A more red-tilted (blue-tilted) spectrum, corresponding to smaller (larger) $n_s$, predict a larger abundance of high (low) mass PBHs, and a lower abundance of low (high) mass PBHs. In order to be compatible with the observations of black hole masses from LVK, a strongly red-tilted spectrum is therefore necessary (see \cite{Franciolini:2022tfm} for further discussion).

The right panel of Figure \ref{fig:mass_function_ns_alpha} shows the mass functions predicted for different profile shapes, corresponding to $\alpha=1.0,1.4,2.0,3.0,4.0,5.0$, and the amplitude of the power spectrum in each case is fixed to give $f_\mathrm{PBH}=1$. Considering different profile shapes can have a large effect on the fiducial value for $\dc$, which has a very large impact on the abundance of PBHs. However, once the amplitude of the power spectrum is adjusted to account for this difference, we conclude that our calculated mass function is robust against changing values for $\alpha$. The peak of the mass function shifts only by a small amount, from $1.7M_\odot$ for $\alpha=5$, to $1.9M_\odot$ for $\alpha=1$.


\section{Conclusions}
\label{sec:conclusions}

PBHs form more easily during cosmic phase transitions and
annihilation epochs than during a pure radiation dominated phase.
This may lead to a dramatic enhancement of the PBH mass function
on the horizon mass scale of such transitions. In particular, PBHs
formed during the cosmic QCD transition may partially contribute to the
LIGO/Virgo observed events of massive black hole mergers.
In this paper we have performed a detailed numerical study of
PBH formation during the QCD transition. The goal of the study was to make 
an accurate derivation of the PBH mass spectrum, as a first step towards
comparing with merger event catalogs of the LIGO/Virgo collaboration. 

We confirm that even though the reduction of $\cs^2$ and $w$ during
the QCD transition is quite small $\sim 10\%$, for scale-invariant,
Gaussian primordial curvature fluctuations of cosmologically interesting amplitude, PBH formation is a factor $\sim 1000$ more likely during the QCD epoch than before or after. This imprints the QCD horizon mass scale into the
PBH mass function. We find for the peak scale 
$M_{\rm PBH}\approx 1.9 M_{\odot}$, with $\sim 90\%$ of the PBHs
having masses between $0.7M_{\odot}$ and $4.5M_{\odot}$. These values
are surprisingly robust with respect to variations in the 
curvature fluctuation shape parameter $\alpha$ and spectral
index $n_s\approx 1$. 

Our study reveals that even in the case of PBH formation with
a varying equation of state during the QCD epoch, critical scaling
approximately holds, albeit with a somewhat changed exponent $\gamma$
and not extending to quite as large $(\delta - \dc)$. We also
find that for the same $(\delta - \dc)$ PBHs formed during the
QCD epoch have higher mass than those formed during radiation domination, when
$w = 1/3$. Furthermore, we find that for a wide range of curvature
fluctuation scales (i.e. mass at horizon entry) the apparent horizon
always appears in conditions when the medium is close to the depth of the
transition (i.e. close to the approximate minimum of $\cs^2$ and $w$). 

During the preparation of this work we shared some of the numerical results of this work with a group of collaborators~\cite{Franciolini:2022tfm}, i.e. the threshold and the scaling law behaviour for $\alpha=3$, consistent with a nearly scale invariant power spectrum with $\ns=1$. Using a similar approach for the computation of the mass distribution as done here, it was found that the LIGO/Virgo observations do not allow the majority of the dark matter to be in the form of stellar mass PBHs. However a sub population of PBHs is compatible with the gravitational wave signals we have, and would help to explain some events like GW190814 where the secondary of the binary system is falling in the mass gap.

When comparing to a prior study \cite{Escriva:2022bwe}, this work is much more detailed and complete, which is essential for having a consistent computation of the mass distribution and abundance of PBHs from a given power spectrum of cosmological perturbations. The present numerical analysis we have done is giving a variation of the threshold of about $30\%$ with respect to what was obtained before, (see Appendix \ref{sec:comparison}). In particular we made a full computation of the scaling law behaviour during the QCD transition, which has not been done previously. All of this significantly affects the computation of the mass distribution, and abundance, as we show in Appendix~\ref{sec:comparison} making a detailed comparison between our results and the ones obtained in~\cite{Escriva:2022bwe}.

Finally, we have contemplated PBH formation during the
$e^+e^-$ annihilation epoch leading to massive 
$M_{\rm PBH}\sim 10^5 M_{\odot}$ black holes. Contrary to current
belief, we argue that PBH formation during this period may actually be
suppressed due to neutrino diffusion/free-streaming. Only a detailed
study taking neutrinos into account may provide a definite answer. Such
a study is beyond the scope of the present paper.

\begin{acknowledgments}
We warmly thank G.~Franciolini, P.~Pani, A.~Urbano, S.~Clesse, A. Escriva, A.J.~Iovino and J.~Miller for useful discussion and comments. 
The numerical computations were performed at the Sapienza University of Rome on the Vera cluster of the Amaldi Research Center funded by the MIUR program “Dipartimento di Eccellenza” (CUP: B81I18001170001).
The work of I.M. has received funding from the European Union’s Horizon2020 research and innovation programme under the Marie Skłodowska-Curie grant agreement No 754496. SY is supported by a Marie Curie-Sklodowska research fellowship. 
I.M. is very grateful to Tomohiro Harada for hospitality at the Rikkyo University of Tokyo where this work has been finalised.

\end{acknowledgments}

\appendix
\section{The cosmic time slicing}
\subsection{Misner-Sharp-Hernandez equations}
\label{sec:MSH}
Here we present the Misner–Sharp–Hernandez equations~\cite{Misner:1964je} which we have used to derive the initial conditions in gradient expansion (see Section \ref{sec:gradient_expansion}), used for the numerical simulation in Section \ref{sec:Numerical_results}. 

Consider the ‘cosmic time’ metric given by equation \eqref{eq:metric_MS} with the definitions of $U$, $\Gamma$ and $M$ given in equations \eqref{eq:U&Gamma} and \eqref{eq:MS_mass} and a perfect fluid with a diagonal stress energy tensor given by 
\begin{equation}
    T^{\mu\nu} = (\rho+p) u^\mu u^\nu + pg_{\mu\nu}
\end{equation}
where $\rho$ is the total energy density and $p$ is the pressure. Then the Misner–Sharp–Hernandez hydrodynamic equations obtained from the Einstein equations and the conservation of the stress energy tensor are:
\begin{align}
   & D_tU = - \left[ \frac{\Gamma}{\rho+p}D_rp + \frac{M}{R^2} + 
    4\pi R^2 p \right]  \label{eq:DtU} \\
   & D_t\rho_0 = - \frac{\rho_0}{\Gamma R^2} D_r(R^2U) \label{eq:continuity} \\
   & D_t\rho = \frac{\rho+p}{\rho_0} D_t\rho_0  \label{eq:energy} \\
   & D_rA = - \frac{A}{\rho+p}D_rp \\
   & D_rM = 4\pi R^2 \Gamma\rho  
\end{align}
where $\rho_0$ in equations \eqref{eq:continuity} and \eqref{eq:energy} is the rest mass density (or the compression factor for a fluid of particles without rest mass). These form the basic set, together with the Hamiltonian constraint given by equation \eqref{eq:Hamiltonian},
 \begin{equation}
 \Gamma^2 = 1 + U^2 - \frac{2M}{R} 
\end{equation}
Two other useful expressions coming from the Einstein equations are
\begin{align}
    D_t\Gamma = - \frac{U}{\rho+p} D_rp \,, \\
    D_tM = - 4\pi R^2Up \,. \label{eq:DtM}
\end{align}

In order to solve this set of equations we need to supply an equation of state specifying the relation between the pressure and the different components of the energy density. For a simple ideal particle gas, we have that
\begin{equation}
    p = (\gamma - 1)\rho_0\varepsilon,
\end{equation}
where $\varepsilon$ is the specific internal energy, related to the velocity dispersion (temperature) of the fluid particles and $\gamma$ is the adiabatic index. The total energy density $\rho_0$ is the sum of the rest mass density and the internal energy density:
\begin{equation}
    \rho = \rho_0(1+\varepsilon)
\end{equation}
When the contribution of the rest mass of the particles to the total energy density is negligible ($\rho\gg\rho_0$, $\varepsilon\gg1$) we get the standard (one-parameter) equation of state used for a cosmological fluid
\begin{equation}
    p = \omega\rho
\end{equation}
A pressureless fluid ($\omega=0$) corresponds to the case where the specific internal energy $\varepsilon$ is effectively zero. In this paper we are considering an equation of state as given by \eqref{eq:EoS} with $\omega$ being a function of temperature, to describe the QCD transition,  as discussed in Section \ref{sec:EoS}.

\subsection{The quasi-homogeneous solution}
\label{sec:QHS}
In the gradient expansion approach~\cite{Shibata:1999zs,Tomita:1975kj,Salopek:1990jq,Polnarev:2006aa,Harada:2015yda} one makes a perturbative expansion of the MSH-equations in the regime of small pressure gradients, which allows consideration of a time independent comoving curvature profile $\zeta(r)$~\cite{Lyth:2004gb}, also for $\zeta\sim1$. 

To simplify the calculation, it is convenient to consider a pure growing mode. In this case the first non-zero order of the expansion is $\mathcal{O}(\epsilon^2)$\cite{Tanaka:2006zp}, where the small parameter $\epsilon$ has already been defined in equation~\eqref{eq:epsilon}, 
\begin{equation}
    \epsilon = \frac{1}{aH\crm e^{\zeta(\crm)}}\,, \quad\quad \frac{\dot{\epsilon}}{\epsilon} = \frac{1+3w(\rho)}{2}H \,. 
\end{equation}
 The second expression for its first time-derivative is obtained using the equations describing the behavior of a FLRW Universe 
\begin{align}
     & H^2 = \frac{8\pi}{3} \rhob \nonumber, \\
     & \dot{\rho} + 3H\lp\rho+p\rp = 0, \\
     &\frac{\ddot a}{a}=\frac{4\pi}{3}\lp\rho+3p\rp, \nonumber
\end{align}
combined with the equation of state for the QCD transition, which we have seen in Section~\ref{sec:EoS}.

Following the same computation as in~\cite{Polnarev:2006aa}, after some manipulation of the MSH equations, one gets the following set of differential equations to solve for the radial component of the perturbation, indicated with the corresponding tilda-variables:
\begin{align}
    & \tilde{\rho} = \frac{r\zeta^\prime(r)}{1+r\zeta^\prime(r)}\tilde M + \frac{1}{3r^2}\lp r^3\tilde M \rp^\prime \nonumber, \\
    & \tilde U = \frac{1}{2} \left[ \tilde M + \zeta^\prime(r) \lp\frac{2}{r} + \zeta^\prime(r)\rp\frac{\crm^2e^{2\zeta(\crm)}}{e^{2\zeta(r)}} \right] \nonumber,\\
    &\tilde{M} + \frac{\d\tilde{M}}{\d\xi} = -3 \left[1+3w(\rhob)\right] \tilde U \nonumber, \\
    & \tilde{A} = - \frac{\cs^2}{1+w(\rho_b)} \tilde{\rho}, \\
    & \left[ 1 + 3w(\rhob)\right] \tilde B + 
    \frac{\d\tilde B}{\d\xi} = - \frac{r}{1+r\zeta^\prime(r)} \tilde A^\prime \nonumber, \\
    & \left[ 1 + 3w(\rhob)\right] \tilde R + 
    \frac{\d\tilde R}{\d\xi} = \tilde A + \tilde U \nonumber, 
\end{align}
where $\d/\d\xi \equiv H^{-1}\d/dt$ and in the super-horizon regime the sound speed can be expressed in terms of background quantities as
\begin{equation}
   \cs^2 = w(\rhob) + \rhob\frac{\d w(\rhob)}{\d\rhob} \,.
\end{equation}
The solution of this system of differential equations gives the initial conditions for the numerical simulations, which for the energy density and the velocity field are 
\begin{align}
    & \frac{\delta\rho}{\rho_b} =
    - \frac{4}{3} \Phi \left(\frac{1}{aH}\right)^2 e^{-5\zeta(r)/2} 
    \nabla^2 e^{\zeta(r)/2}, \\
    & \frac{\delta U}{U} = 
    \lp\Phi-1\rp \left(\frac{1}{aH}\right)^2 \zeta^\prime(r) 
    \lp \frac{2}{r} + \zeta^\prime(r) \rp e^{-2\zeta(r)}. 
    \end{align}
The other variables can be written as a linear combination of these
\begin{align}
    & \frac{\delta M}{M} = -
    \frac{1}{2} \lp\Phi-1\rp \frac{\delta U}{U} \nonumber, \\
    & \delta A = - \frac{\cs^2}{1+w(\rho_b)} \frac{\delta\rho}{\rho_b} \nonumber, \\
    & \frac{\delta B}{B} = \frac{I_1}{\Phi} r\lp \frac{\delta\rho}{\rho_b} \rp^\prime,  \\
    & \frac{\delta R}{R_\mathrm{b}} = - \frac{I_1}{\Phi} \frac{\delta\rho}{\rho_b} + \frac{I_2}{\Phi-1} \frac{\delta U}{U}, \nonumber 
\end{align}
where the coefficients $\Phi$, $I_1$ and $I_2$ depend on the moment of the horizon crossing with respect to the QCD transition. These are obtained by solving the following differential equations, written with respect to the cosmological horizon used as a measure of time

\begin{figure}[t!]
\centering
\vspace{-1cm}
\includegraphics[width=0.5\textwidth]{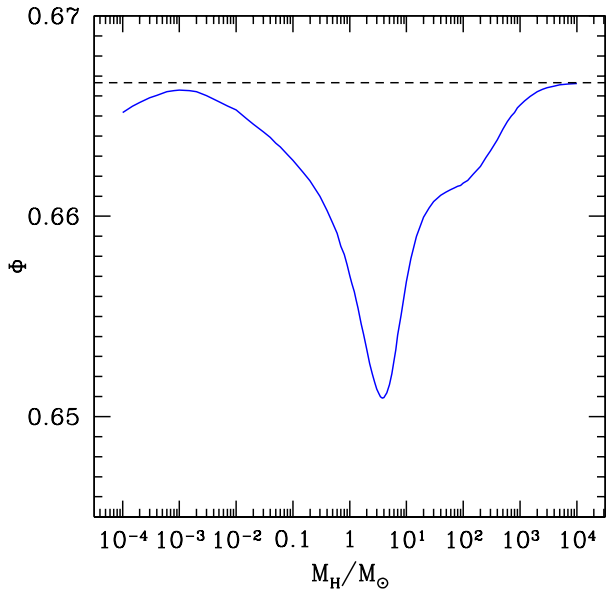}
\vspace{-2.3cm}
\caption{\label{fig:Phi} {\bf The behaviour of $\Phi(t)$.} The solution of equation \eqref{eq:Phi} is plotted against the mass of the cosmological horizon $\MH$ measured at horizon crossing, normalized with respect the solar mass $M_\odot$. The dashed line shows the value of $\bar{\Phi}=2/3$ when $\bar{w}=1/3$.}
\end{figure}

\begin{figure*}[t!]
\centering
\vspace{-1.cm}
\includegraphics[width=0.495\textwidth]{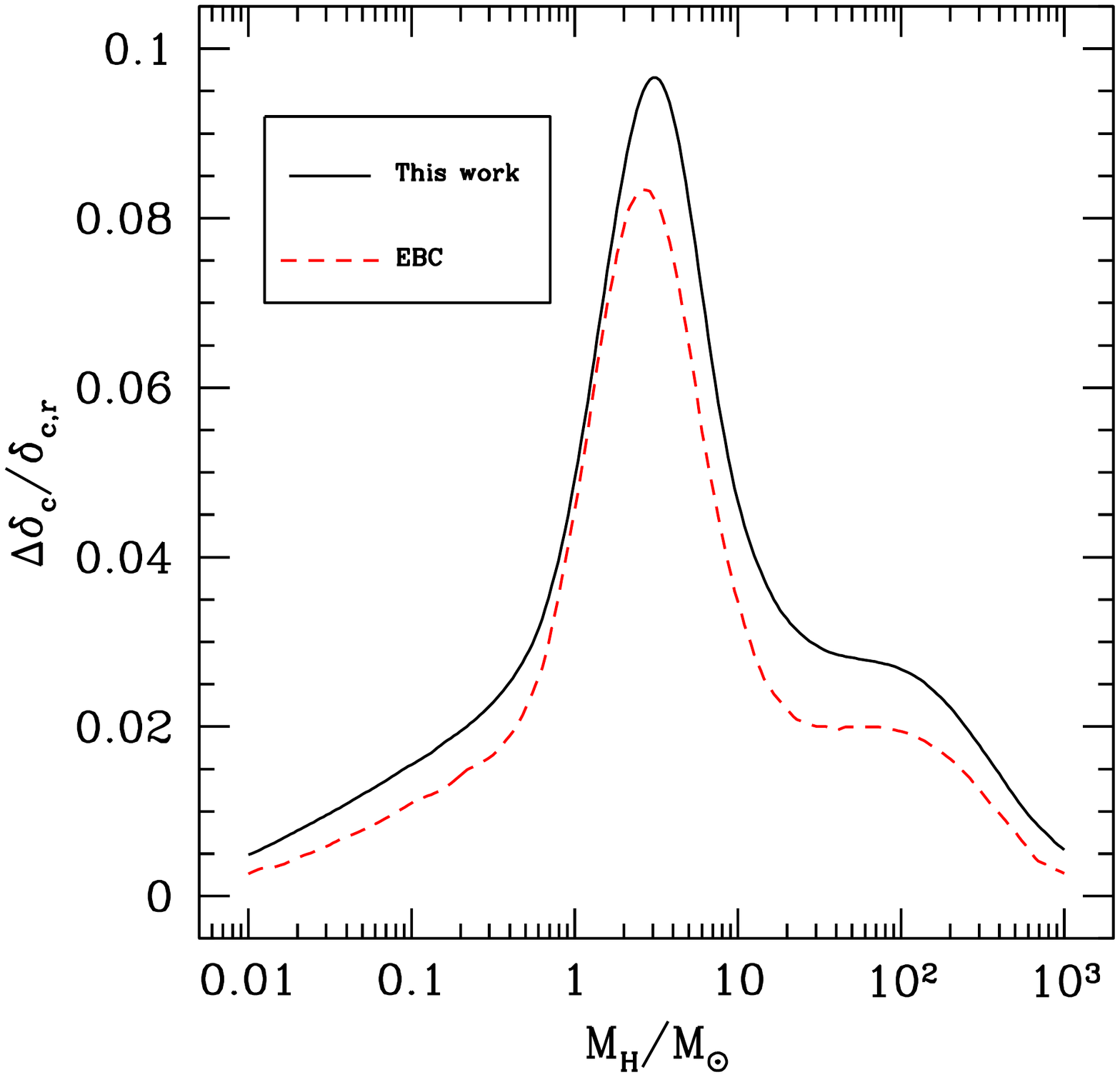}
\includegraphics[width=0.495\textwidth]{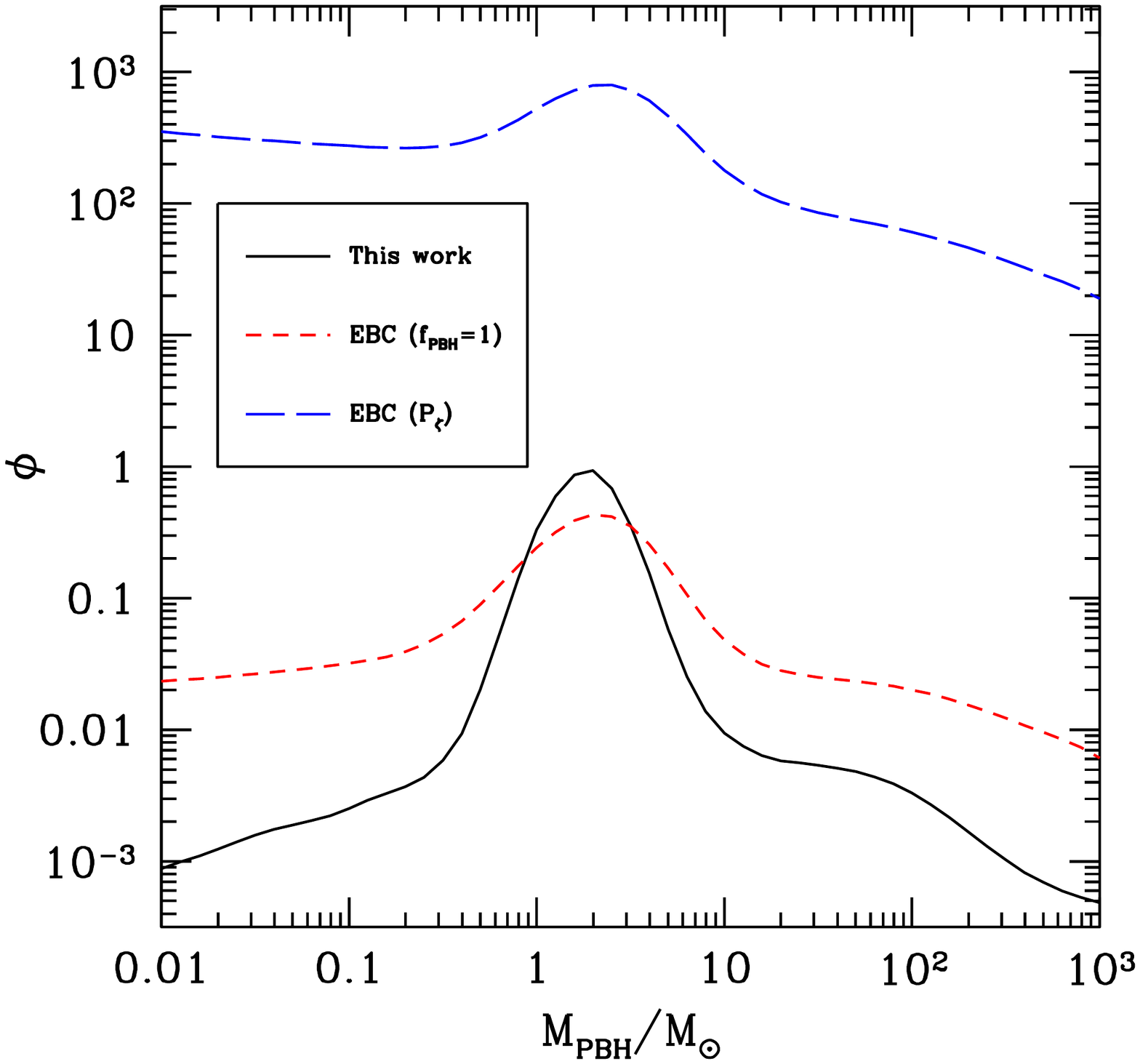}
\vspace{-2.3cm}
\caption{{\bf PBH threshold and Mass functions with different approaches.} In the left panel the fractional change in the critical value is plotted for our work and EBC. Our results correspond to the shape given by $\alpha=3$, whilst the results from EBC correspond to $\alpha=q=2.5$. We see approximately a $25\%$ larger effect from the phase transition in our results. In the right panel the mass function calculated here is compared with that calculated using the methodology from EBC. The solid black line shows our calculation. The dashed red line shows the calculation from EBC, but with the amplitude of the power spectrum normalised to give $f_\mathrm{PBH}=1$. The dashed blue line shows the calculation from EBC, using the same power spectrum as the solid black line.}
\label{fig:EBC}
\end{figure*}

\begin{align}
    & \RH \frac{\d\Phi}{\d\RH} + \frac{5+3w}{3(1+w)} \Phi = 1 \nonumber, \\
    & \RH \frac{\d I_1}{\d\RH} + \frac{2(1+3w)}{3(1+w)} I_1 = 
    \frac{2\cs^2}{3\lp1+w\rp^2}\Phi \label{eq:PhiI1I2}, \\
    & \RH \frac{\d I_2}{\d\RH} + \frac{2(1+3w)}{3(1+w)} I_2 = 
    \frac{2\lp\Phi-1\rp}{3\lp1+w\rp}, \nonumber
\end{align}
In the limit of $w=\mathrm{const.}$, the solution of these equations is given by the averaged values 
\begin{align}
    & \bar \Phi = \frac{3(1+w)}{5+3w}  \nonumber, \\
    & \bar{I}_1 = \frac{3w}{(1+3w)(5+3w)},\label{eq:soluntion_PhiI1I2} \\
    & \bar{I}_2 = - \frac{2}{(1+3w)(5+3w)} \nonumber,
\end{align}
which is is an attractor solution of Eqs.~\eqref{eq:PhiI1I2}, i.e. if $w(t)$ slowly varies in time, ${\d \Phi(t)}/{\d t} = {\d I_{1,2}(t)}/{\d t} \simeq 0$ and the evolution of $\Phi, I_{1,2}$ approaches the averaged values (for $w=1/3$ this gives $\Phi=2/3$, $I_1=1/12$ and $I_2=-1/6$).

The behavior of $\Phi$ across the QCD transition, entering in the computation for the threshold $\dc$, is shown in Fig.~\ref{fig:Phi}: this differs from the averaged values particularly in the region where $w$ and $\cs^2$ are quickly varying with respect the mass of the cosmological horizon $\MH=1/2H$. Although the relative change of $\Phi$, with respect to the constant value of $\bar{\Phi}$ is only of order a few percent, this gives a non-negligible contribution to the modified value of the threshold during the QCD transition which, as we we have seen in Section \ref{sec:Numerical_results}, has an overall change of about $10\%$.

\section{Comparison with previous literature}
\label{sec:comparison}
Several months prior to the publication of this paper, another paper exploring the formation of PBHs during the QCD was released - Escriva, Bagui and Clesse \cite{Escriva:2022bwe} (hereafter, referred to as EBC). Whilst the results from their simulations are broadly in line with the results presented here, our calculations provide a significant advance in accuracy and methodology.

\subsection{The threshold}
Whilst significant, the changes to the calculated values of $\dc$ are relatively minor - as shown in Figure \ref{fig:EBC}.
To make the comparison, we have used the values for $\dc$ from our paper corresponding to a shape parameter $\alpha=3$ (corresponding to the expected profile shape for a broad power spectrum), and used the closest fit, $q=\alpha=2.5$ from EBC. This small difference means that we slightly underestimate the differences in $\dc$ between the papers, which would have been at least $25\%$ if comparing the same value of $\alpha$ (see Figure~\ref{fig:EBC}).

In this paper the curvature profile used to set the initial conditions is using an exponential basis in terms of $\alpha$, with the compaction function given by
\begin{equation}
\mathcal{C}(\tilde r) = \mathcal{A} \,\tilde r^2 \exp\left[ -\frac{1}{\alpha} \left( \frac{\tilde r}{\tilde r_\textrm{m}} \right)^{2\alpha} \right]
\end{equation}
where $\tilde r = r e^{\zeta(r)}$, and $\mathcal{A}$ is a parameter varying the peak of the compaction function, i.e. the perturbation amplitude. Although this parameterisation gives a profile of $\delta\rho/\rho_b$ with a spiky behavior in the centre for $\alpha\leq0.5$  (the first derivatives goes to infinity), the profile of $\delta\rho/\rho_b$ is compensated for any values of $\alpha$, i.e. the overdensity is always compensated by an underdensity, keeping the total energy density of the perturbed region equal to the total energy density of the unperturbed Universe.

In EBC instead, the authors has preferred to relax this aspect, using a polynomial form to set up their initial conditions, to have always a smooth behaviour in the centre. This has the drawback of not having compensated perturbations for $\alpha\leq0.5$. Because the smoothing of the central behaviour is in any case obtained by the numerical evolution as an effect of the pressure gradients after a few time steps, we argue that our choice of an exponential parameterisation, justified by the energy conservation, is preferable. Debating more about this different approach in setting up the initial conditions with respect to the one used by EBC is beyond the scope of this work.

It is also worth mentioning that the authors could not see the behaviour of the scaling law that we have analysed in Section~\ref{sec:scaling}, not being able to consider very small values of $\delta-\dc$ as done in this paper: to get down to a very small value of $\delta-\dc$ it is crucial to have a code using an AMR scheme, describing with enough accuracy the formation of regions with large density gradients, characterised by strong compression waves propagating outwards~\cite{Musco:2008hv}.

\subsection{The Mass distribution}
Our paper does, however, make significant improvements on the calculation of the PBH abundance and mass function. We employ a peaks theory calculation of the PBH abundance, rather than a Press-Schechter approach, and we also account for the non-linearity of the density $\delta$ relative to the curvature perturbation $\zeta$. Overall, this means that the power spectrum calculated by EBC for giving $f_\mathrm{PBH}=1$ is around 50\% smaller than that calculated here (or alternatively, that the calculated PBH abundance differs by a few orders of magnitude for the same power spectrum).

Figure \ref{fig:EBC} shows the difference in the calculated mass functions between our paper and EBC. Note that we have taken the fiducial threshold value for collapse (during radiation domination) to be the same, to avoid spurious effects for the abundance - which is extremely sensitive to small changes in $\dc$. The dashed red line shows the calculation from our paper - and is the same as that plotted in Figure \ref{fig:mass function}. The solid black line on the plot is the mass function which would be calculated by EBC using the same power spectrum - and is different by several orders of magnitude. This is expected, and is due to our inclusion of the non-linear effects in the density contrast. The dotted blue line shows the EBC mass function, with the amplitude of the power spectrum modified such that $f_\mathrm{PBH}=1$ for PBHs in the range $0.01<M_\mathrm{PBH}/M_\odot<100$ (the same criterion used throughout this paper). Here, we see that the peak of the mass function calculated by EBC is significantly broader and lower than we calculate. This is mostly due to using peaks theory to perform the calculation, but is also a result of the differences in the value of $\dc$ during the transition. Using peaks theory predicts a smaller power spectrum to produce the same number of PBHs (or equivalently, a higher abundance of PBHs for the same power spectrum). The result is that the PBH abundance is more sensitive to changes in $\dc$, resulting in a sharper peak to the mass function.

$$$$
\begin{figure*}[t!]
	\centering
    \vspace{-1.cm}
	\includegraphics[width=0.495\textwidth]{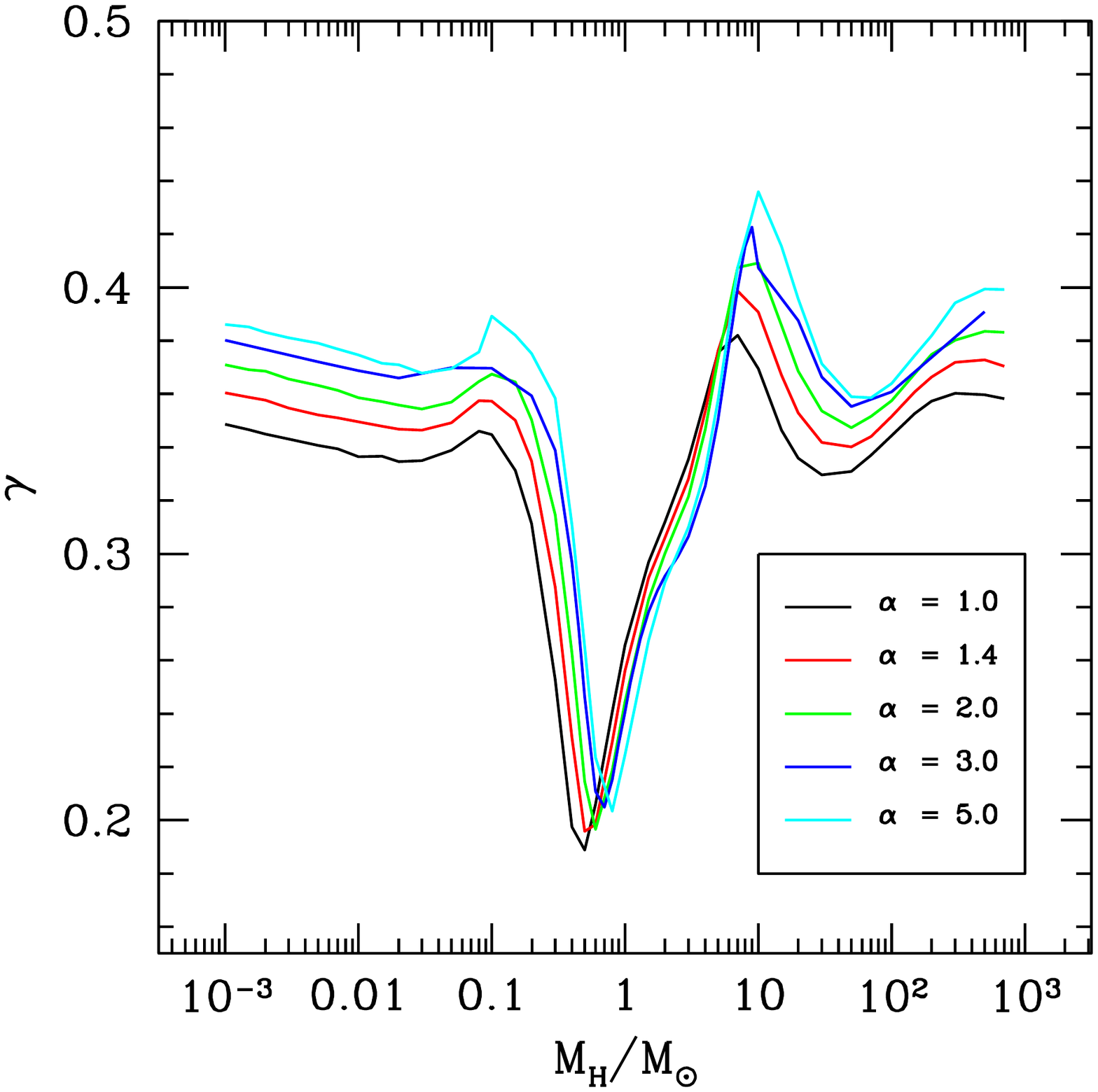}
	\includegraphics[width=0.495\textwidth]{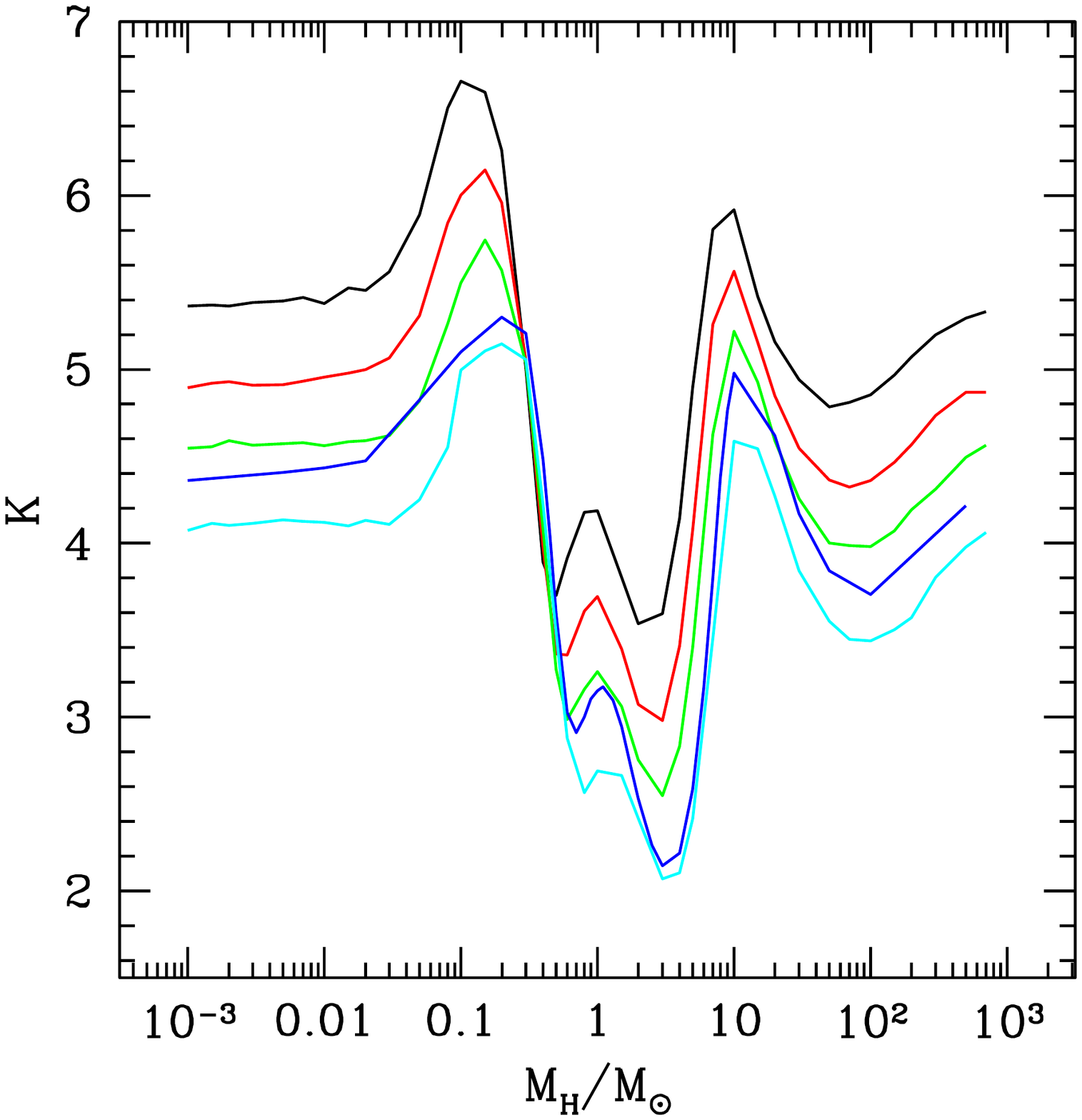}
\vspace{-2.3cm}
\caption{{\bf The scaling parameters during the QCD transition.} This figure shows how the parameters $\gamma$ and $K$ are modified during the QCD transition, plotted as functions of $M_\mathrm{H}/M_\odot$ for $\alpha=1,1.4,2,3,5$, indicated using different line colours.}
\label{fig:Kgamma_QCD}
\end{figure*}
$$$$

\subsection{The scaling law parameters}
In~\cite{Franciolini:2022tfm}, for the purpose of simplifying the calculation for the Bayesian inference analysis performed to compare the LIGO/Virgo catalog with the mass distribution of PBH obtained with these numerical results, the parameters $\gamma$ and $K$ of the scaling law obtained in Section~\ref{sec:scaling} have been numerically fitted by computing an averaged value through the whole region of $\delta-\dc$, for the case $\alpha=3$. This assumes that the scaling law remains valid, with $\gamma$ and $K$ being then just a simple function of $\MH$. This is an approximation which we did not need to make in this work when computing the mass distribution of PBHs, where we have instead taken into account the complete data set concerning the final PBH mass obtained from the numerical simulations.

Nevertheless it is interesting to generalize the approximation used in~\cite{Franciolini:2022tfm} also for the different shapes considered in this work. This is summarised in Figure~\ref{fig:Kgamma_QCD} where the behaviour of the averaged values of $\gamma$ and $K$, respectively in the left and right panel, has been plotted for $1\leq\alpha\leq5$. 

The same qualitative behaviour is shown for all of the different shapes, where the effect of the transition is delayed to larger horizon masses for larger $\alpha$, consistent with the minimum of $\dc$ being delayed towards larger masses when 
$\alpha$ is increasing (see Figure~\ref{fig:delta_c_QCD}). Because of the approximation of the numerical fit, in Figure~\ref{fig:Kgamma_QCD} the averaged values of $\gamma$ do not converge exactly to $\gamma\simeq0.36$ when $w=1/3$ (the value of $\gamma$ for a radiation dominated medium) keeping instead always a slight dependence on the shape.

\bibliography{PBHs_QCD}

\begin{thebibliography}{79}%
\makeatletter
\providecommand \@ifxundefined [1]{%
 \@ifx{#1\undefined}
}%
\providecommand \@ifnum [1]{%
 \ifnum #1\expandafter \@firstoftwo
 \else \expandafter \@secondoftwo
 \fi
}%
\providecommand \@ifx [1]{%
 \ifx #1\expandafter \@firstoftwo
 \else \expandafter \@secondoftwo
 \fi
}%
\providecommand \natexlab [1]{#1}%
\providecommand \enquote  [1]{``#1''}%
\providecommand \bibnamefont  [1]{#1}%
\providecommand \bibfnamefont [1]{#1}%
\providecommand \citenamefont [1]{#1}%
\providecommand \href@noop [0]{\@secondoftwo}%
\providecommand \href [0]{\begingroup \@sanitize@url \@href}%
\providecommand \@href[1]{\@@startlink{#1}\@@href}%
\providecommand \@@href[1]{\endgroup#1\@@endlink}%
\providecommand \@sanitize@url [0]{\catcode `\\12\catcode `\$12\catcode
  `\&12\catcode `\#12\catcode `\^12\catcode `\_12\catcode `\%12\relax}%
\providecommand \@@startlink[1]{}%
\providecommand \@@endlink[0]{}%
\providecommand \url  [0]{\begingroup\@sanitize@url \@url }%
\providecommand \@url [1]{\endgroup\@href {#1}{\urlprefix }}%
\providecommand \urlprefix  [0]{URL }%
\providecommand \Eprint [0]{\href }%
\providecommand \doibase [0]{http://dx.doi.org/}%
\providecommand \selectlanguage [0]{\@gobble}%
\providecommand \bibinfo  [0]{\@secondoftwo}%
\providecommand \bibfield  [0]{\@secondoftwo}%
\providecommand \translation [1]{[#1]}%
\providecommand \BibitemOpen [0]{}%
\providecommand \bibitemStop [0]{}%
\providecommand \bibitemNoStop [0]{.\EOS\space}%
\providecommand \EOS [0]{\spacefactor3000\relax}%
\providecommand \BibitemShut  [1]{\csname bibitem#1\endcsname}%
\let\auto@bib@innerbib\@empty
\bibitem [{\citenamefont {Abbott}\ \emph {et~al.}(2019)\citenamefont {Abbott}
  \emph {et~al.}}]{LIGOScientific:2018mvr}%
  \BibitemOpen
  \bibfield  {author} {\bibinfo {author} {\bibfnamefont {B.~P.}\ \bibnamefont
  {Abbott}} \emph {et~al.} (\bibinfo {collaboration} {LIGO Scientific,
  Virgo}),\ }\href {\doibase 10.1103/PhysRevX.9.031040} {\bibfield  {journal}
  {\bibinfo  {journal} {Phys. Rev. X}\ }\textbf {\bibinfo {volume} {9}},\
  \bibinfo {pages} {031040} (\bibinfo {year} {2019})},\ \Eprint
  {http://arxiv.org/abs/1811.12907} {arXiv:1811.12907 [astro-ph.HE]}
  \BibitemShut {NoStop}%
\bibitem [{\citenamefont {Abbott}\ \emph
  {et~al.}(2021{\natexlab{a}})\citenamefont {Abbott} \emph
  {et~al.}}]{LIGOScientific:2020ibl}%
  \BibitemOpen
  \bibfield  {author} {\bibinfo {author} {\bibfnamefont {R.}~\bibnamefont
  {Abbott}} \emph {et~al.} (\bibinfo {collaboration} {LIGO Scientific,
  Virgo}),\ }\href {\doibase 10.1103/PhysRevX.11.021053} {\bibfield  {journal}
  {\bibinfo  {journal} {Phys. Rev. X}\ }\textbf {\bibinfo {volume} {11}},\
  \bibinfo {pages} {021053} (\bibinfo {year} {2021}{\natexlab{a}})},\ \Eprint
  {http://arxiv.org/abs/2010.14527} {arXiv:2010.14527 [gr-qc]} \BibitemShut
  {NoStop}%
\bibitem [{\citenamefont {Abbott}\ \emph
  {et~al.}(2021{\natexlab{b}})\citenamefont {Abbott} \emph
  {et~al.}}]{LIGOScientific:2021djp}%
  \BibitemOpen
  \bibfield  {author} {\bibinfo {author} {\bibfnamefont {R.}~\bibnamefont
  {Abbott}} \emph {et~al.} (\bibinfo {collaboration} {LIGO-VIRGO-KAGRA}),\
  }\href@noop {} {\  (\bibinfo {year} {2021}{\natexlab{b}})},\ \Eprint
  {http://arxiv.org/abs/2111.03606} {arXiv:2111.03606 [gr-qc]} \BibitemShut
  {NoStop}%
\bibitem [{\citenamefont {Zel'dovich}\ and\ \citenamefont
  {Novikov}(1967)}]{Zeldovich:1967}%
  \BibitemOpen
  \bibfield  {author} {\bibinfo {author} {\bibfnamefont {Y.~B.}\ \bibnamefont
  {Zel'dovich}}\ and\ \bibinfo {author} {\bibfnamefont {I.}~\bibnamefont
  {Novikov}},\ }\href@noop {} {\bibfield  {journal} {\bibinfo  {journal}
  {Soviet Astronomy}\ }\textbf {\bibinfo {volume} {10}},\ \bibinfo {pages}
  {602} (\bibinfo {year} {1967})}\BibitemShut {NoStop}%
\bibitem [{\citenamefont {Hawking}(1971)}]{Hawking:1971ei}%
  \BibitemOpen
  \bibfield  {author} {\bibinfo {author} {\bibfnamefont {S.}~\bibnamefont
  {Hawking}},\ }\href@noop {} {\bibfield  {journal} {\bibinfo  {journal} {Mon.
  Not. Roy. Astron. Soc.}\ }\textbf {\bibinfo {volume} {152}},\ \bibinfo
  {pages} {75} (\bibinfo {year} {1971})}\BibitemShut {NoStop}%
\bibitem [{\citenamefont {Carr}(1975)}]{Carr:1975qj}%
  \BibitemOpen
  \bibfield  {author} {\bibinfo {author} {\bibfnamefont {B.~J.}\ \bibnamefont
  {Carr}},\ }\href {\doibase 10.1086/153853} {\bibfield  {journal} {\bibinfo
  {journal} {Astrophys. J.}\ }\textbf {\bibinfo {volume} {201}},\ \bibinfo
  {pages} {1} (\bibinfo {year} {1975})}\BibitemShut {NoStop}%
\bibitem [{\citenamefont {Khlopov}(2010)}]{Khlopov:2008qy}%
  \BibitemOpen
  \bibfield  {author} {\bibinfo {author} {\bibfnamefont {M.~Y.}\ \bibnamefont
  {Khlopov}},\ }\href {\doibase 10.1088/1674-4527/10/6/001} {\bibfield
  {journal} {\bibinfo  {journal} {Res. Astron. Astrophys.}\ }\textbf {\bibinfo
  {volume} {10}},\ \bibinfo {pages} {495} (\bibinfo {year} {2010})},\ \Eprint
  {http://arxiv.org/abs/0801.0116} {arXiv:0801.0116 [astro-ph]} \BibitemShut
  {NoStop}%
\bibitem [{\citenamefont {Niemeyer}\ and\ \citenamefont
  {Jedamzik}(1998)}]{Niemeyer:1997mt}%
  \BibitemOpen
  \bibfield  {author} {\bibinfo {author} {\bibfnamefont {J.~C.}\ \bibnamefont
  {Niemeyer}}\ and\ \bibinfo {author} {\bibfnamefont {K.}~\bibnamefont
  {Jedamzik}},\ }\href {\doibase 10.1103/PhysRevLett.80.5481} {\bibfield
  {journal} {\bibinfo  {journal} {Phys. Rev. Lett.}\ }\textbf {\bibinfo
  {volume} {80}},\ \bibinfo {pages} {5481} (\bibinfo {year} {1998})},\ \Eprint
  {http://arxiv.org/abs/astro-ph/9709072} {arXiv:astro-ph/9709072} \BibitemShut
  {NoStop}%
\bibitem [{\citenamefont {Niemeyer}\ and\ \citenamefont
  {Jedamzik}(1999)}]{Niemeyer:1999ak}%
  \BibitemOpen
  \bibfield  {author} {\bibinfo {author} {\bibfnamefont {J.~C.}\ \bibnamefont
  {Niemeyer}}\ and\ \bibinfo {author} {\bibfnamefont {K.}~\bibnamefont
  {Jedamzik}},\ }\href {\doibase 10.1103/PhysRevD.59.124013} {\bibfield
  {journal} {\bibinfo  {journal} {Phys. Rev. D}\ }\textbf {\bibinfo {volume}
  {59}},\ \bibinfo {pages} {124013} (\bibinfo {year} {1999})},\ \Eprint
  {http://arxiv.org/abs/astro-ph/9901292} {arXiv:astro-ph/9901292} \BibitemShut
  {NoStop}%
\bibitem [{\citenamefont {Musco}\ \emph {et~al.}(2005)\citenamefont {Musco},
  \citenamefont {Miller},\ and\ \citenamefont {Rezzolla}}]{Musco:2004ak}%
  \BibitemOpen
  \bibfield  {author} {\bibinfo {author} {\bibfnamefont {I.}~\bibnamefont
  {Musco}}, \bibinfo {author} {\bibfnamefont {J.~C.}\ \bibnamefont {Miller}}, \
  and\ \bibinfo {author} {\bibfnamefont {L.}~\bibnamefont {Rezzolla}},\ }\href
  {\doibase 10.1088/0264-9381/22/7/013} {\bibfield  {journal} {\bibinfo
  {journal} {Class. Quant. Grav.}\ }\textbf {\bibinfo {volume} {22}},\ \bibinfo
  {pages} {1405} (\bibinfo {year} {2005})},\ \Eprint
  {http://arxiv.org/abs/gr-qc/0412063} {arXiv:gr-qc/0412063} \BibitemShut
  {NoStop}%
\bibitem [{\citenamefont {Chapline}(1975)}]{Chapline:1975tn}%
  \BibitemOpen
  \bibfield  {author} {\bibinfo {author} {\bibfnamefont {G.~F.}\ \bibnamefont
  {Chapline}},\ }\href {\doibase 10.1103/PhysRevD.12.2949} {\bibfield
  {journal} {\bibinfo  {journal} {Phys. Rev. D}\ }\textbf {\bibinfo {volume}
  {12}},\ \bibinfo {pages} {2949} (\bibinfo {year} {1975})}\BibitemShut
  {NoStop}%
\bibitem [{\citenamefont {Jedamzik}(1997)}]{Jedamzik:1996mr}%
  \BibitemOpen
  \bibfield  {author} {\bibinfo {author} {\bibfnamefont {K.}~\bibnamefont
  {Jedamzik}},\ }\href {\doibase 10.1103/PhysRevD.55.R5871} {\bibfield
  {journal} {\bibinfo  {journal} {Phys. Rev. D}\ }\textbf {\bibinfo {volume}
  {55}},\ \bibinfo {pages} {5871} (\bibinfo {year} {1997})},\ \Eprint
  {http://arxiv.org/abs/astro-ph/9605152} {arXiv:astro-ph/9605152} \BibitemShut
  {NoStop}%
\bibitem [{\citenamefont {Jedamzik}(1998)}]{Jedamzik:1998hc}%
  \BibitemOpen
  \bibfield  {author} {\bibinfo {author} {\bibfnamefont {K.}~\bibnamefont
  {Jedamzik}},\ }\href {\doibase 10.1016/S0370-1573(98)00067-2} {\bibfield
  {journal} {\bibinfo  {journal} {Phys. Rept.}\ }\textbf {\bibinfo {volume}
  {307}},\ \bibinfo {pages} {155} (\bibinfo {year} {1998})},\ \Eprint
  {http://arxiv.org/abs/astro-ph/9805147} {arXiv:astro-ph/9805147} \BibitemShut
  {NoStop}%
\bibitem [{\citenamefont {Jedamzik}\ and\ \citenamefont
  {Niemeyer}(1999)}]{Jedamzik:1999am}%
  \BibitemOpen
  \bibfield  {author} {\bibinfo {author} {\bibfnamefont {K.}~\bibnamefont
  {Jedamzik}}\ and\ \bibinfo {author} {\bibfnamefont {J.~C.}\ \bibnamefont
  {Niemeyer}},\ }\href {\doibase 10.1103/PhysRevD.59.124014} {\bibfield
  {journal} {\bibinfo  {journal} {Phys. Rev. D}\ }\textbf {\bibinfo {volume}
  {59}},\ \bibinfo {pages} {124014} (\bibinfo {year} {1999})},\ \Eprint
  {http://arxiv.org/abs/astro-ph/9901293} {arXiv:astro-ph/9901293} \BibitemShut
  {NoStop}%
\bibitem [{\citenamefont {Borsanyi}\ \emph {et~al.}(2016)\citenamefont
  {Borsanyi} \emph {et~al.}}]{Borsanyi:2016ksw}%
  \BibitemOpen
  \bibfield  {author} {\bibinfo {author} {\bibfnamefont {S.}~\bibnamefont
  {Borsanyi}} \emph {et~al.},\ }\href {\doibase 10.1038/nature20115} {\bibfield
   {journal} {\bibinfo  {journal} {Nature}\ }\textbf {\bibinfo {volume}
  {539}},\ \bibinfo {pages} {69} (\bibinfo {year} {2016})},\ \Eprint
  {http://arxiv.org/abs/1606.07494} {arXiv:1606.07494 [hep-lat]} \BibitemShut
  {NoStop}%
\bibitem [{\citenamefont {Bhattacharya}\ \emph {et~al.}(2014)\citenamefont
  {Bhattacharya} \emph {et~al.}}]{Bhattacharya:2014ara}%
  \BibitemOpen
  \bibfield  {author} {\bibinfo {author} {\bibfnamefont {T.}~\bibnamefont
  {Bhattacharya}} \emph {et~al.},\ }\href {\doibase
  10.1103/PhysRevLett.113.082001} {\bibfield  {journal} {\bibinfo  {journal}
  {Phys. Rev. Lett.}\ }\textbf {\bibinfo {volume} {113}},\ \bibinfo {pages}
  {082001} (\bibinfo {year} {2014})},\ \Eprint {http://arxiv.org/abs/1402.5175}
  {arXiv:1402.5175 [hep-lat]} \BibitemShut {NoStop}%
\bibitem [{\citenamefont {Byrnes}\ \emph {et~al.}(2018)\citenamefont {Byrnes},
  \citenamefont {Hindmarsh}, \citenamefont {Young},\ and\ \citenamefont
  {Hawkins}}]{Byrnes:2018clq}%
  \BibitemOpen
  \bibfield  {author} {\bibinfo {author} {\bibfnamefont {C.~T.}\ \bibnamefont
  {Byrnes}}, \bibinfo {author} {\bibfnamefont {M.}~\bibnamefont {Hindmarsh}},
  \bibinfo {author} {\bibfnamefont {S.}~\bibnamefont {Young}}, \ and\ \bibinfo
  {author} {\bibfnamefont {M.~R.~S.}\ \bibnamefont {Hawkins}},\ }\href
  {\doibase 10.1088/1475-7516/2018/08/041} {\bibfield  {journal} {\bibinfo
  {journal} {JCAP}\ }\textbf {\bibinfo {volume} {08}},\ \bibinfo {pages} {041}
  (\bibinfo {year} {2018})},\ \Eprint {http://arxiv.org/abs/1801.06138}
  {arXiv:1801.06138 [astro-ph.CO]} \BibitemShut {NoStop}%
\bibitem [{\citenamefont {Carr}\ \emph {et~al.}(2021)\citenamefont {Carr},
  \citenamefont {Clesse}, \citenamefont {Garc\'\i{}a-Bellido},\ and\
  \citenamefont {K\"uhnel}}]{Carr:2019kxo}%
  \BibitemOpen
  \bibfield  {author} {\bibinfo {author} {\bibfnamefont {B.}~\bibnamefont
  {Carr}}, \bibinfo {author} {\bibfnamefont {S.}~\bibnamefont {Clesse}},
  \bibinfo {author} {\bibfnamefont {J.}~\bibnamefont {Garc\'\i{}a-Bellido}}, \
  and\ \bibinfo {author} {\bibfnamefont {F.}~\bibnamefont {K\"uhnel}},\ }\href
  {\doibase 10.1016/j.dark.2020.100755} {\bibfield  {journal} {\bibinfo
  {journal} {Phys. Dark Univ.}\ }\textbf {\bibinfo {volume} {31}},\ \bibinfo
  {pages} {100755} (\bibinfo {year} {2021})},\ \Eprint
  {http://arxiv.org/abs/1906.08217} {arXiv:1906.08217 [astro-ph.CO]}
  \BibitemShut {NoStop}%
\bibitem [{\citenamefont {Sobrinho}\ and\ \citenamefont
  {Augusto}(2020)}]{Sobrinho:2020cco}%
  \BibitemOpen
  \bibfield  {author} {\bibinfo {author} {\bibfnamefont {J.}~\bibnamefont
  {Sobrinho}}\ and\ \bibinfo {author} {\bibfnamefont {P.}~\bibnamefont
  {Augusto}},\ }\href@noop {} {\  (\bibinfo {year} {2020})},\ \Eprint
  {http://arxiv.org/abs/2005.10037} {arXiv:2005.10037 [astro-ph.CO]}
  \BibitemShut {NoStop}%
\bibitem [{\citenamefont {Sasaki}\ \emph {et~al.}(2016)\citenamefont {Sasaki},
  \citenamefont {Suyama}, \citenamefont {Tanaka},\ and\ \citenamefont
  {Yokoyama}}]{Sasaki:2016jop}%
  \BibitemOpen
  \bibfield  {author} {\bibinfo {author} {\bibfnamefont {M.}~\bibnamefont
  {Sasaki}}, \bibinfo {author} {\bibfnamefont {T.}~\bibnamefont {Suyama}},
  \bibinfo {author} {\bibfnamefont {T.}~\bibnamefont {Tanaka}}, \ and\ \bibinfo
  {author} {\bibfnamefont {S.}~\bibnamefont {Yokoyama}},\ }\href {\doibase
  10.1103/PhysRevLett.121.059901, 10.1103/PhysRevLett.117.061101} {\bibfield
  {journal} {\bibinfo  {journal} {Phys. Rev. Lett.}\ }\textbf {\bibinfo
  {volume} {117}},\ \bibinfo {pages} {061101} (\bibinfo {year} {2016})},\
  \bibinfo {note} {[erratum: Phys. Rev. Lett.121,no.5,059901(2018)]},\ \Eprint
  {http://arxiv.org/abs/1603.08338} {arXiv:1603.08338 [astro-ph.CO]}
  \BibitemShut {NoStop}%
\bibitem [{\citenamefont {Raidal}\ \emph {et~al.}(2019)\citenamefont {Raidal},
  \citenamefont {Spethmann}, \citenamefont {Vaskonen},\ and\ \citenamefont
  {Veermäe}}]{Raidal:2018bbj}%
  \BibitemOpen
  \bibfield  {author} {\bibinfo {author} {\bibfnamefont {M.}~\bibnamefont
  {Raidal}}, \bibinfo {author} {\bibfnamefont {C.}~\bibnamefont {Spethmann}},
  \bibinfo {author} {\bibfnamefont {V.}~\bibnamefont {Vaskonen}}, \ and\
  \bibinfo {author} {\bibfnamefont {H.}~\bibnamefont {Veermäe}},\ }\href
  {\doibase 10.1088/1475-7516/2019/02/018} {\bibfield  {journal} {\bibinfo
  {journal} {JCAP}\ }\textbf {\bibinfo {volume} {02}},\ \bibinfo {pages} {018}
  (\bibinfo {year} {2019})},\ \Eprint {http://arxiv.org/abs/1812.01930}
  {arXiv:1812.01930 [astro-ph.CO]} \BibitemShut {NoStop}%
\bibitem [{\citenamefont {Jedamzik}(2020)}]{Jedamzik:2020ypm}%
  \BibitemOpen
  \bibfield  {author} {\bibinfo {author} {\bibfnamefont {K.}~\bibnamefont
  {Jedamzik}},\ }\href {\doibase 10.1088/1475-7516/2020/09/022} {\bibfield
  {journal} {\bibinfo  {journal} {JCAP}\ }\textbf {\bibinfo {volume} {09}},\
  \bibinfo {pages} {022} (\bibinfo {year} {2020})},\ \Eprint
  {http://arxiv.org/abs/2006.11172} {arXiv:2006.11172 [astro-ph.CO]}
  \BibitemShut {NoStop}%
\bibitem [{\citenamefont {Young}\ and\ \citenamefont
  {Hamers}(2020)}]{Young:2020scc}%
  \BibitemOpen
  \bibfield  {author} {\bibinfo {author} {\bibfnamefont {S.}~\bibnamefont
  {Young}}\ and\ \bibinfo {author} {\bibfnamefont {A.~S.}\ \bibnamefont
  {Hamers}},\ }\href {\doibase 10.1088/1475-7516/2020/10/036} {\bibfield
  {journal} {\bibinfo  {journal} {JCAP}\ }\textbf {\bibinfo {volume} {10}},\
  \bibinfo {pages} {036} (\bibinfo {year} {2020})},\ \Eprint
  {http://arxiv.org/abs/2006.15023} {arXiv:2006.15023 [astro-ph.CO]}
  \BibitemShut {NoStop}%
\bibitem [{\citenamefont {Hütsi}\ \emph {et~al.}(2021)\citenamefont {Hütsi},
  \citenamefont {Raidal}, \citenamefont {Vaskonen},\ and\ \citenamefont
  {Veermäe}}]{Hutsi:2020sol}%
  \BibitemOpen
  \bibfield  {author} {\bibinfo {author} {\bibfnamefont {G.}~\bibnamefont
  {Hütsi}}, \bibinfo {author} {\bibfnamefont {M.}~\bibnamefont {Raidal}},
  \bibinfo {author} {\bibfnamefont {V.}~\bibnamefont {Vaskonen}}, \ and\
  \bibinfo {author} {\bibfnamefont {H.}~\bibnamefont {Veermäe}},\ }\href
  {\doibase 10.1088/1475-7516/2021/03/068} {\bibfield  {journal} {\bibinfo
  {journal} {JCAP}\ }\textbf {\bibinfo {volume} {2103}},\ \bibinfo {pages}
  {068} (\bibinfo {year} {2021})},\ \Eprint {http://arxiv.org/abs/2012.02786}
  {arXiv:2012.02786 [astro-ph.CO]} \BibitemShut {NoStop}%
\bibitem [{\citenamefont {Juan}\ \emph {et~al.}(2022)\citenamefont {Juan},
  \citenamefont {Serpico},\ and\ \citenamefont {Abell\'an}}]{Juan:2022mir}%
  \BibitemOpen
  \bibfield  {author} {\bibinfo {author} {\bibfnamefont {J.~I.}\ \bibnamefont
  {Juan}}, \bibinfo {author} {\bibfnamefont {P.}~\bibnamefont {Serpico}}, \
  and\ \bibinfo {author} {\bibfnamefont {G.~F.}\ \bibnamefont {Abell\'an}},\
  }\href@noop {} {\  (\bibinfo {year} {2022})},\ \Eprint
  {http://arxiv.org/abs/2204.07027} {arXiv:2204.07027 [astro-ph.CO]}
  \BibitemShut {NoStop}%
\bibitem [{\citenamefont {Franciolini}\ \emph
  {et~al.}(2022{\natexlab{a}})\citenamefont {Franciolini}, \citenamefont
  {Musco}, \citenamefont {Pani},\ and\ \citenamefont
  {Urbano}}]{Franciolini:2022tfm}%
  \BibitemOpen
  \bibfield  {author} {\bibinfo {author} {\bibfnamefont {G.}~\bibnamefont
  {Franciolini}}, \bibinfo {author} {\bibfnamefont {I.}~\bibnamefont {Musco}},
  \bibinfo {author} {\bibfnamefont {P.}~\bibnamefont {Pani}}, \ and\ \bibinfo
  {author} {\bibfnamefont {A.}~\bibnamefont {Urbano}},\ }\href {\doibase
  10.1103/PhysRevD.106.123526} {\bibfield  {journal} {\bibinfo  {journal}
  {Phys. Rev. D}\ }\textbf {\bibinfo {volume} {106}},\ \bibinfo {pages}
  {123526} (\bibinfo {year} {2022}{\natexlab{a}})},\ \Eprint
  {http://arxiv.org/abs/2209.05959} {arXiv:2209.05959 [astro-ph.CO]}
  \BibitemShut {NoStop}%
\bibitem [{\citenamefont {Clesse}\ and\ \citenamefont
  {Garc\'\i{}a-Bellido}(2017)}]{Clesse:2016vqa}%
  \BibitemOpen
  \bibfield  {author} {\bibinfo {author} {\bibfnamefont {S.}~\bibnamefont
  {Clesse}}\ and\ \bibinfo {author} {\bibfnamefont {J.}~\bibnamefont
  {Garc\'\i{}a-Bellido}},\ }\href {\doibase 10.1016/j.dark.2016.10.002}
  {\bibfield  {journal} {\bibinfo  {journal} {Phys. Dark Univ.}\ }\textbf
  {\bibinfo {volume} {15}},\ \bibinfo {pages} {142} (\bibinfo {year} {2017})},\
  \Eprint {http://arxiv.org/abs/1603.05234} {arXiv:1603.05234 [astro-ph.CO]}
  \BibitemShut {NoStop}%
\bibitem [{\citenamefont {Calcino}\ \emph {et~al.}(2018)\citenamefont
  {Calcino}, \citenamefont {Garcia-Bellido},\ and\ \citenamefont
  {Davis}}]{Calcino:2018mwh}%
  \BibitemOpen
  \bibfield  {author} {\bibinfo {author} {\bibfnamefont {J.}~\bibnamefont
  {Calcino}}, \bibinfo {author} {\bibfnamefont {J.}~\bibnamefont
  {Garcia-Bellido}}, \ and\ \bibinfo {author} {\bibfnamefont {T.~M.}\
  \bibnamefont {Davis}},\ }\href {\doibase 10.1093/mnras/sty1368} {\bibfield
  {journal} {\bibinfo  {journal} {Mon. Not. Roy. Astron. Soc.}\ }\textbf
  {\bibinfo {volume} {479}},\ \bibinfo {pages} {2889} (\bibinfo {year}
  {2018})},\ \Eprint {http://arxiv.org/abs/1803.09205} {arXiv:1803.09205
  [astro-ph.CO]} \BibitemShut {NoStop}%
\bibitem [{\citenamefont {Peta\v{c}}\ \emph {et~al.}(2022)\citenamefont
  {Peta\v{c}}, \citenamefont {Lavalle},\ and\ \citenamefont
  {Jedamzik}}]{Petac:2022rio}%
  \BibitemOpen
  \bibfield  {author} {\bibinfo {author} {\bibfnamefont {M.}~\bibnamefont
  {Peta\v{c}}}, \bibinfo {author} {\bibfnamefont {J.}~\bibnamefont {Lavalle}},
  \ and\ \bibinfo {author} {\bibfnamefont {K.}~\bibnamefont {Jedamzik}},\
  }\href {\doibase 10.1103/PhysRevD.105.083520} {\bibfield  {journal} {\bibinfo
   {journal} {Phys. Rev. D}\ }\textbf {\bibinfo {volume} {105}},\ \bibinfo
  {pages} {083520} (\bibinfo {year} {2022})},\ \Eprint
  {http://arxiv.org/abs/2201.02521} {arXiv:2201.02521 [astro-ph.CO]}
  \BibitemShut {NoStop}%
\bibitem [{\citenamefont {Gorton}\ and\ \citenamefont
  {Green}(2022)}]{Gorton:2022fyb}%
  \BibitemOpen
  \bibfield  {author} {\bibinfo {author} {\bibfnamefont {M.}~\bibnamefont
  {Gorton}}\ and\ \bibinfo {author} {\bibfnamefont {A.~M.}\ \bibnamefont
  {Green}},\ }\href {\doibase 10.1088/1475-7516/2022/08/035} {\bibfield
  {journal} {\bibinfo  {journal} {JCAP}\ }\textbf {\bibinfo {volume} {08}},\
  \bibinfo {pages} {035} (\bibinfo {year} {2022})},\ \Eprint
  {http://arxiv.org/abs/2203.04209} {arXiv:2203.04209 [astro-ph.CO]}
  \BibitemShut {NoStop}%
\bibitem [{\citenamefont {De~Luca}\ \emph {et~al.}(2022)\citenamefont
  {De~Luca}, \citenamefont {Franciolini}, \citenamefont {Riotto},\ and\
  \citenamefont {Veerm\"ae}}]{DeLuca:2022uvz}%
  \BibitemOpen
  \bibfield  {author} {\bibinfo {author} {\bibfnamefont {V.}~\bibnamefont
  {De~Luca}}, \bibinfo {author} {\bibfnamefont {G.}~\bibnamefont
  {Franciolini}}, \bibinfo {author} {\bibfnamefont {A.}~\bibnamefont {Riotto}},
  \ and\ \bibinfo {author} {\bibfnamefont {H.}~\bibnamefont {Veerm\"ae}},\
  }\href@noop {} {\  (\bibinfo {year} {2022})},\ \Eprint
  {http://arxiv.org/abs/2208.01683} {arXiv:2208.01683 [astro-ph.CO]}
  \BibitemShut {NoStop}%
\bibitem [{\citenamefont {Bird}\ \emph {et~al.}(2016)\citenamefont {Bird},
  \citenamefont {Cholis}, \citenamefont {Muñoz}, \citenamefont {Ali-Haïmoud},
  \citenamefont {Kamionkowski}, \citenamefont {Kovetz}, \citenamefont
  {Raccanelli},\ and\ \citenamefont {Riess}}]{Bird:2016dcv}%
  \BibitemOpen
  \bibfield  {author} {\bibinfo {author} {\bibfnamefont {S.}~\bibnamefont
  {Bird}}, \bibinfo {author} {\bibfnamefont {I.}~\bibnamefont {Cholis}},
  \bibinfo {author} {\bibfnamefont {J.~B.}\ \bibnamefont {Muñoz}}, \bibinfo
  {author} {\bibfnamefont {Y.}~\bibnamefont {Ali-Haïmoud}}, \bibinfo {author}
  {\bibfnamefont {M.}~\bibnamefont {Kamionkowski}}, \bibinfo {author}
  {\bibfnamefont {E.~D.}\ \bibnamefont {Kovetz}}, \bibinfo {author}
  {\bibfnamefont {A.}~\bibnamefont {Raccanelli}}, \ and\ \bibinfo {author}
  {\bibfnamefont {A.~G.}\ \bibnamefont {Riess}},\ }\href {\doibase
  10.1103/PhysRevLett.116.201301} {\bibfield  {journal} {\bibinfo  {journal}
  {Phys. Rev. Lett.}\ }\textbf {\bibinfo {volume} {116}},\ \bibinfo {pages}
  {201301} (\bibinfo {year} {2016})},\ \Eprint
  {http://arxiv.org/abs/1603.00464} {arXiv:1603.00464 [astro-ph.CO]}
  \BibitemShut {NoStop}%
\bibitem [{\citenamefont {Eroshenko}(2018)}]{Eroshenko:2016hmn}%
  \BibitemOpen
  \bibfield  {author} {\bibinfo {author} {\bibfnamefont {Y.~N.}\ \bibnamefont
  {Eroshenko}},\ }\href {\doibase 10.1088/1742-6596/1051/1/012010} {\bibfield
  {journal} {\bibinfo  {journal} {J. Phys. Conf. Ser.}\ }\textbf {\bibinfo
  {volume} {1051}},\ \bibinfo {pages} {012010} (\bibinfo {year} {2018})},\
  \Eprint {http://arxiv.org/abs/1604.04932} {arXiv:1604.04932 [astro-ph.CO]}
  \BibitemShut {NoStop}%
\bibitem [{\citenamefont {Wang}\ \emph {et~al.}(2018)\citenamefont {Wang},
  \citenamefont {Wang}, \citenamefont {Huang},\ and\ \citenamefont
  {Li}}]{Wang:2016ana}%
  \BibitemOpen
  \bibfield  {author} {\bibinfo {author} {\bibfnamefont {S.}~\bibnamefont
  {Wang}}, \bibinfo {author} {\bibfnamefont {Y.-F.}\ \bibnamefont {Wang}},
  \bibinfo {author} {\bibfnamefont {Q.-G.}\ \bibnamefont {Huang}}, \ and\
  \bibinfo {author} {\bibfnamefont {T.~G.~F.}\ \bibnamefont {Li}},\ }\href
  {\doibase 10.1103/PhysRevLett.120.191102} {\bibfield  {journal} {\bibinfo
  {journal} {Phys. Rev. Lett.}\ }\textbf {\bibinfo {volume} {120}},\ \bibinfo
  {pages} {191102} (\bibinfo {year} {2018})},\ \Eprint
  {http://arxiv.org/abs/1610.08725} {arXiv:1610.08725 [astro-ph.CO]}
  \BibitemShut {NoStop}%
\bibitem [{\citenamefont {Ali-Haïmoud}\ \emph {et~al.}(2017)\citenamefont
  {Ali-Haïmoud}, \citenamefont {Kovetz},\ and\ \citenamefont
  {Kamionkowski}}]{Ali-Haimoud:2017rtz}%
  \BibitemOpen
  \bibfield  {author} {\bibinfo {author} {\bibfnamefont {Y.}~\bibnamefont
  {Ali-Haïmoud}}, \bibinfo {author} {\bibfnamefont {E.~D.}\ \bibnamefont
  {Kovetz}}, \ and\ \bibinfo {author} {\bibfnamefont {M.}~\bibnamefont
  {Kamionkowski}},\ }\href {\doibase 10.1103/PhysRevD.96.123523} {\bibfield
  {journal} {\bibinfo  {journal} {Phys. Rev.}\ }\textbf {\bibinfo {volume}
  {D96}},\ \bibinfo {pages} {123523} (\bibinfo {year} {2017})},\ \Eprint
  {http://arxiv.org/abs/1709.06576} {arXiv:1709.06576 [astro-ph.CO]}
  \BibitemShut {NoStop}%
\bibitem [{\citenamefont {Chen}\ and\ \citenamefont
  {Huang}(2018)}]{Chen:2018czv}%
  \BibitemOpen
  \bibfield  {author} {\bibinfo {author} {\bibfnamefont {Z.-C.}\ \bibnamefont
  {Chen}}\ and\ \bibinfo {author} {\bibfnamefont {Q.-G.}\ \bibnamefont
  {Huang}},\ }\href {\doibase 10.3847/1538-4357/aad6e2} {\bibfield  {journal}
  {\bibinfo  {journal} {Astrophys. J.}\ }\textbf {\bibinfo {volume} {864}},\
  \bibinfo {pages} {61} (\bibinfo {year} {2018})},\ \Eprint
  {http://arxiv.org/abs/1801.10327} {arXiv:1801.10327 [astro-ph.CO]}
  \BibitemShut {NoStop}%
\bibitem [{\citenamefont {Liu}\ \emph {et~al.}(2019)\citenamefont {Liu},
  \citenamefont {Guo},\ and\ \citenamefont {Cai}}]{Liu:2019rnx}%
  \BibitemOpen
  \bibfield  {author} {\bibinfo {author} {\bibfnamefont {L.}~\bibnamefont
  {Liu}}, \bibinfo {author} {\bibfnamefont {Z.-K.}\ \bibnamefont {Guo}}, \ and\
  \bibinfo {author} {\bibfnamefont {R.-G.}\ \bibnamefont {Cai}},\ }\href
  {\doibase 10.1140/epjc/s10052-019-7227-0} {\bibfield  {journal} {\bibinfo
  {journal} {Eur. Phys. J.}\ }\textbf {\bibinfo {volume} {C79}},\ \bibinfo
  {pages} {717} (\bibinfo {year} {2019})},\ \Eprint
  {http://arxiv.org/abs/1901.07672} {arXiv:1901.07672 [astro-ph.CO]}
  \BibitemShut {NoStop}%
\bibitem [{\citenamefont {H\"utsi}\ \emph {et~al.}(2019)\citenamefont
  {H\"utsi}, \citenamefont {Raidal},\ and\ \citenamefont
  {Veerm\"ae}}]{Hutsi:2019hlw}%
  \BibitemOpen
  \bibfield  {author} {\bibinfo {author} {\bibfnamefont {G.}~\bibnamefont
  {H\"utsi}}, \bibinfo {author} {\bibfnamefont {M.}~\bibnamefont {Raidal}}, \
  and\ \bibinfo {author} {\bibfnamefont {H.}~\bibnamefont {Veerm\"ae}},\ }\href
  {\doibase 10.1103/PhysRevD.100.083016} {\bibfield  {journal} {\bibinfo
  {journal} {Phys. Rev. D}\ }\textbf {\bibinfo {volume} {100}},\ \bibinfo
  {pages} {083016} (\bibinfo {year} {2019})},\ \Eprint
  {http://arxiv.org/abs/1907.06533} {arXiv:1907.06533 [astro-ph.CO]}
  \BibitemShut {NoStop}%
\bibitem [{\citenamefont {Vaskonen}\ and\ \citenamefont
  {Veerm\"ae}(2020)}]{Vaskonen:2019jpv}%
  \BibitemOpen
  \bibfield  {author} {\bibinfo {author} {\bibfnamefont {V.}~\bibnamefont
  {Vaskonen}}\ and\ \bibinfo {author} {\bibfnamefont {H.}~\bibnamefont
  {Veerm\"ae}},\ }\href {\doibase 10.1103/PhysRevD.101.043015} {\bibfield
  {journal} {\bibinfo  {journal} {Phys. Rev. D}\ }\textbf {\bibinfo {volume}
  {101}},\ \bibinfo {pages} {043015} (\bibinfo {year} {2020})},\ \Eprint
  {http://arxiv.org/abs/1908.09752} {arXiv:1908.09752 [astro-ph.CO]}
  \BibitemShut {NoStop}%
\bibitem [{\citenamefont {Gow}\ \emph {et~al.}(2020)\citenamefont {Gow},
  \citenamefont {Byrnes}, \citenamefont {Hall},\ and\ \citenamefont
  {Peacock}}]{Gow:2019pok}%
  \BibitemOpen
  \bibfield  {author} {\bibinfo {author} {\bibfnamefont {A.~D.}\ \bibnamefont
  {Gow}}, \bibinfo {author} {\bibfnamefont {C.~T.}\ \bibnamefont {Byrnes}},
  \bibinfo {author} {\bibfnamefont {A.}~\bibnamefont {Hall}}, \ and\ \bibinfo
  {author} {\bibfnamefont {J.~A.}\ \bibnamefont {Peacock}},\ }\href {\doibase
  10.1088/1475-7516/2020/01/031} {\bibfield  {journal} {\bibinfo  {journal}
  {JCAP}\ }\textbf {\bibinfo {volume} {01}},\ \bibinfo {pages} {031} (\bibinfo
  {year} {2020})},\ \Eprint {http://arxiv.org/abs/1911.12685} {arXiv:1911.12685
  [astro-ph.CO]} \BibitemShut {NoStop}%
\bibitem [{\citenamefont {Wu}(2020)}]{Wu:2020drm}%
  \BibitemOpen
  \bibfield  {author} {\bibinfo {author} {\bibfnamefont {Y.}~\bibnamefont
  {Wu}},\ }\href {\doibase 10.1103/PhysRevD.101.083008} {\bibfield  {journal}
  {\bibinfo  {journal} {Phys. Rev.}\ }\textbf {\bibinfo {volume} {D101}},\
  \bibinfo {pages} {083008} (\bibinfo {year} {2020})},\ \Eprint
  {http://arxiv.org/abs/2001.03833} {arXiv:2001.03833 [astro-ph.CO]}
  \BibitemShut {NoStop}%
\bibitem [{\citenamefont {De~Luca}\ \emph {et~al.}(2020)\citenamefont
  {De~Luca}, \citenamefont {Franciolini}, \citenamefont {Pani},\ and\
  \citenamefont {Riotto}}]{DeLuca:2020qqa}%
  \BibitemOpen
  \bibfield  {author} {\bibinfo {author} {\bibfnamefont {V.}~\bibnamefont
  {De~Luca}}, \bibinfo {author} {\bibfnamefont {G.}~\bibnamefont
  {Franciolini}}, \bibinfo {author} {\bibfnamefont {P.}~\bibnamefont {Pani}}, \
  and\ \bibinfo {author} {\bibfnamefont {A.}~\bibnamefont {Riotto}},\ }\href
  {\doibase 10.1088/1475-7516/2020/06/044} {\bibfield  {journal} {\bibinfo
  {journal} {JCAP}\ }\textbf {\bibinfo {volume} {06}},\ \bibinfo {pages} {044}
  (\bibinfo {year} {2020})},\ \Eprint {http://arxiv.org/abs/2005.05641}
  {arXiv:2005.05641 [astro-ph.CO]} \BibitemShut {NoStop}%
\bibitem [{\citenamefont {Jedamzik}(2021)}]{Jedamzik:2020omx}%
  \BibitemOpen
  \bibfield  {author} {\bibinfo {author} {\bibfnamefont {K.}~\bibnamefont
  {Jedamzik}},\ }\href {\doibase 10.1103/PhysRevLett.126.051302} {\bibfield
  {journal} {\bibinfo  {journal} {Phys. Rev. Lett.}\ }\textbf {\bibinfo
  {volume} {126}},\ \bibinfo {pages} {051302} (\bibinfo {year} {2021})},\
  \Eprint {http://arxiv.org/abs/2007.03565} {arXiv:2007.03565 [astro-ph.CO]}
  \BibitemShut {NoStop}%
\bibitem [{\citenamefont {Hall}\ \emph {et~al.}(2020)\citenamefont {Hall},
  \citenamefont {Gow},\ and\ \citenamefont {Byrnes}}]{Hall:2020daa}%
  \BibitemOpen
  \bibfield  {author} {\bibinfo {author} {\bibfnamefont {A.}~\bibnamefont
  {Hall}}, \bibinfo {author} {\bibfnamefont {A.~D.}\ \bibnamefont {Gow}}, \
  and\ \bibinfo {author} {\bibfnamefont {C.~T.}\ \bibnamefont {Byrnes}},\
  }\href {\doibase 10.1103/PhysRevD.102.123524} {\bibfield  {journal} {\bibinfo
   {journal} {Phys. Rev. D}\ }\textbf {\bibinfo {volume} {102}},\ \bibinfo
  {pages} {123524} (\bibinfo {year} {2020})},\ \Eprint
  {http://arxiv.org/abs/2008.13704} {arXiv:2008.13704 [astro-ph.CO]}
  \BibitemShut {NoStop}%
\bibitem [{\citenamefont {Wong}\ \emph {et~al.}(2021)\citenamefont {Wong},
  \citenamefont {Franciolini}, \citenamefont {De~Luca}, \citenamefont
  {Baibhav}, \citenamefont {Berti}, \citenamefont {Pani},\ and\ \citenamefont
  {Riotto}}]{Wong:2020yig}%
  \BibitemOpen
  \bibfield  {author} {\bibinfo {author} {\bibfnamefont {K.~W.~K.}\
  \bibnamefont {Wong}}, \bibinfo {author} {\bibfnamefont {G.}~\bibnamefont
  {Franciolini}}, \bibinfo {author} {\bibfnamefont {V.}~\bibnamefont
  {De~Luca}}, \bibinfo {author} {\bibfnamefont {V.}~\bibnamefont {Baibhav}},
  \bibinfo {author} {\bibfnamefont {E.}~\bibnamefont {Berti}}, \bibinfo
  {author} {\bibfnamefont {P.}~\bibnamefont {Pani}}, \ and\ \bibinfo {author}
  {\bibfnamefont {A.}~\bibnamefont {Riotto}},\ }\href {\doibase
  10.1103/PhysRevD.103.023026} {\bibfield  {journal} {\bibinfo  {journal}
  {Phys. Rev.}\ }\textbf {\bibinfo {volume} {D103}},\ \bibinfo {pages} {023026}
  (\bibinfo {year} {2021})},\ \Eprint {http://arxiv.org/abs/2011.01865}
  {arXiv:2011.01865 [gr-qc]} \BibitemShut {NoStop}%
\bibitem [{\citenamefont {Kritos}\ \emph {et~al.}(2021)\citenamefont {Kritos},
  \citenamefont {De~Luca}, \citenamefont {Franciolini}, \citenamefont
  {Kehagias},\ and\ \citenamefont {Riotto}}]{Kritos:2020wcl}%
  \BibitemOpen
  \bibfield  {author} {\bibinfo {author} {\bibfnamefont {K.}~\bibnamefont
  {Kritos}}, \bibinfo {author} {\bibfnamefont {V.}~\bibnamefont {De~Luca}},
  \bibinfo {author} {\bibfnamefont {G.}~\bibnamefont {Franciolini}}, \bibinfo
  {author} {\bibfnamefont {A.}~\bibnamefont {Kehagias}}, \ and\ \bibinfo
  {author} {\bibfnamefont {A.}~\bibnamefont {Riotto}},\ }\href {\doibase
  10.1088/1475-7516/2021/05/039} {\bibfield  {journal} {\bibinfo  {journal}
  {JCAP}\ }\textbf {\bibinfo {volume} {05}},\ \bibinfo {pages} {039} (\bibinfo
  {year} {2021})},\ \Eprint {http://arxiv.org/abs/2012.03585} {arXiv:2012.03585
  [gr-qc]} \BibitemShut {NoStop}%
\bibitem [{\citenamefont {Franciolini}\ \emph
  {et~al.}(2022{\natexlab{b}})\citenamefont {Franciolini}, \citenamefont
  {Cotesta}, \citenamefont {Loutrel}, \citenamefont {Berti}, \citenamefont
  {Pani},\ and\ \citenamefont {Riotto}}]{Franciolini:2021xbq}%
  \BibitemOpen
  \bibfield  {author} {\bibinfo {author} {\bibfnamefont {G.}~\bibnamefont
  {Franciolini}}, \bibinfo {author} {\bibfnamefont {R.}~\bibnamefont
  {Cotesta}}, \bibinfo {author} {\bibfnamefont {N.}~\bibnamefont {Loutrel}},
  \bibinfo {author} {\bibfnamefont {E.}~\bibnamefont {Berti}}, \bibinfo
  {author} {\bibfnamefont {P.}~\bibnamefont {Pani}}, \ and\ \bibinfo {author}
  {\bibfnamefont {A.}~\bibnamefont {Riotto}},\ }\href {\doibase
  10.1103/PhysRevD.105.063510} {\bibfield  {journal} {\bibinfo  {journal}
  {Phys. Rev. D}\ }\textbf {\bibinfo {volume} {105}},\ \bibinfo {pages}
  {063510} (\bibinfo {year} {2022}{\natexlab{b}})},\ \Eprint
  {http://arxiv.org/abs/2112.10660} {arXiv:2112.10660 [astro-ph.CO]}
  \BibitemShut {NoStop}%
\bibitem [{\citenamefont {Bavera}\ \emph {et~al.}(2021)\citenamefont {Bavera},
  \citenamefont {Franciolini}, \citenamefont {Cusin}, \citenamefont {Riotto},
  \citenamefont {Zevin},\ and\ \citenamefont {Fragos}}]{Bavera:2021wmw}%
  \BibitemOpen
  \bibfield  {author} {\bibinfo {author} {\bibfnamefont {S.~S.}\ \bibnamefont
  {Bavera}}, \bibinfo {author} {\bibfnamefont {G.}~\bibnamefont {Franciolini}},
  \bibinfo {author} {\bibfnamefont {G.}~\bibnamefont {Cusin}}, \bibinfo
  {author} {\bibfnamefont {A.}~\bibnamefont {Riotto}}, \bibinfo {author}
  {\bibfnamefont {M.}~\bibnamefont {Zevin}}, \ and\ \bibinfo {author}
  {\bibfnamefont {T.}~\bibnamefont {Fragos}},\ }\href@noop {} {\  (\bibinfo
  {year} {2021})},\ \Eprint {http://arxiv.org/abs/2109.05836} {arXiv:2109.05836
  [astro-ph.CO]} \BibitemShut {NoStop}%
\bibitem [{\citenamefont {Escriv\`a}\ \emph {et~al.}(2022)\citenamefont
  {Escriv\`a}, \citenamefont {Bagui},\ and\ \citenamefont
  {Clesse}}]{Escriva:2022bwe}%
  \BibitemOpen
  \bibfield  {author} {\bibinfo {author} {\bibfnamefont {A.}~\bibnamefont
  {Escriv\`a}}, \bibinfo {author} {\bibfnamefont {E.}~\bibnamefont {Bagui}}, \
  and\ \bibinfo {author} {\bibfnamefont {S.}~\bibnamefont {Clesse}},\
  }\href@noop {} {\  (\bibinfo {year} {2022})},\ \Eprint
  {http://arxiv.org/abs/2209.06196} {arXiv:2209.06196 [astro-ph.CO]}
  \BibitemShut {NoStop}%
\bibitem [{\citenamefont {Musco}(2019)}]{Musco:2018rwt}%
  \BibitemOpen
  \bibfield  {author} {\bibinfo {author} {\bibfnamefont {I.}~\bibnamefont
  {Musco}},\ }\href {\doibase 10.1103/PhysRevD.100.123524} {\bibfield
  {journal} {\bibinfo  {journal} {Phys. Rev. D}\ }\textbf {\bibinfo {volume}
  {100}},\ \bibinfo {pages} {123524} (\bibinfo {year} {2019})},\ \Eprint
  {http://arxiv.org/abs/1809.02127} {arXiv:1809.02127 [gr-qc]} \BibitemShut
  {NoStop}%
\bibitem [{\citenamefont {Bardeen}\ \emph
  {et~al.}(1986{\natexlab{a}})\citenamefont {Bardeen}, \citenamefont {Bond},
  \citenamefont {Kaiser},\ and\ \citenamefont {Szalay}}]{bbks}%
  \BibitemOpen
  \bibfield  {author} {\bibinfo {author} {\bibfnamefont {J.~M.}\ \bibnamefont
  {Bardeen}}, \bibinfo {author} {\bibfnamefont {J.}~\bibnamefont {Bond}},
  \bibinfo {author} {\bibfnamefont {N.}~\bibnamefont {Kaiser}}, \ and\ \bibinfo
  {author} {\bibfnamefont {A.}~\bibnamefont {Szalay}},\ }\href {\doibase
  10.1086/164143} {\bibfield  {journal} {\bibinfo  {journal} {Astrophys. J.}\
  }\textbf {\bibinfo {volume} {304}},\ \bibinfo {pages} {15} (\bibinfo {year}
  {1986}{\natexlab{a}})}\BibitemShut {NoStop}%
\bibitem [{\citenamefont {Misner}\ and\ \citenamefont
  {Sharp}(1964)}]{Misner:1964je}%
  \BibitemOpen
  \bibfield  {author} {\bibinfo {author} {\bibfnamefont {C.~W.}\ \bibnamefont
  {Misner}}\ and\ \bibinfo {author} {\bibfnamefont {D.~H.}\ \bibnamefont
  {Sharp}},\ }\href {\doibase 10.1103/PhysRev.136.B571} {\bibfield  {journal}
  {\bibinfo  {journal} {Phys. Rev.}\ }\textbf {\bibinfo {volume} {136}},\
  \bibinfo {pages} {B571} (\bibinfo {year} {1964})}\BibitemShut {NoStop}%
\bibitem [{\citenamefont {Shibata}\ and\ \citenamefont
  {Sasaki}(1999)}]{Shibata:1999zs}%
  \BibitemOpen
  \bibfield  {author} {\bibinfo {author} {\bibfnamefont {M.}~\bibnamefont
  {Shibata}}\ and\ \bibinfo {author} {\bibfnamefont {M.}~\bibnamefont
  {Sasaki}},\ }\href {\doibase 10.1103/PhysRevD.60.084002} {\bibfield
  {journal} {\bibinfo  {journal} {Phys. Rev. D}\ }\textbf {\bibinfo {volume}
  {60}},\ \bibinfo {pages} {084002} (\bibinfo {year} {1999})},\ \Eprint
  {http://arxiv.org/abs/gr-qc/9905064} {arXiv:gr-qc/9905064} \BibitemShut
  {NoStop}%
\bibitem [{\citenamefont {Tomita}(1975)}]{Tomita:1975kj}%
  \BibitemOpen
  \bibfield  {author} {\bibinfo {author} {\bibfnamefont {K.}~\bibnamefont
  {Tomita}},\ }\href {\doibase 10.1143/PTP.54.730} {\bibfield  {journal}
  {\bibinfo  {journal} {Prog. Theor. Phys.}\ }\textbf {\bibinfo {volume}
  {54}},\ \bibinfo {pages} {730} (\bibinfo {year} {1975})}\BibitemShut
  {NoStop}%
\bibitem [{\citenamefont {Salopek}\ and\ \citenamefont
  {Bond}(1990)}]{Salopek:1990jq}%
  \BibitemOpen
  \bibfield  {author} {\bibinfo {author} {\bibfnamefont {D.~S.}\ \bibnamefont
  {Salopek}}\ and\ \bibinfo {author} {\bibfnamefont {J.~R.}\ \bibnamefont
  {Bond}},\ }\href {\doibase 10.1103/PhysRevD.42.3936} {\bibfield  {journal}
  {\bibinfo  {journal} {Phys. Rev. D}\ }\textbf {\bibinfo {volume} {42}},\
  \bibinfo {pages} {3936} (\bibinfo {year} {1990})}\BibitemShut {NoStop}%
\bibitem [{\citenamefont {Polnarev}\ and\ \citenamefont
  {Musco}(2007)}]{Polnarev:2006aa}%
  \BibitemOpen
  \bibfield  {author} {\bibinfo {author} {\bibfnamefont {A.~G.}\ \bibnamefont
  {Polnarev}}\ and\ \bibinfo {author} {\bibfnamefont {I.}~\bibnamefont
  {Musco}},\ }\href {\doibase 10.1088/0264-9381/24/6/003} {\bibfield  {journal}
  {\bibinfo  {journal} {Class. Quant. Grav.}\ }\textbf {\bibinfo {volume}
  {24}},\ \bibinfo {pages} {1405} (\bibinfo {year} {2007})},\ \Eprint
  {http://arxiv.org/abs/gr-qc/0605122} {arXiv:gr-qc/0605122} \BibitemShut
  {NoStop}%
\bibitem [{\citenamefont {Harada}\ \emph {et~al.}(2015)\citenamefont {Harada},
  \citenamefont {Yoo}, \citenamefont {Nakama},\ and\ \citenamefont
  {Koga}}]{Harada:2015yda}%
  \BibitemOpen
  \bibfield  {author} {\bibinfo {author} {\bibfnamefont {T.}~\bibnamefont
  {Harada}}, \bibinfo {author} {\bibfnamefont {C.-M.}\ \bibnamefont {Yoo}},
  \bibinfo {author} {\bibfnamefont {T.}~\bibnamefont {Nakama}}, \ and\ \bibinfo
  {author} {\bibfnamefont {Y.}~\bibnamefont {Koga}},\ }\href {\doibase
  10.1103/PhysRevD.91.084057} {\bibfield  {journal} {\bibinfo  {journal} {Phys.
  Rev. D}\ }\textbf {\bibinfo {volume} {91}},\ \bibinfo {pages} {084057}
  (\bibinfo {year} {2015})},\ \Eprint {http://arxiv.org/abs/1503.03934}
  {arXiv:1503.03934 [gr-qc]} \BibitemShut {NoStop}%
\bibitem [{\citenamefont {Lyth}\ \emph {et~al.}(2005)\citenamefont {Lyth},
  \citenamefont {Malik},\ and\ \citenamefont {Sasaki}}]{Lyth:2004gb}%
  \BibitemOpen
  \bibfield  {author} {\bibinfo {author} {\bibfnamefont {D.~H.}\ \bibnamefont
  {Lyth}}, \bibinfo {author} {\bibfnamefont {K.~A.}\ \bibnamefont {Malik}}, \
  and\ \bibinfo {author} {\bibfnamefont {M.}~\bibnamefont {Sasaki}},\ }\href
  {\doibase 10.1088/1475-7516/2005/05/004} {\bibfield  {journal} {\bibinfo
  {journal} {JCAP}\ }\textbf {\bibinfo {volume} {05}},\ \bibinfo {pages} {004}
  (\bibinfo {year} {2005})},\ \Eprint {http://arxiv.org/abs/astro-ph/0411220}
  {arXiv:astro-ph/0411220} \BibitemShut {NoStop}%
\bibitem [{\citenamefont {Yoo}\ \emph {et~al.}(2021)\citenamefont {Yoo},
  \citenamefont {Harada}, \citenamefont {Hirano},\ and\ \citenamefont
  {Kohri}}]{Yoo:2020dkz}%
  \BibitemOpen
  \bibfield  {author} {\bibinfo {author} {\bibfnamefont {C.-M.}\ \bibnamefont
  {Yoo}}, \bibinfo {author} {\bibfnamefont {T.}~\bibnamefont {Harada}},
  \bibinfo {author} {\bibfnamefont {S.}~\bibnamefont {Hirano}}, \ and\ \bibinfo
  {author} {\bibfnamefont {K.}~\bibnamefont {Kohri}},\ }\href {\doibase
  10.1093/ptep/ptaa155} {\bibfield  {journal} {\bibinfo  {journal} {PTEP}\
  }\textbf {\bibinfo {volume} {2021}},\ \bibinfo {pages} {013E02} (\bibinfo
  {year} {2021})},\ \Eprint {http://arxiv.org/abs/2008.02425} {arXiv:2008.02425
  [astro-ph.CO]} \BibitemShut {NoStop}%
\bibitem [{\citenamefont {Helou}\ \emph {et~al.}(2017)\citenamefont {Helou},
  \citenamefont {Musco},\ and\ \citenamefont {Miller}}]{Helou:2016xyu}%
  \BibitemOpen
  \bibfield  {author} {\bibinfo {author} {\bibfnamefont {A.}~\bibnamefont
  {Helou}}, \bibinfo {author} {\bibfnamefont {I.}~\bibnamefont {Musco}}, \ and\
  \bibinfo {author} {\bibfnamefont {J.~C.}\ \bibnamefont {Miller}},\ }\href
  {\doibase 10.1088/1361-6382/aa6d8f} {\bibfield  {journal} {\bibinfo
  {journal} {Class. Quant. Grav.}\ }\textbf {\bibinfo {volume} {34}},\ \bibinfo
  {pages} {135012} (\bibinfo {year} {2017})},\ \Eprint
  {http://arxiv.org/abs/1601.05109} {arXiv:1601.05109 [gr-qc]} \BibitemShut
  {NoStop}%
\bibitem [{\citenamefont {Bardeen}\ \emph
  {et~al.}(1986{\natexlab{b}})\citenamefont {Bardeen}, \citenamefont {Bond},
  \citenamefont {Kaiser},\ and\ \citenamefont {Szalay}}]{Bardeen:1985tr}%
  \BibitemOpen
  \bibfield  {author} {\bibinfo {author} {\bibfnamefont {J.~M.}\ \bibnamefont
  {Bardeen}}, \bibinfo {author} {\bibfnamefont {J.~R.}\ \bibnamefont {Bond}},
  \bibinfo {author} {\bibfnamefont {N.}~\bibnamefont {Kaiser}}, \ and\ \bibinfo
  {author} {\bibfnamefont {A.~S.}\ \bibnamefont {Szalay}},\ }\href {\doibase
  10.1086/164143} {\bibfield  {journal} {\bibinfo  {journal} {Astrophys. J.}\
  }\textbf {\bibinfo {volume} {304}},\ \bibinfo {pages} {15} (\bibinfo {year}
  {1986}{\natexlab{b}})}\BibitemShut {NoStop}%
\bibitem [{\citenamefont {Young}(2022)}]{Young:2022phe}%
  \BibitemOpen
  \bibfield  {author} {\bibinfo {author} {\bibfnamefont {S.}~\bibnamefont
  {Young}},\ }\href {\doibase 10.1088/1475-7516/2022/05/037} {\bibfield
  {journal} {\bibinfo  {journal} {JCAP}\ }\textbf {\bibinfo {volume} {05}},\
  \bibinfo {pages} {037} (\bibinfo {year} {2022})},\ \Eprint
  {http://arxiv.org/abs/2201.13345} {arXiv:2201.13345 [astro-ph.CO]}
  \BibitemShut {NoStop}%
\bibitem [{\citenamefont {Musco}\ \emph {et~al.}(2009)\citenamefont {Musco},
  \citenamefont {Miller},\ and\ \citenamefont {Polnarev}}]{Musco:2008hv}%
  \BibitemOpen
  \bibfield  {author} {\bibinfo {author} {\bibfnamefont {I.}~\bibnamefont
  {Musco}}, \bibinfo {author} {\bibfnamefont {J.~C.}\ \bibnamefont {Miller}}, \
  and\ \bibinfo {author} {\bibfnamefont {A.~G.}\ \bibnamefont {Polnarev}},\
  }\href {\doibase 10.1088/0264-9381/26/23/235001} {\bibfield  {journal}
  {\bibinfo  {journal} {Class. Quant. Grav.}\ }\textbf {\bibinfo {volume}
  {26}},\ \bibinfo {pages} {235001} (\bibinfo {year} {2009})},\ \Eprint
  {http://arxiv.org/abs/0811.1452} {arXiv:0811.1452 [gr-qc]} \BibitemShut
  {NoStop}%
\bibitem [{\citenamefont {Musco}\ and\ \citenamefont
  {Miller}(2013)}]{Musco:2012au}%
  \BibitemOpen
  \bibfield  {author} {\bibinfo {author} {\bibfnamefont {I.}~\bibnamefont
  {Musco}}\ and\ \bibinfo {author} {\bibfnamefont {J.~C.}\ \bibnamefont
  {Miller}},\ }\href {\doibase 10.1088/0264-9381/30/14/145009} {\bibfield
  {journal} {\bibinfo  {journal} {Class. Quant. Grav.}\ }\textbf {\bibinfo
  {volume} {30}},\ \bibinfo {pages} {145009} (\bibinfo {year} {2013})},\
  \Eprint {http://arxiv.org/abs/1201.2379} {arXiv:1201.2379 [gr-qc]}
  \BibitemShut {NoStop}%
\bibitem [{\citenamefont {Hernandez}\ and\ \citenamefont
  {Misner}(1966)}]{Hernandez:1966zia}%
  \BibitemOpen
  \bibfield  {author} {\bibinfo {author} {\bibfnamefont {W.~C.}\ \bibnamefont
  {Hernandez}}\ and\ \bibinfo {author} {\bibfnamefont {C.~W.}\ \bibnamefont
  {Misner}},\ }\href {\doibase 10.1086/148525} {\bibfield  {journal} {\bibinfo
  {journal} {Astrophys. J.}\ }\textbf {\bibinfo {volume} {143}},\ \bibinfo
  {pages} {452} (\bibinfo {year} {1966})}\BibitemShut {NoStop}%
\bibitem [{\citenamefont {Escriv\`a}\ \emph {et~al.}(2020)\citenamefont
  {Escriv\`a}, \citenamefont {Germani},\ and\ \citenamefont
  {Sheth}}]{Escriva:2019phb}%
  \BibitemOpen
  \bibfield  {author} {\bibinfo {author} {\bibfnamefont {A.}~\bibnamefont
  {Escriv\`a}}, \bibinfo {author} {\bibfnamefont {C.}~\bibnamefont {Germani}},
  \ and\ \bibinfo {author} {\bibfnamefont {R.~K.}\ \bibnamefont {Sheth}},\
  }\href {\doibase 10.1103/PhysRevD.101.044022} {\bibfield  {journal} {\bibinfo
   {journal} {Phys. Rev. D}\ }\textbf {\bibinfo {volume} {101}},\ \bibinfo
  {pages} {044022} (\bibinfo {year} {2020})},\ \Eprint
  {http://arxiv.org/abs/1907.13311} {arXiv:1907.13311 [gr-qc]} \BibitemShut
  {NoStop}%
\bibitem [{\citenamefont {Musco}\ \emph {et~al.}(2021)\citenamefont {Musco},
  \citenamefont {De~Luca}, \citenamefont {Franciolini},\ and\ \citenamefont
  {Riotto}}]{Musco:2020jjb}%
  \BibitemOpen
  \bibfield  {author} {\bibinfo {author} {\bibfnamefont {I.}~\bibnamefont
  {Musco}}, \bibinfo {author} {\bibfnamefont {V.}~\bibnamefont {De~Luca}},
  \bibinfo {author} {\bibfnamefont {G.}~\bibnamefont {Franciolini}}, \ and\
  \bibinfo {author} {\bibfnamefont {A.}~\bibnamefont {Riotto}},\ }\href
  {\doibase 10.1103/PhysRevD.103.063538} {\bibfield  {journal} {\bibinfo
  {journal} {Phys. Rev. D}\ }\textbf {\bibinfo {volume} {103}},\ \bibinfo
  {pages} {063538} (\bibinfo {year} {2021})},\ \Eprint
  {http://arxiv.org/abs/2011.03014} {arXiv:2011.03014 [astro-ph.CO]}
  \BibitemShut {NoStop}%
\bibitem [{\citenamefont {Neilsen}\ and\ \citenamefont
  {Choptuik}(2000)}]{Neilsen:1998qc}%
  \BibitemOpen
  \bibfield  {author} {\bibinfo {author} {\bibfnamefont {D.~W.}\ \bibnamefont
  {Neilsen}}\ and\ \bibinfo {author} {\bibfnamefont {M.~W.}\ \bibnamefont
  {Choptuik}},\ }\href {\doibase 10.1088/0264-9381/17/4/303} {\bibfield
  {journal} {\bibinfo  {journal} {Class. Quant. Grav.}\ }\textbf {\bibinfo
  {volume} {17}},\ \bibinfo {pages} {761} (\bibinfo {year} {2000})},\ \Eprint
  {http://arxiv.org/abs/gr-qc/9812053} {arXiv:gr-qc/9812053} \BibitemShut
  {NoStop}%
\bibitem [{\citenamefont {Escriv\`a}\ and\ \citenamefont
  {Romano}(2021)}]{Escriva:2021pmf}%
  \BibitemOpen
  \bibfield  {author} {\bibinfo {author} {\bibfnamefont {A.}~\bibnamefont
  {Escriv\`a}}\ and\ \bibinfo {author} {\bibfnamefont {A.~E.}\ \bibnamefont
  {Romano}},\ }\href {\doibase 10.1088/1475-7516/2021/05/066} {\bibfield
  {journal} {\bibinfo  {journal} {JCAP}\ }\textbf {\bibinfo {volume} {05}},\
  \bibinfo {pages} {066} (\bibinfo {year} {2021})},\ \Eprint
  {http://arxiv.org/abs/2103.03867} {arXiv:2103.03867 [gr-qc]} \BibitemShut
  {NoStop}%
\bibitem [{\citenamefont {Young}\ \emph {et~al.}(2019)\citenamefont {Young},
  \citenamefont {Musco},\ and\ \citenamefont {Byrnes}}]{Young:2019yug}%
  \BibitemOpen
  \bibfield  {author} {\bibinfo {author} {\bibfnamefont {S.}~\bibnamefont
  {Young}}, \bibinfo {author} {\bibfnamefont {I.}~\bibnamefont {Musco}}, \ and\
  \bibinfo {author} {\bibfnamefont {C.~T.}\ \bibnamefont {Byrnes}},\ }\href
  {\doibase 10.1088/1475-7516/2019/11/012} {\bibfield  {journal} {\bibinfo
  {journal} {JCAP}\ }\textbf {\bibinfo {volume} {11}},\ \bibinfo {pages} {012}
  (\bibinfo {year} {2019})},\ \Eprint {http://arxiv.org/abs/1904.00984}
  {arXiv:1904.00984 [astro-ph.CO]} \BibitemShut {NoStop}%
\bibitem [{\citenamefont {Figueroa}\ \emph {et~al.}(2021)\citenamefont
  {Figueroa}, \citenamefont {Raatikainen}, \citenamefont {Rasanen},\ and\
  \citenamefont {Tomberg}}]{Figueroa:2020jkf}%
  \BibitemOpen
  \bibfield  {author} {\bibinfo {author} {\bibfnamefont {D.~G.}\ \bibnamefont
  {Figueroa}}, \bibinfo {author} {\bibfnamefont {S.}~\bibnamefont
  {Raatikainen}}, \bibinfo {author} {\bibfnamefont {S.}~\bibnamefont
  {Rasanen}}, \ and\ \bibinfo {author} {\bibfnamefont {E.}~\bibnamefont
  {Tomberg}},\ }\href {\doibase 10.1103/PhysRevLett.127.101302} {\bibfield
  {journal} {\bibinfo  {journal} {Phys. Rev. Lett.}\ }\textbf {\bibinfo
  {volume} {127}},\ \bibinfo {pages} {101302} (\bibinfo {year} {2021})},\
  \Eprint {http://arxiv.org/abs/2012.06551} {arXiv:2012.06551 [astro-ph.CO]}
  \BibitemShut {NoStop}%
\bibitem [{\citenamefont {Biagetti}\ \emph {et~al.}(2021)\citenamefont
  {Biagetti}, \citenamefont {De~Luca}, \citenamefont {Franciolini},
  \citenamefont {Kehagias},\ and\ \citenamefont {Riotto}}]{Biagetti:2021eep}%
  \BibitemOpen
  \bibfield  {author} {\bibinfo {author} {\bibfnamefont {M.}~\bibnamefont
  {Biagetti}}, \bibinfo {author} {\bibfnamefont {V.}~\bibnamefont {De~Luca}},
  \bibinfo {author} {\bibfnamefont {G.}~\bibnamefont {Franciolini}}, \bibinfo
  {author} {\bibfnamefont {A.}~\bibnamefont {Kehagias}}, \ and\ \bibinfo
  {author} {\bibfnamefont {A.}~\bibnamefont {Riotto}},\ }\href {\doibase
  10.1016/j.physletb.2021.136602} {\bibfield  {journal} {\bibinfo  {journal}
  {Phys. Lett. B}\ }\textbf {\bibinfo {volume} {820}},\ \bibinfo {pages}
  {136602} (\bibinfo {year} {2021})},\ \Eprint
  {http://arxiv.org/abs/2105.07810} {arXiv:2105.07810 [astro-ph.CO]}
  \BibitemShut {NoStop}%
\bibitem [{\citenamefont {Ferrante}\ \emph {et~al.}(2023)\citenamefont
  {Ferrante}, \citenamefont {Franciolini}, \citenamefont {Iovino},\ and\
  \citenamefont {Urbano}}]{Ferrante:2022mui}%
  \BibitemOpen
  \bibfield  {author} {\bibinfo {author} {\bibfnamefont {G.}~\bibnamefont
  {Ferrante}}, \bibinfo {author} {\bibfnamefont {G.}~\bibnamefont
  {Franciolini}}, \bibinfo {author} {\bibfnamefont {A.}~\bibnamefont {Iovino},
  \bibfnamefont {Junior.}}, \ and\ \bibinfo {author} {\bibfnamefont
  {A.}~\bibnamefont {Urbano}},\ }\href {\doibase 10.1103/PhysRevD.107.043520}
  {\bibfield  {journal} {\bibinfo  {journal} {Phys. Rev. D}\ }\textbf {\bibinfo
  {volume} {107}},\ \bibinfo {pages} {043520} (\bibinfo {year} {2023})},\
  \Eprint {http://arxiv.org/abs/2211.01728} {arXiv:2211.01728 [astro-ph.CO]}
  \BibitemShut {NoStop}%
\bibitem [{\citenamefont {Gow}\ \emph {et~al.}(2023)\citenamefont {Gow},
  \citenamefont {Assadullahi}, \citenamefont {Jackson}, \citenamefont {Koyama},
  \citenamefont {Vennin},\ and\ \citenamefont {Wands}}]{Gow:2022jfb}%
  \BibitemOpen
  \bibfield  {author} {\bibinfo {author} {\bibfnamefont {A.~D.}\ \bibnamefont
  {Gow}}, \bibinfo {author} {\bibfnamefont {H.}~\bibnamefont {Assadullahi}},
  \bibinfo {author} {\bibfnamefont {J.~H.~P.}\ \bibnamefont {Jackson}},
  \bibinfo {author} {\bibfnamefont {K.}~\bibnamefont {Koyama}}, \bibinfo
  {author} {\bibfnamefont {V.}~\bibnamefont {Vennin}}, \ and\ \bibinfo {author}
  {\bibfnamefont {D.}~\bibnamefont {Wands}},\ }\href {\doibase
  10.1209/0295-5075/acd417} {\bibfield  {journal} {\bibinfo  {journal} {EPL}\
  }\textbf {\bibinfo {volume} {142}},\ \bibinfo {pages} {49001} (\bibinfo
  {year} {2023})},\ \Eprint {http://arxiv.org/abs/2211.08348} {arXiv:2211.08348
  [astro-ph.CO]} \BibitemShut {NoStop}%
\bibitem [{\citenamefont {Young}\ \emph {et~al.}(2014)\citenamefont {Young},
  \citenamefont {Byrnes},\ and\ \citenamefont {Sasaki}}]{Young:2014ana}%
  \BibitemOpen
  \bibfield  {author} {\bibinfo {author} {\bibfnamefont {S.}~\bibnamefont
  {Young}}, \bibinfo {author} {\bibfnamefont {C.~T.}\ \bibnamefont {Byrnes}}, \
  and\ \bibinfo {author} {\bibfnamefont {M.}~\bibnamefont {Sasaki}},\ }\href
  {\doibase 10.1088/1475-7516/2014/07/045} {\bibfield  {journal} {\bibinfo
  {journal} {JCAP}\ }\textbf {\bibinfo {volume} {07}},\ \bibinfo {pages} {045}
  (\bibinfo {year} {2014})},\ \Eprint {http://arxiv.org/abs/1405.7023}
  {arXiv:1405.7023 [gr-qc]} \BibitemShut {NoStop}%
\bibitem [{\citenamefont {Young}(2019)}]{Young:2019osy}%
  \BibitemOpen
  \bibfield  {author} {\bibinfo {author} {\bibfnamefont {S.}~\bibnamefont
  {Young}},\ }\href {\doibase 10.1142/S0218271820300025} {\bibfield  {journal}
  {\bibinfo  {journal} {Int. J. Mod. Phys. D}\ }\textbf {\bibinfo {volume}
  {29}},\ \bibinfo {pages} {2030002} (\bibinfo {year} {2019})},\ \Eprint
  {http://arxiv.org/abs/1905.01230} {arXiv:1905.01230 [astro-ph.CO]}
  \BibitemShut {NoStop}%
\bibitem [{\citenamefont {Kopp}\ \emph {et~al.}(2011)\citenamefont {Kopp},
  \citenamefont {Hofmann},\ and\ \citenamefont {Weller}}]{Kopp:2010sh}%
  \BibitemOpen
  \bibfield  {author} {\bibinfo {author} {\bibfnamefont {M.}~\bibnamefont
  {Kopp}}, \bibinfo {author} {\bibfnamefont {S.}~\bibnamefont {Hofmann}}, \
  and\ \bibinfo {author} {\bibfnamefont {J.}~\bibnamefont {Weller}},\ }\href
  {\doibase 10.1103/PhysRevD.83.124025} {\bibfield  {journal} {\bibinfo
  {journal} {Phys. Rev. D}\ }\textbf {\bibinfo {volume} {83}},\ \bibinfo
  {pages} {124025} (\bibinfo {year} {2011})},\ \Eprint
  {http://arxiv.org/abs/1012.4369} {arXiv:1012.4369 [astro-ph.CO]} \BibitemShut
  {NoStop}%
\bibitem [{\citenamefont {Nakama}\ \emph {et~al.}(2017)\citenamefont {Nakama},
  \citenamefont {Silk},\ and\ \citenamefont {Kamionkowski}}]{Nakama:2016gzw}%
  \BibitemOpen
  \bibfield  {author} {\bibinfo {author} {\bibfnamefont {T.}~\bibnamefont
  {Nakama}}, \bibinfo {author} {\bibfnamefont {J.}~\bibnamefont {Silk}}, \ and\
  \bibinfo {author} {\bibfnamefont {M.}~\bibnamefont {Kamionkowski}},\ }\href
  {\doibase 10.1103/PhysRevD.95.043511} {\bibfield  {journal} {\bibinfo
  {journal} {Phys. Rev. D}\ }\textbf {\bibinfo {volume} {95}},\ \bibinfo
  {pages} {043511} (\bibinfo {year} {2017})},\ \Eprint
  {http://arxiv.org/abs/1612.06264} {arXiv:1612.06264 [astro-ph.CO]}
  \BibitemShut {NoStop}%
\bibitem [{\citenamefont {Tanaka}\ and\ \citenamefont
  {Sasaki}(2007)}]{Tanaka:2006zp}%
  \BibitemOpen
  \bibfield  {author} {\bibinfo {author} {\bibfnamefont {Y.}~\bibnamefont
  {Tanaka}}\ and\ \bibinfo {author} {\bibfnamefont {M.}~\bibnamefont
  {Sasaki}},\ }\href {\doibase 10.1143/PTP.117.633} {\bibfield  {journal}
  {\bibinfo  {journal} {Prog. Theor. Phys.}\ }\textbf {\bibinfo {volume}
  {117}},\ \bibinfo {pages} {633} (\bibinfo {year} {2007})},\ \Eprint
  {http://arxiv.org/abs/gr-qc/0612191} {arXiv:gr-qc/0612191} \BibitemShut
  {NoStop}%
\end{thebibliography}%

\end{document}